\documentclass[aps,prd,twocolumn,nofootinbib,superscriptaddress]{revtex4-2}

\usepackage{lmodern}
\usepackage{amsmath}
\usepackage{amssymb}
\usepackage{amsfonts}
\usepackage{amsbsy}
\usepackage{mathrsfs}
\usepackage{latexsym}
\usepackage{bm}
\usepackage{wasysym}
\usepackage{mathbbol}
\usepackage{bigints}
\usepackage{physics}
\usepackage{booktabs}
\allowdisplaybreaks
\usepackage{placeins}

\usepackage{graphicx}
\usepackage{subcaption} 
\usepackage[normalem]{ulem}
\usepackage[dvipsnames]{xcolor}
\usepackage{multirow}
\usepackage{csquotes}
\usepackage{float} 
\usepackage{makecell}
\usepackage{tikz}
\usetikzlibrary{arrows.meta,positioning,calc}

\usepackage[paperwidth=210mm,paperheight=297mm,centering,hmargin=2cm,vmargin=2.6cm]{geometry}

\usepackage[T1]{fontenc}
\usepackage{tgpagella}
\usepackage{natbib}
\usepackage{comment}
\usepackage{tikz-3dplot}

\usepackage{orcidlink} 

\definecolor{napiergreen}{rgb}{0.16, 0.5, 0.0}

\begin{document}
\title{Out-of-time-order Correlators in Volcano potentials}

\author{Aritra Banerjee\,\orcidlink{0000-0002-9619-7682}}\email{aritra.banerjee@pilani.bits-pilani.ac.in}
\affiliation{Department of Physics, Birla Institute of Technology and Science - Pilani, Rajasthan, 333031, India}
\affiliation{Asia Pacific Center for Theoretical Physics, Postech, Pohang 37673, Korea}
\author{Reetik Jangir\,\orcidlink{0009-0007-8734-1332}}\email{p20250036@pilani.bits-pilani.ac.in}
\affiliation{Department of Physics, Birla Institute of Technology and Science - Pilani, Rajasthan, 333031, India}
\author{Rishant Sharma, \orcidlink{0009-0009-8765-8086}}\email{f20220814@pilani.bits-pilani.ac.in }
\affiliation{Department of Physics, Birla Institute of Technology and Science - Pilani, Rajasthan, 333031, India}

\begin{abstract}
   The out-of-time-order correlator (OTOC) has emerged as an important diagnostic for quantum information scrambling and a key signature of quantum chaos. In this work, we discuss out-of-time-order correlators, analytically and numerically, for the so-called Volcano potentials. This family of potentials parametrically interpolates between bounded wells and confining walls, and includes Pöschl-Teller as a specific case. Through a combination of numerical diagonalization and semiclassical analysis, we investigate various regimes of the potential parameter  regimes containing regions of both positive and negative curvatures. Moreover, we demonstrate in this case that short-time exponential growth of the OTOC does not necessarily signal global quantum chaos; instead, it  correlates with the sampling of negative curvature regions by excited states.
\end{abstract}

\maketitle

\section{Introduction}
In recent years, a consensus has emerged across physics communities that the Out-of-Time-Ordered Correlator (OTOC) serves as a widely-used, real-time diagnostic for quantum information scrambling \cite{Rozenbaum:2016mmv,Garcia-Mata:2018slr, Swingle:2018ekw, Gharibyan:2018fax,Lewis-Swan:2018sdr,Xu:2019lhc,Fortes:2019frf,Pilatowsky-Cameo:2019qxt, Hashimoto:2017oit, Bhattacharyya:2019txx, Akutagawa:2020qbj,Romatschke:2024mxr}. Unlike traditional probes such as spectral statistics or level-spacing distributions \cite{Bohigas1984,Haake2019}, which require global knowledge of the energy spectrum, the OTOC—defined formally as the expectation value of the squared commutator, directly quantifies the growth of an initially localized operator under unitary Heisenberg evolution. It effectively quantifies the spatial and operator-space growth of a Heisenberg operator and provides a diagnostic of the quantum butterfly effect: the sensitivity of the subsequent dynamics to a local perturbation.

This can be put in juxtaposition to what happens in classical chaos, where the hallmark is the presence of sensitive dependence on initial conditions, wherein infinitesimally separated trajectories in phase space diverge exponentially in time \cite{LichtenbergLieberman1992, strogatz2018nonlinear}. Translating this geometric picture into the framework of quantum mechanics, however, presents a profound conceptual challenge \cite{MichaelBerry_1989}. The Hilbert-space distance between quantum states remains constant under Hamiltonian evolution, which appears to preclude any direct quantum analogue of classical trajectory divergence \cite{Peres1984}. The dynamics of operators, however, provides a different perspective. While the Schrödinger evolution of state vectors is unitary and preserves overlaps, the commutator algebra of operators evolves non-trivially. This is where the usefulness of OTOCs become apparent, providing a  direct measure of how a local perturbation at time zero affects a measurement at a later time\footnote{As a matter of fact, OTOCs have been experimentally measured in a myriad of setups, including that of trapped ions, see \cite{Garttner:2016mqj} for example.}. 

However, when it comes to discerning (non-)integrability and \textit{global} chaos through OTOCs, there are a few caveats. The most prominent of these (in one-dimensions) is the deceptive simplicity of the inverted harmonic oscillator \cite{Hashimoto:2020xfr}. This system, though perfectly integrable and exactly solvable, serves as the prototype for exponential OTOC growth simply by virtue of its local geometry \cite{Ali:2019zcj,Qu:2021ius, Morita:2021syq}. In fact, whenever in a complicated potential one encounters states straddling such local instabilities, the OTOC can give rise to transient hyperbolic growth, even when the underlying one-dimensional system remains globally integrable, calling for caution in such studies. This local sensitivity is seen both in classical and quantum dynamics, giving a powerful link between the two regimes. Such situations have been discussed, for a few physically interesting potentials  \cite{Morita:2021syq, Li:2023ukv}, but the general crossover between locally unstable and globally regular sectors of a bound state spectrum remains comparatively unexplored.

To demonstrate this effect better, one needs to study systems whose classical phase space contains both locally unstable and stable regions. In this work, we turn to a very important quantum system: the multi-parameter family of Volcano Potentials \cite{Nieto:2000ik,Koley:2006ku}. 
This structure is of particular interest because it naturally contains distributed regions of negative curvature, which, as we will show, give rise to transient exponential sensitivity of the OTOC. The physical parameters of the potential thus tune not only the asymptotic confinement but also the extent and location of these locally unstable regions. Hence, the Volcano potential is an ideal working example for investigating the interplay between local geometry and OTOC-based sensitivity.
In fact, for certain regions of the parameter space, this potential becomes identical to the famous \textit{Pöschl-Teller potential} \cite{Poschl:1933zz,CooperKhareSukhatme1995}, appearing in a wide spectrum of physical situations. This makes our choice even more exciting. 

The Volcano potential, with its tunable negative curvature regions, allows us to compute OTOCs starting from initial bound states and isolate the effects of local instability from those of global non-integrability. We plot the early and late time growth of the OTOC in a parameter space populated by enough number of bound states and show how the onset of exponential sensitivity correlates with the transition from positive to negative curvature. We further investigate both the classical and quantum notion of this localized sensitivity, as the way states sample the curvature of the potential acquires a distinct interpretation in the latter case, due to \textit{Airy corrections} in turning points.

Beyond numerics, we analytically check this interplay of semiclassical and Airy-corrected semiclassical sensitivities via the exactly solvable case of Pöschl-Teller potential. This is a finite attractive well with a discrete bound state sector separated from the continuum. We find out clean expressions for the crossover criterion of states across curvature boundaries for this case. We also work out a rigorous roadmap to calculating matrix elements in the Pöschl-Teller spectrum by projecting onto the bound state manifold for analytic tractability. One can then find a closed, albeit messy, form for the OTOC in the bound state \textit{projected sector}. Although this is not the full OTOC, we explicitly check this against the same continuum-decoupled observable calculated from numerical methods, which agree well with each other.

Finally, the Volcano potential family contains the effective potentials governing field localization in warped extra dimensions \cite{BL0,BL1,BL2,BL3,BL4,BL5,BL6,BL7,BL8,BL9,BL10}\footnote{This list is in no way exhaustive, please see a review like \cite{Maartens:2003tw} for more references and discussions.}, such as those found in the Randall–Sundrum (RS) braneworld \cite{Randall:1999ee}. Volcano potentials in these cases are especially found for localized graviton zero modes where effective dynamics of metric fluctuations reduce to a Schrödinger equation with such a potential. The local depth of such a potential dictates the probability of getting graviton bound states given certain asymptotic conditions. The tails of the graviton wavefunction, stretches into the bulk, and are responsible for sub-leading corrections to Newtonian gravity at short distances. We comment on the equivalent problem of this type in our setup. 

The rest of the paper is organized in the following manner: In Sec.\eqref{sec2} we will introduce the potential and the parameter space we are interested in. In Sec.\eqref{sec3} we will quickly introduce the OTOC computation. In Sec.\eqref{sec4} we elucidate on sensitivity and curvature dependence for early time growth of OTOCs in these potentials. We also discuss how quantum corrections affect the sensitivity of bound states. The analytically tractable case of Pöschl-Teller potential will be discussed in detail in Sec.\eqref{secPT}. We will then discuss about the implications of field localization on Braneworld models in Sec.\eqref{secbrane}. We conclude our deliberation in Sec.\eqref{sec6} with some future outlook. Appendices contain further calculations, including analytical structure of wavefunctions, details of mathematical computation of matrix elements and details on numerical convergence.

\section{Volcano Potential}\label{sec2}

In this work, we consider a family of potentials given by\footnote{In general one can take \begin{equation*}
  V(x)
  =
  -\left(
  c_1 e^{2g(x)}
  +
  c_2 g''(x)
  \right),
\end{equation*}
with the particularly useful choice $g(x)=\nu\ln\cosh x$,
we can transform to the form used in this paper.}
\cite{Koley:2006ku}
\begin{equation}\label{volca}
V(x) = -\left(a_1 \cosh^{2\nu} x + a_2 \text{sech}^2 x \right)\end{equation}

where $a_1$ and $a_2$ are real parameters and $\nu$ controls the asymptotic structure of the potential. The Volcano potential derives its name from its characteristic shape in the confining regime: a deep central well, a barrier-like flank, and steep rising walls — resembling a volcanic crater. The $\text{sech}^2 x$ term generates a smooth attractive well localized around $x=0$ and its competition with $\cosh^{2\nu} x$ term determines the potential behaviour.

If we look at such potentials closely, note that at the origin $V(0)=-(a_1+a_2)$, and for small $x$ we can expand,
\begin{equation}
  V(x)
  =
  -(a_1+a_2)
  +
  (a_2-\nu a_1)x^2
  +O(x^4).
  \label{eq:V-small-x}
\end{equation}
Thus the origin is a local minimum whenever  $a_2>\nu a_1$, assuming $\nu a_1>0$. The nonzero extrema therefore obey the condition
\begin{equation}
  \cosh^{2\nu+2}x_b
  =
  \frac{a_2}{\nu a_1}
\implies
  x_b
  =
  \cosh^{-1}
  \left[
    \left(
      \frac{a_2}{\nu a_1}
    \right)^{1/(2\nu+2)}
  \right],
  \label{eq:xb-explicit}
\end{equation}

which makes the potential extrema of the form $V_b=-(1+\nu)a_1  \left( \frac{a_2}{\nu a_1}\right)^{\nu/(\nu+1)}$. The case $\nu=-1$ is exceptional, since it makes both terms in the potential $\propto\sech^2 x$, and needs separate discussion.

Naturally, for a well supporting many long-lived bound states (i.e. normalizable eigenstate below continuum threshold), one wants the well depth to be large compared with the characteristic level spacing.\footnote{Precisely, via harmonic approximation, the frequency at the bottom of the well would be $\omega_0= \sqrt{\frac{2(a_2-\nu a_1)}{\mu}}$, leading to a bound population below the barrier of $ N_{\rm well} \sim\frac{\Delta V}{\hbar\omega_0}$.} So it appears the best regime where one can probe bound states would be set by the depth/width ratio $a_2/\nu a_1$. But herein, we should also discuss the asymptotic signs of the potential. In this regard, we look at three qualitative regimes of the potential, with distinct features:
\begin{itemize}
    \item \textbf{For $\nu= 0$ :} the hyperbolic cosine term reduces to a constant, and the potential becomes a shifted Pöschl--Teller(PT hereafter) well:
\begin{equation}
V(x) = -a_1 - a_2 \operatorname{sech}^2 x.
\end{equation}
This is a well-studied exactly solvable system with a finite number of bound states with $a_2 > 0$. Note that for pure PT well we have $a_1=0$, i.e. a pure $\sech^2x$ term with $a_2 = \frac{\Lambda(\Lambda+1)}{2}.$
    
    \item \textbf{For $\nu<0$ :} The $\cosh^{2\nu}x$ term decays as $|x| \to \infty$, effectively behaving like a $\operatorname{sech}$-type tail. The potential therefore remains a bounded well, with asymptotic decay to a constant value. This regime retains the qualitative features of the shifted PT case but introduces a modified spectral structure controlled by $\nu$.

    \item \textbf{For $\nu > 0$ :} The $\cosh^{2\nu}x$ term grows exponentially as $|x| \to \infty$. In this case, the asymptotic behaviour of the potential depends critically on the sign of $a_1$. If $a_1 > 0$, the potential becomes unbounded from below as $|x| \to \infty$. Proper self-adjoint realization therefore requires $a_1 < 0$, which flips the asymptotic sign and produces steep confining walls that grow as $|a_1|\cosh^{2\nu}x$. This ensures a discrete, real spectrum and makes the system suitable for studying a large number of bound states.

    \end{itemize}
Hence, to summarize asymptotic behaviour, if we fix $a_2>0$:
\begin{widetext}

\begin{equation}
\boxed{
\begin{array}{c|cc}
 &a_1>0&a_1<0\\\hline
\nu>0&V\to-\infty ~(no~global~bound~states)&V\to+\infty ~(confining)\\
\nu<0&V\to0^-~(finite~bound~well)&V\to0^+~(finite~bound~well)
\end{array}}
\end{equation}
\end{widetext}

\begin{figure*}[htp!]
    \centering
    \includegraphics[width=\textwidth]{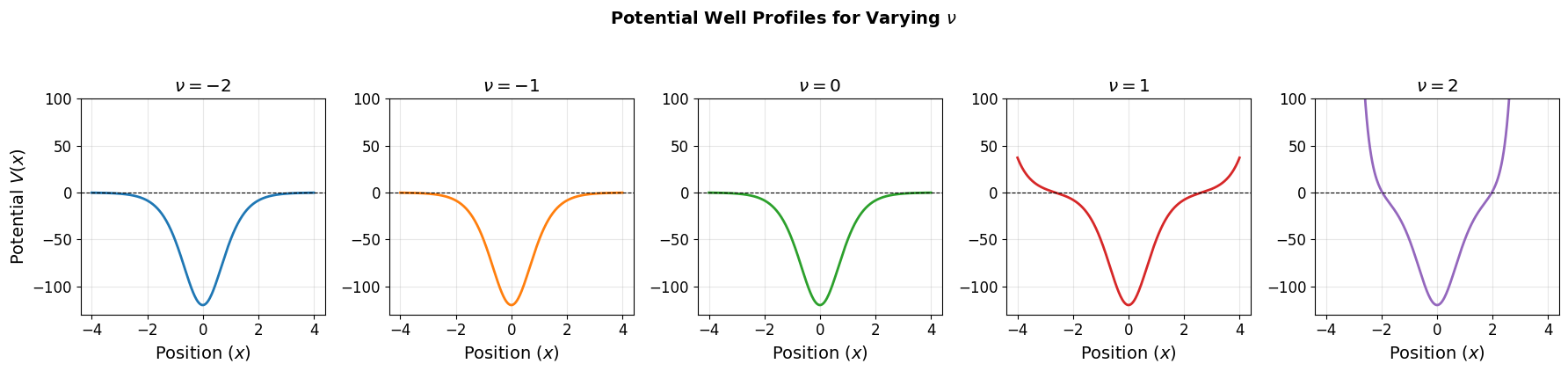} 
    \caption{Volcano potential well profiles $V(x)$ for for various values of $\nu$ with $a_1=-0.05$ and $a_2=120$. The first three are variations on the PT potential and the succession can be written as $\text{deformed PT}\ \longrightarrow\ \text{PT}\ \longrightarrow\ \text{shifted PT}\ \longrightarrow\ \text{soft confinement}\ \longrightarrow\ \text{hard confinement}.$ Note that since $a_1$ is small, the variants of PT have same eigenfunctions (up to energy shift/perturbative corrections).}
    \label{fig:potentials}
\end{figure*}

    One can see FIG. \eqref{fig:potentials} for a qualitative idea of how the potential behaves at various values of $\nu$. In this study, our primary focus is the behaviour of the OTOC of Volcano Potential and its dependence on $\nu$. To isolate this effect, we would like to go from open-localized well (like PT) to confined walls in what follows. Thus,  we fix $a_1$ and $a_2$ in the beginning and only change the values of $\nu$, which we choose to be $\{-2,-1,0,1,2\}$ for our discussions. Crucially, we must select a negative value for $a_1$ to prevent the potential from diverging to negative infinity for positive values of $\nu$, thereby avoiding the non-Hermiticity issues. 
    
    Furthermore, a rigorous analysis of OTOC dynamics requires that a sufficient number of bound states exist\footnote{There are, however, isolated bound states in other regions of parameter space (even in the runaway case) that can be found analytically, see Appendix.\eqref{Appendix_D} for a discussion involving different regimes. }. Therefore, the chosen values of $a_1$ and $a_2$ must correspond to a potential having a sufficient number of bound states. One such pair is $a_1 = -0.05$, $a_2 = 120$, which we will be using here, unless otherwise specified. Note that $|a_1|\ll|a_2|$ here which makes sure the condition $a_2\gg |\nu a_1|$ is satisfied for sufficiently small $\nu\in \mathbb{Z}$, and also makes the shifts away from PT for $\nu<0$ case rather small.  Additionally, not all parameter spaces exhibit a negative curvature region for considered values of $\nu$, a feature essential for our subsequent analysis. This provides further support to our choice of parameter space, which exhibits both negative curvature regions along with bound states supported within that region. A summary for suitable values in parameter space is given in tabular form in Appendix.\eqref{AppA} for the clarity of the reader.

\section{The setup}\label{sec3}
\subsection{OTOC}
The OTOC, as discussed before, has emerged as a powerful diagnostic tool for characterizing quantum information scrambling and distinguishing between integrable and chaotic quantum systems. It quantifies the sensitivity of quantum dynamics to small perturbations, playing a role analogous to the Lyapunov exponent in classical chaos \cite{Shenker:2014cwa,Bohrdt:2016vhv,shen2017out}.

For a quantum system with a Hamiltonian $\hat{H}(x,p)$ prepared in an energy eigenstate $|n\rangle$, the OTOC intensity $C_n(t)$ is defined as the expectation value of the squared commutator of two operators \cite{Hashimoto:2017oit}, typically position and momentum operators :
\begin{equation}C_n(t) = \langle n| [\hat{x}(t), \hat{p}(0)]^\dagger [\hat{x}(t), \hat{p}(0)]|n \rangle\end{equation}

where the time evolution in the Heisenberg picture is
\begin{equation}
\hat{x}(t) = U^\dagger(t) \hat{x}(0) U(t), \quad U(t) = e^{-i\hat{H}t/\hbar}
\end{equation}
This observable is more like a time dependent four-point correlator.
Chaotic systems are characterized by an sustained early-time exponential growth of the OTOC ($C(t) \sim e^{\lambda t}$), whereas integrable systems show polynomial growth or bounded oscillatory behaviour $\sim \cos^2(\omega t)$ \cite{Hashimoto:2017oit, Morita:2021syq, Ali:2019zcj}.

Now, Expanding in the energy eigenbasis $\{|m\rangle\}$ of the Hamiltonian $\hat{H}|m\rangle = E_m |m\rangle$, the time evolution becomes 
\begin{equation}\hat{x}(t) = \sum_{m,k} e^{i(E_m - E_k)t/\hbar} x_{mk} |m\rangle \langle k|\end{equation}
where $x_{mk} = \langle m | \hat{x}(0) | k \rangle$. The OTOC can therefore be expressed in terms of matrix elements and energy differences,
\begin{widetext}
\begin{equation}
C_n(t) = \sum_k \left| \sum_m \left(e^{i(E_k -E_m)t/\hbar} x_{km} p_{mn} 
- e^{i(E_m-E_n)t/\hbar} p_{km} x_{mn} \right) \right|^2.
\end{equation}
\end{widetext}

This microcanonical representation makes explicit the role of the energy spectrum and operator matrix elements in determining the temporal structure of the OTOC. We will focus on computing this quantity, and will not discuss, for example, OTOC for systems coupled to a heat bath \cite{Hashimoto:2017oit}.

For our analysis, the Hamiltonian has the usual structure of $H = p^2/2\mu+V(x)$, where the potential is given by the generic Volcano potential discussed before. We will now evaluate the OTOC numerically for the volcano potential for aforementioned values of the parameter $\nu$, which controls the asymptotic structure of the system. This allows us to investigate how the spectral structure and asymptotic behaviour influence operator growth and information scrambling.

\subsection{Implementation}
Numerically, the calculation is performed within a Discrete Variable Representation (DVR) framework \cite{harris1965calculation}. The spatial domain is discretised into $N$ grid points over a sufficiently large interval $[-L,L]$ to ensure convergence. For the present study, we use $L=10$ and $N=1000$ for all numerical calculations, unless stated otherwise. The Hamiltonian $\hat H = \hat T + \hat V$ is constructed as an $N \times N$ matrix, where the kinetic term is approximated using a finite-difference scheme and the potential term is diagonal in position space. The resulting Hamiltonian matrix is exact-diagonalized to obtain full energy spectrum $E_n$ and eigenvectors $|n\rangle$. Time evolution is implemented directly in the energy eigenbasis. Rather than computationally constructing full dense matrix for Heisenberg-evolved operator $\hat{x}(t)$ at each time step, the position and momentum operators are transformed into the energy basis, and the time evolution $U(t) = e^{-i\hat{H}t/\hbar}$ is applied via complex phase arrays. This allows the action of the commutator on the target eigenstate, \begin{equation}|\psi_c(t)\rangle = [\hat{x}(t), \hat{p}(0)]|n\rangle\end{equation} to be computed using optimized, vectorized matrix-vector multiplication.
Finally $C_n(t)$
is computed for different eigenstates and parameter values of $\nu$. The full time dependence of $C_n(t)$ is thus obtained, enabling a systematic comparison of operator growth across the various dynamical regimes of the Volcano potential. 

Note that the finite-grid discretization does not distinguish between bound and continuum states intrinsically — it simply returns the eigenvalues of a finite matrix. While bound states are square integrable and box-size independent, discretizing states in the continuum is subtle. The physical continuum
is converted by the finite box into a discrete set of normalizable
standing-wave states. Their level spacing and normalization depend directly on $L$, and as $L$ increases, the spacing decreases, providing an increasingly dense discretization of the continuum. 

The cases of the Volcano family we consider in this work are quite distinct in this regard. For $\nu<0$, where $V(x)\to0$, a physical bound state is
identified by $E<0$ with dissociation to continuum above $E=0$. For $\nu=0$, where
$V(x)\to-a_1$, the continuum threshold is just shifted. For the confining cases $\nu>0$ with $a_1<0$, the potential diverges
to $+\infty$ and the physical spectrum is purely discrete; here the role of the finite box is only to approximate the infinite spatial domain, and convergence requires that the eigenfunctions under study be exponentially small near $x=\pm L$.

The distinction is particularly important for the OTOC. Even when
the initial state $|n\rangle$ is a true bound state, the full
spectral representation of the OTOC contains intermediate states from both the bound and continuum
sectors, because neither $x$ nor $p$ preserves the bound state
subspace on their own. In a finite box, continuum integrals are therefore replaced
by sums over the box-discretized standing-wave states. Individual
continuum eigenvalues and matrix elements become box dependent, but the
combined contribution to a well-converged observable should approach an $L$-independent result as the continuum discretization becomes
denser.\footnote{Note that the individual discretized continuum levels are not expected to
converge as physical eigenstates; it is the continuum contribution to
the complete observable that should become approximately independent of the
numerical box.}. The reader is directed to the Appendix.\eqref{AppConv} for more comments on the convergence, using explicit potentials. As a further numerical consistency check, we verified that the initial-time OTOC for the full spectrum in all of the cases satisfies $C_n(0) \approx \hbar^2$.

\begin{figure*}[htbp]
    \centering
   
    \begin{subfigure}{0.485\textwidth}
        \includegraphics[width=\linewidth]{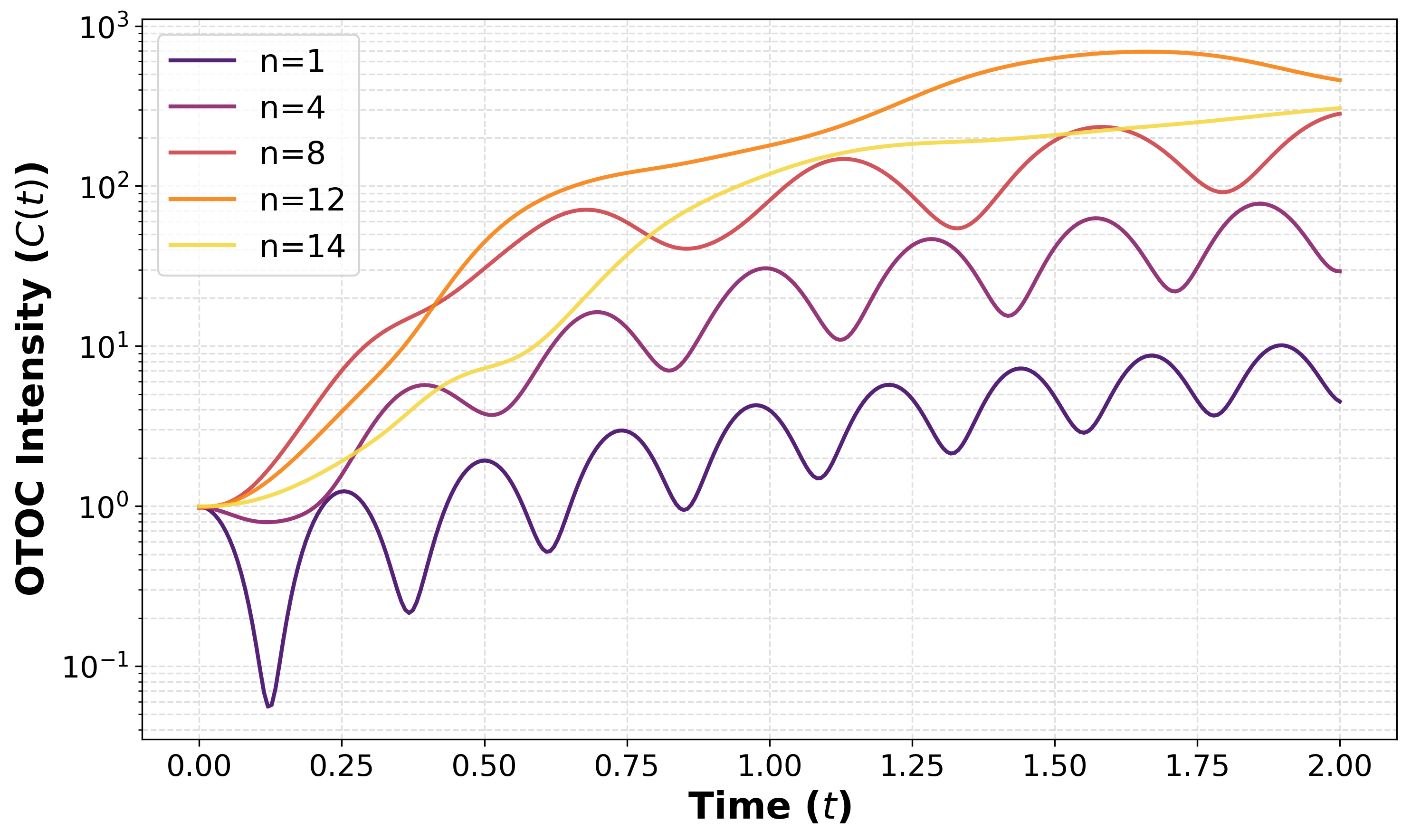}
        \caption{$\nu = -2$}
    \end{subfigure}\hfill
    \begin{subfigure}{0.485\textwidth}
       
        \includegraphics[width=\linewidth]{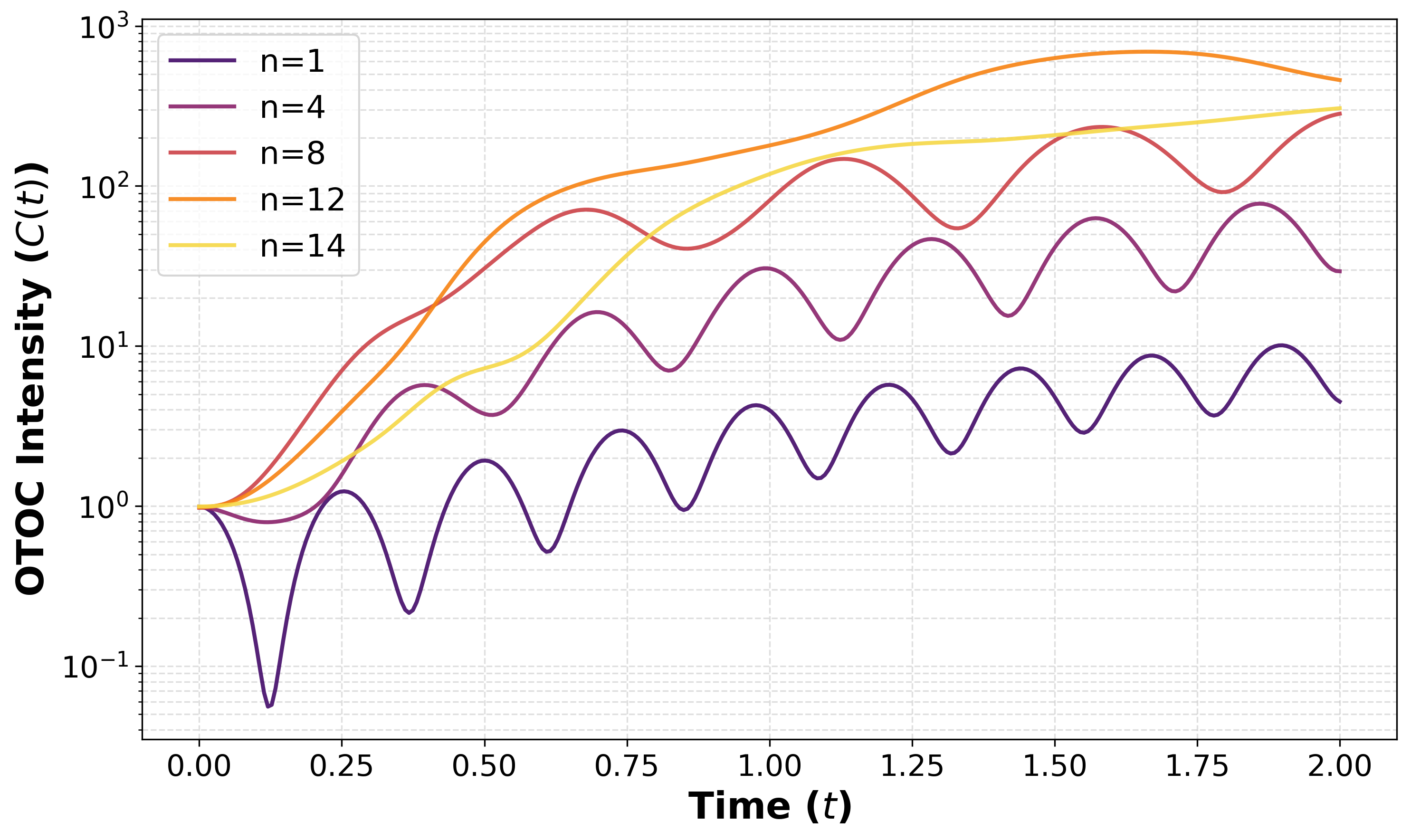}
                     
        \caption{$\nu = -1$}
    \end{subfigure}
    
    \vspace{0.1cm} 

    \begin{subfigure}{0.485\textwidth}
      
        \includegraphics[width=\linewidth]{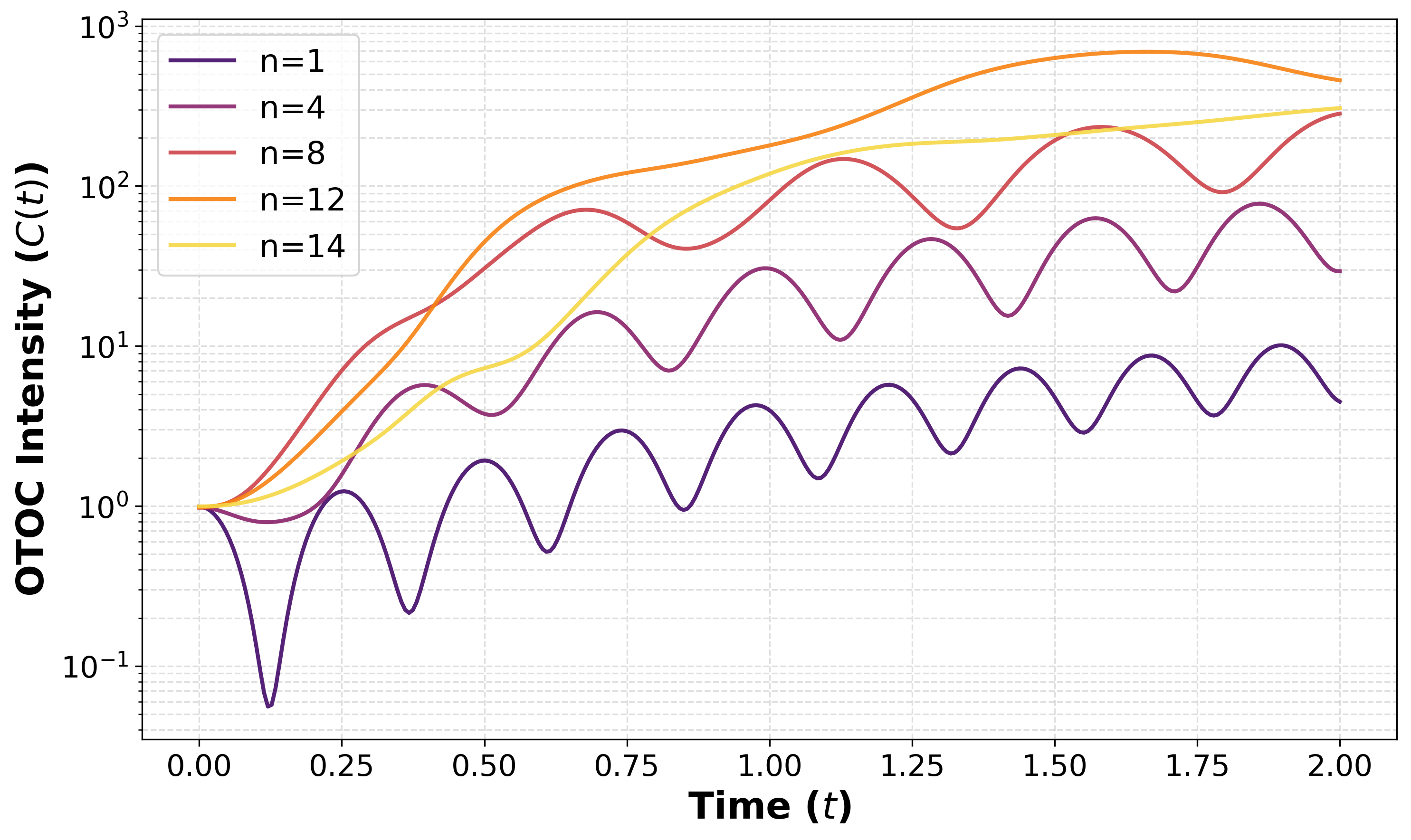}
                   
        \caption{$\nu = 0$}
    \end{subfigure}\hfill
    \begin{subfigure}{0.485\textwidth}
       
        \includegraphics[width=\linewidth]{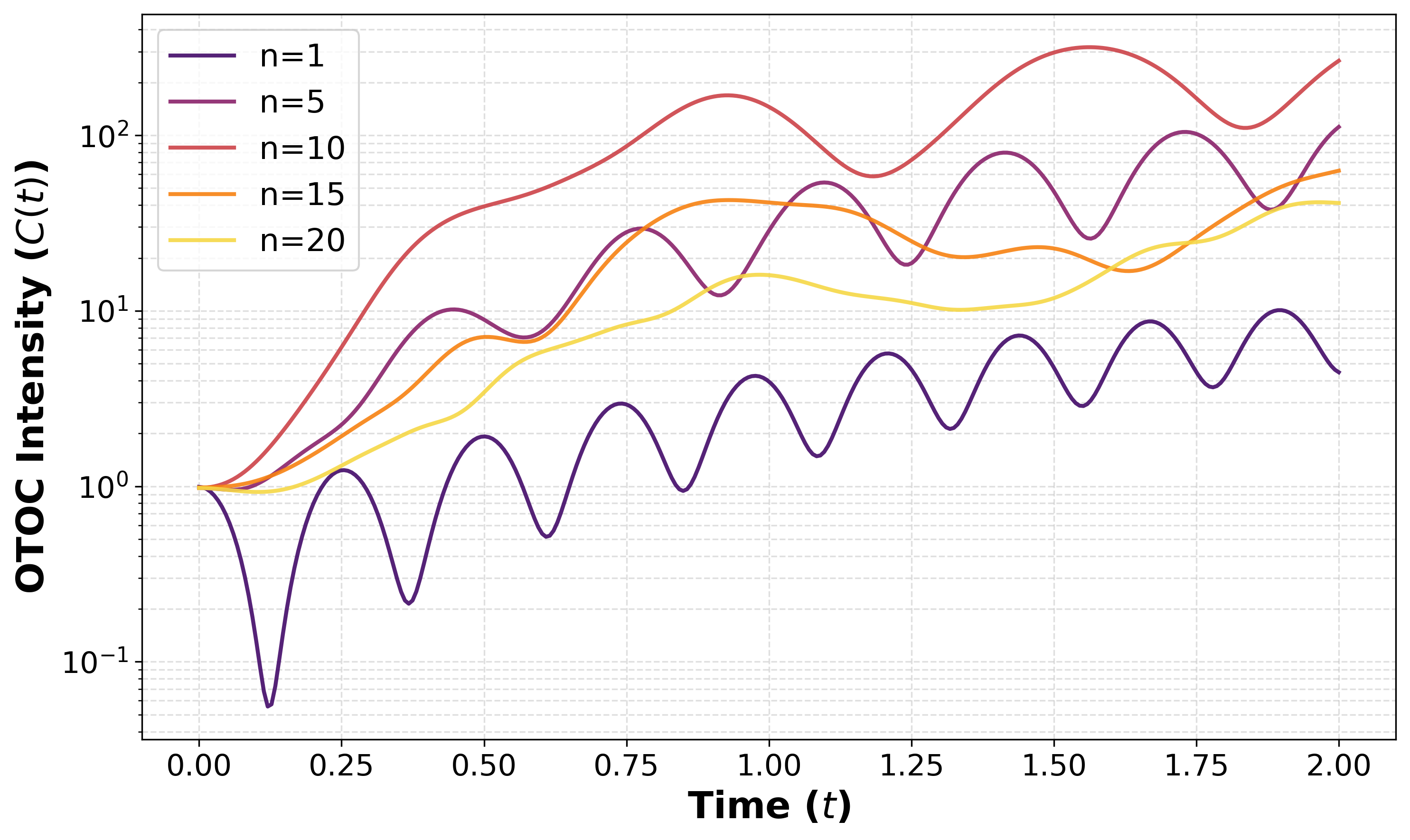}

        \caption{$\nu = 1$}
    \end{subfigure}
    
    \vspace{0.1cm}

    \begin{subfigure}{0.485\textwidth}
        
        \includegraphics[width=1\linewidth]{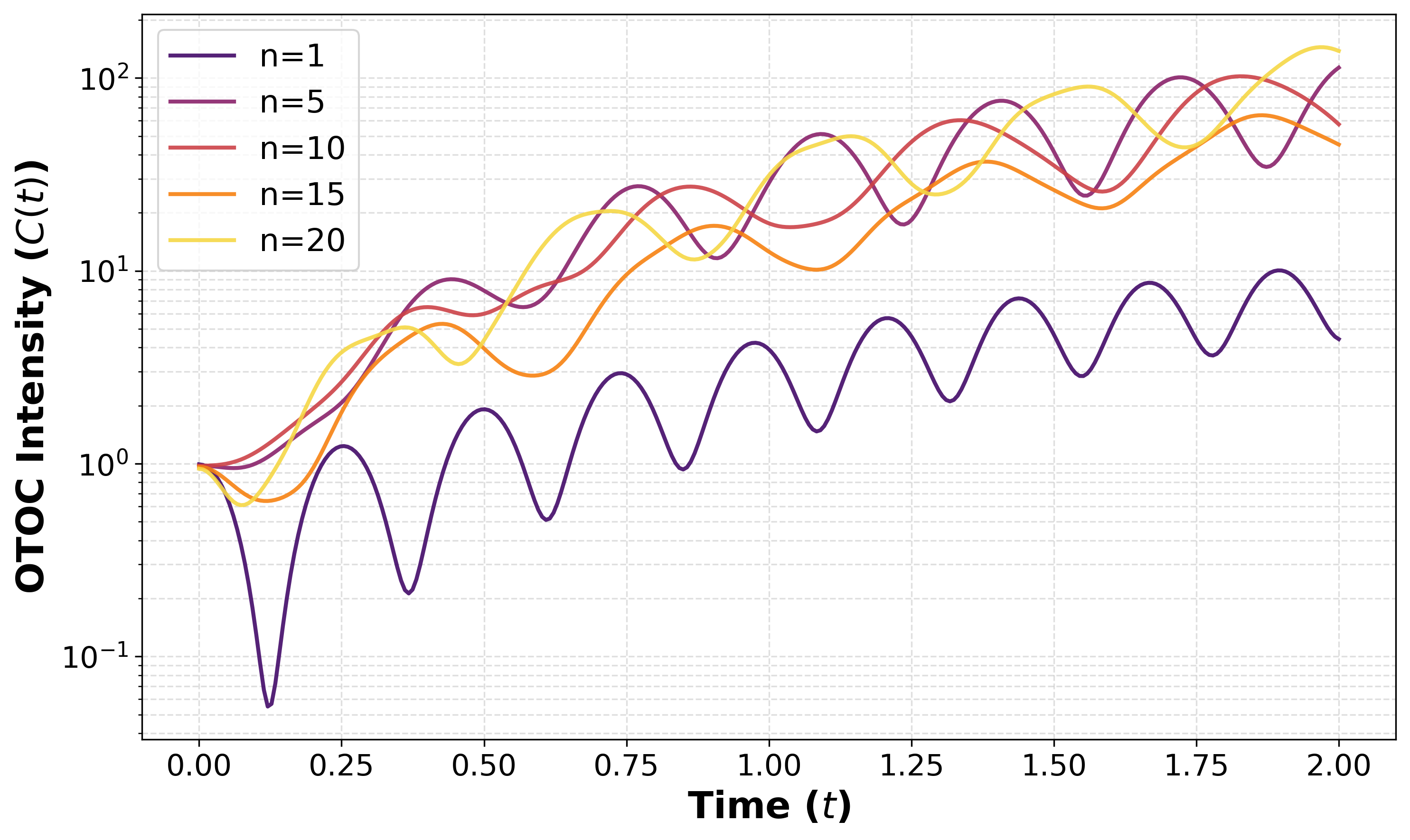}

        \caption{$\nu = 2$}
    \end{subfigure}

    \caption{Short-time exponential-like growth and subsequent oscillation of the OTOC ($C_n(t)$) in logarithmic scale for various values of $\nu$ with $a_1=-0.05$ and $a_2=120$. Note that higher states tend to start out with more rapid irregular growth and then slowly stabilize. Oscillations for higher states also dephase more sharply.}
    \label{fig:otoc_short_part1}
\end{figure*}

\begin{figure*}[htbp]
    \centering
   
    \begin{subfigure}{0.485\textwidth}
    
        \includegraphics[width=1\linewidth]{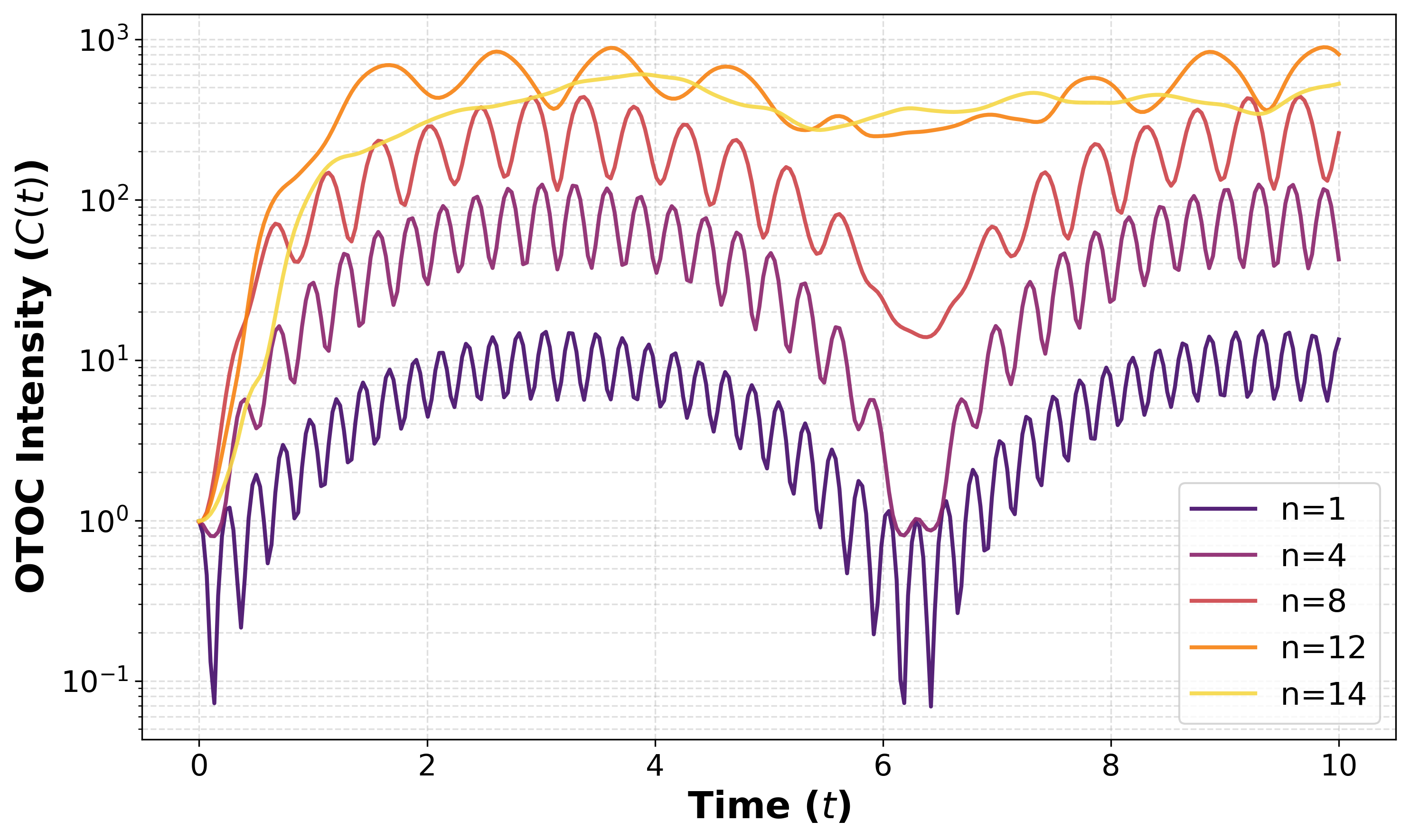}
        
        \caption{$\nu = -2$}
    \end{subfigure}\hfill
    \begin{subfigure}{0.485\textwidth}
       
        \includegraphics[width=1\linewidth]{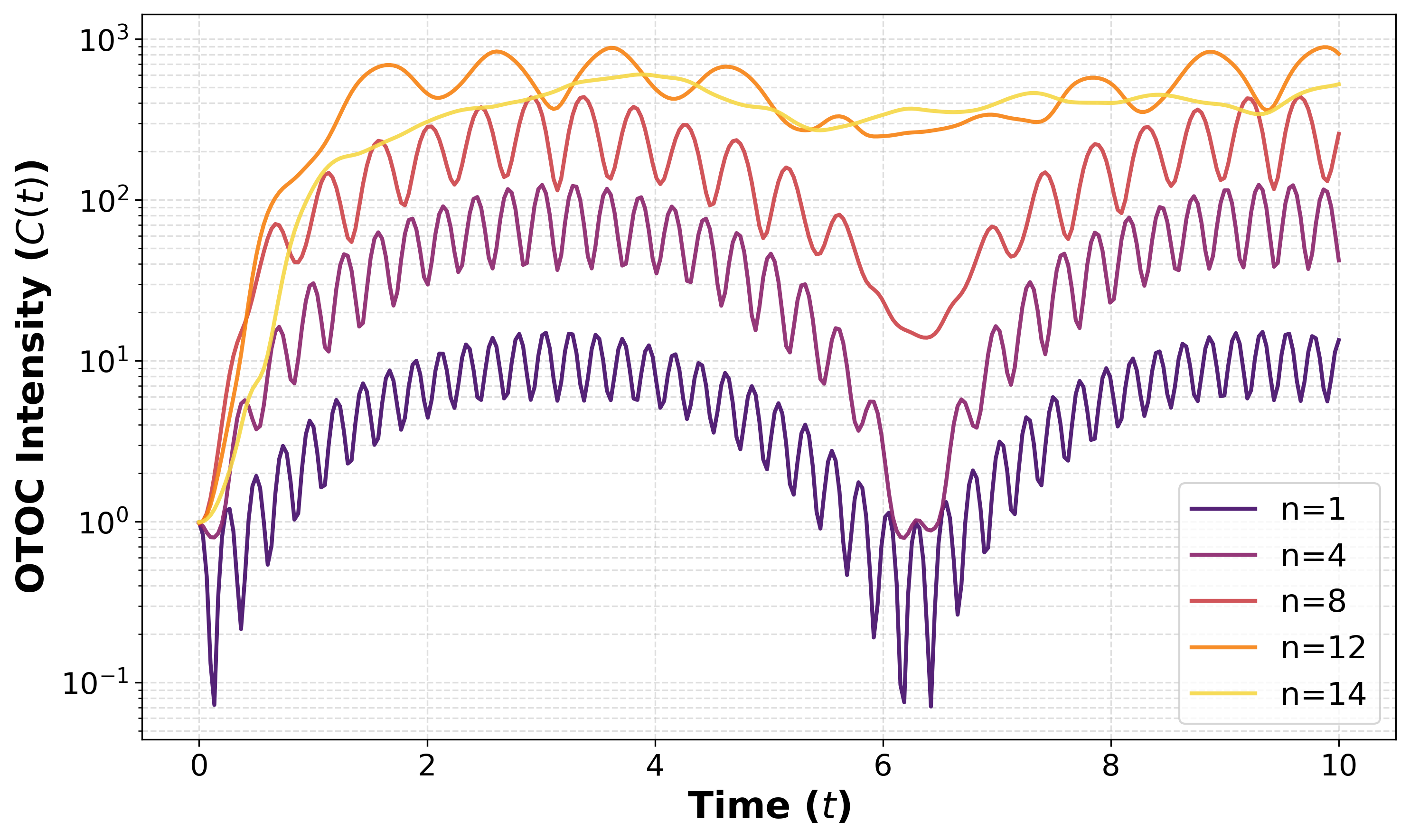}
   
        \caption{$\nu = -1$}
    \end{subfigure}
    
    \vspace{0.1cm} 

    \begin{subfigure}{0.485\textwidth}
       
        \includegraphics[width=1\linewidth]{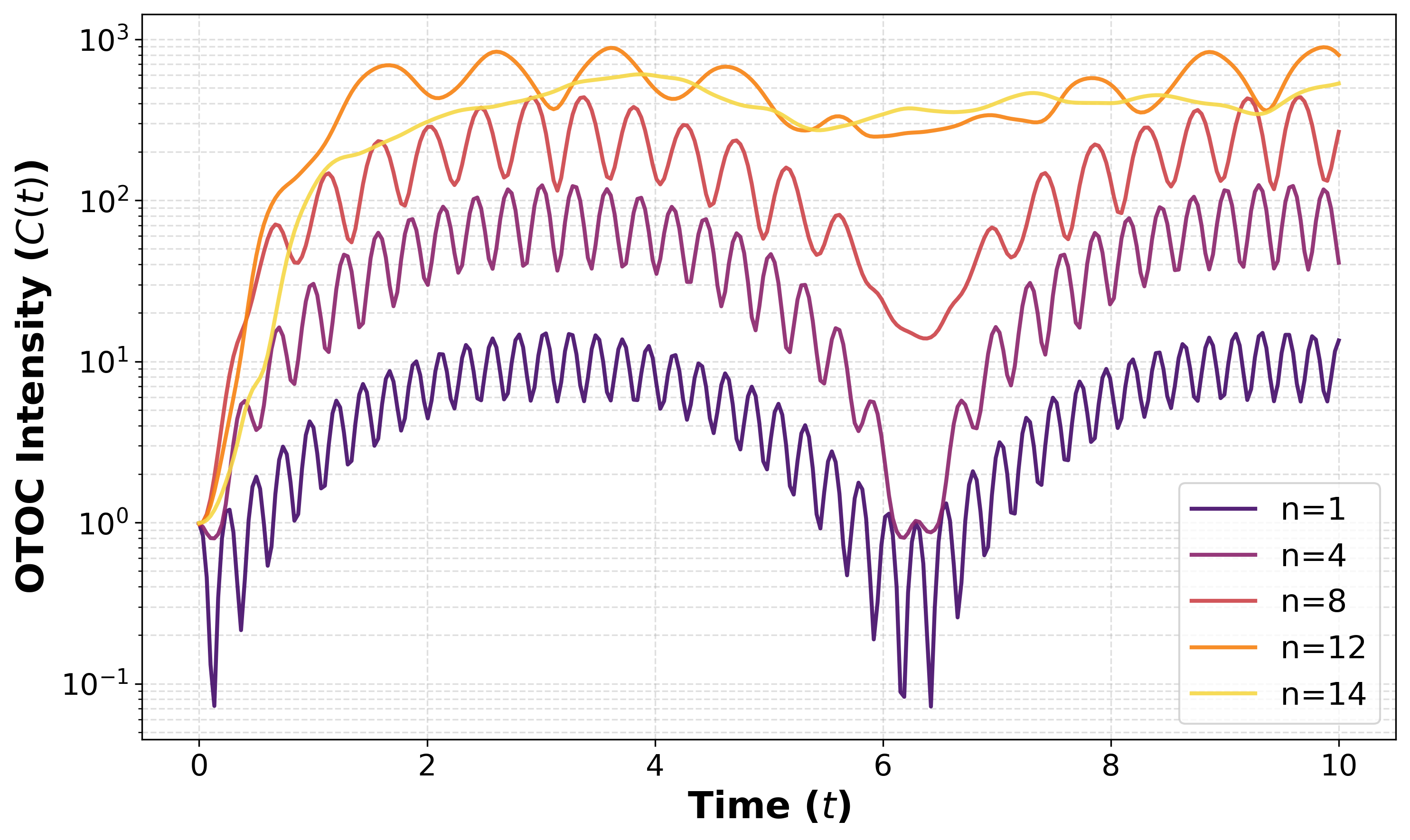}
       
        \caption{$\nu = 0$}
    \end{subfigure}\hfill
    \begin{subfigure}{0.485\textwidth}
        
           \includegraphics[width=1\linewidth]{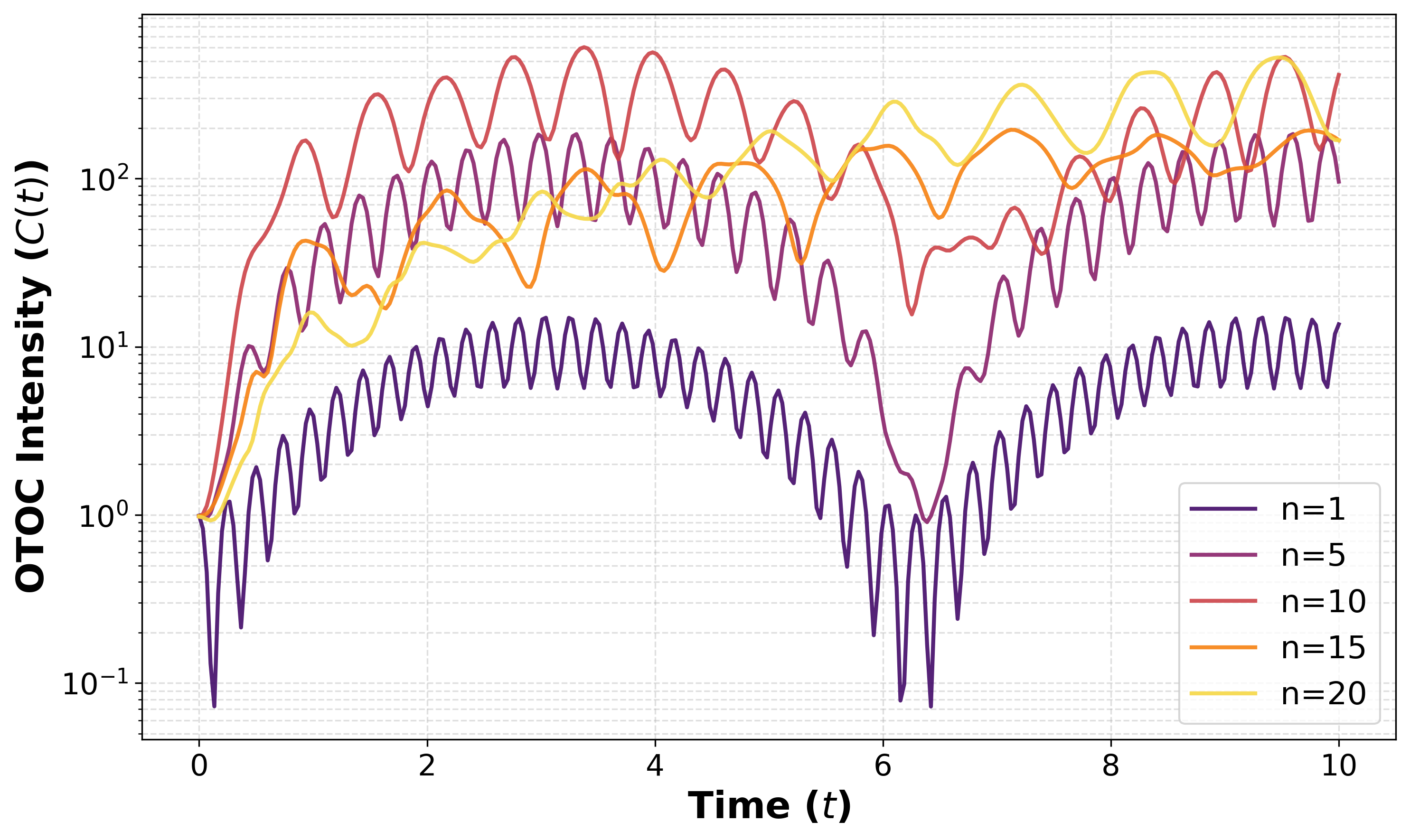}
           
        \caption{$\nu = 1$}
    \end{subfigure}
    
    \vspace{0.1cm}

    \begin{subfigure}{0.485\textwidth}
        
        \includegraphics[width=1\linewidth]{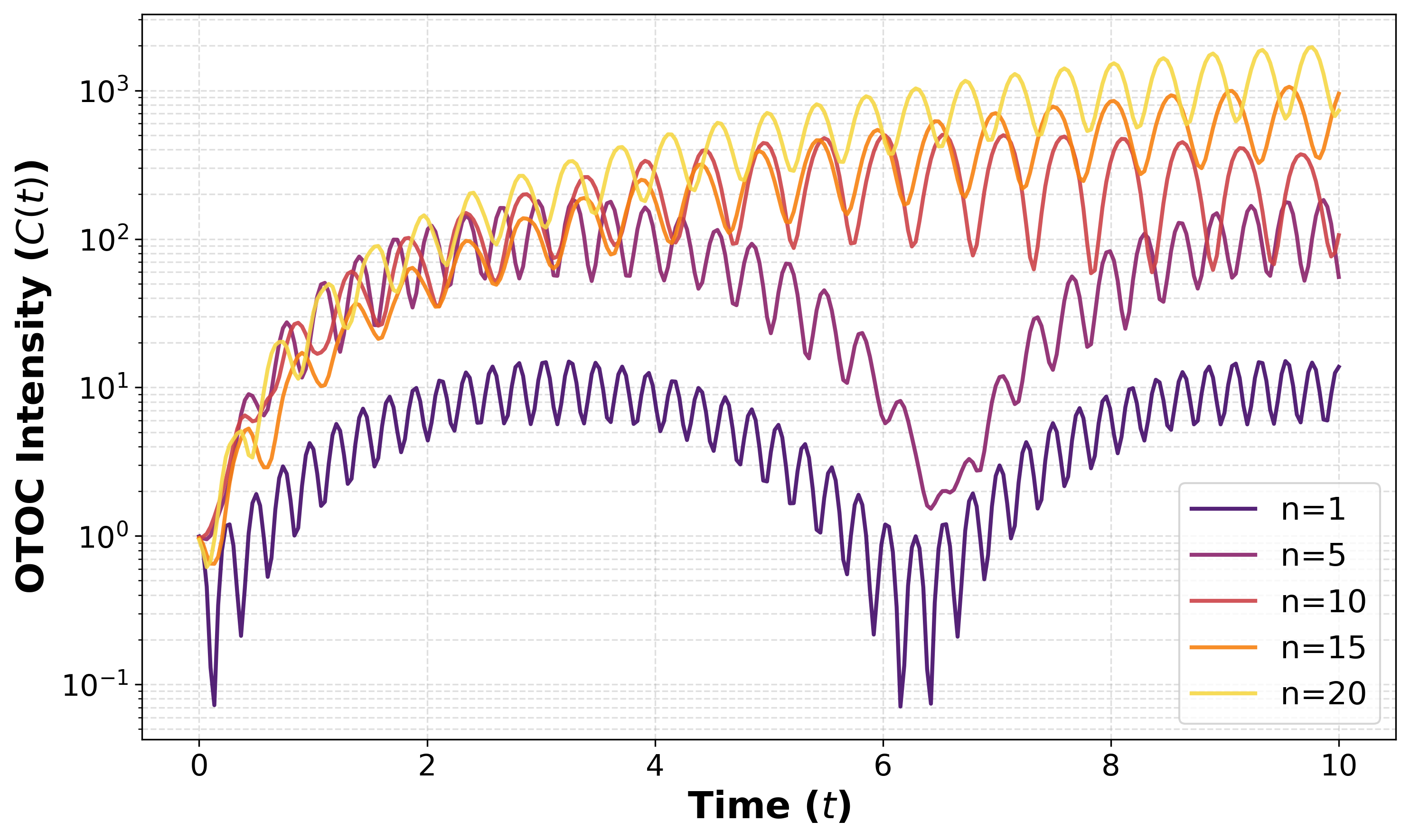}
       \caption{$\nu = 2$}
    \end{subfigure}

    \caption{Long-time growth and saturation of the OTOC ($C_n(t)$) in logarithmic scale for various values of $\nu$ with $a_1=-0.05$ and $a_2=120$. The temporal region of rapid non-oscillatory growth increases for higher states. But definitely some higher states show more oscillatory growth than others.}
    \label{fig:otoc_long_part1}
\end{figure*}

With the technical details explained, we go forward to numerically plot the OTOC for different $\nu$. Note again we will be considering only bound initial states across the $\nu$ parameter space, and the number of approximate bound states can be given by the semiclassical counting function\footnote{For PT and shifted PT case one can also use the exact energy formula $E_n \sim -(\Lambda-n)^2$.}. In the case of $\nu\in\{-2,-1,0\}$ the number can be found to be around $N_{bound}\sim 15$ \footnote{In numerics and plots, we will always consistently use units of $\hbar=\mu=1$, unless stated otherwise. Analytical calculations in the main text will contain $\hbar,\mu$, unless otherwise specified. }, while for the confining cases $\nu \in \{1,2\}$ there are effectively infinite number of bound states. The early time growth for all these cases (see FIG.\eqref{fig:otoc_short_part1}) clearly shows oscillatory or short bursts of exponential sensitivity, the latter especially pronounced for higher bound eigenstates. These persist for late time growth 
(see FIG.\eqref{fig:otoc_long_part1}), but the overall structure looks more or less fixed by an oscillatory envelope. 

Further, while the lowest-lying states are predominantly characterized by regular oscillatory behaviour, the dynamics becomes varyingly sensitive to the local instability as one moves up in the spectrum.  Thus, the OTOC provides a clear indication that the local instability is not uniformly imprinted across the spectrum: its dynamical signature varies for states with increasing excitation energy. In what follows, we will delve more into the physics of these curves.

\section{Curvature Dependence of OTOC}\label{sec4}
As discussed before, we test whether the observed short-time growth is associated with local negative curvature. For any smooth potential $V(x)$, the local dynamical behaviour is governed by the sign of its curvature $V''(x)$. The region of positive curvature locally resembles a stable harmonic oscillator, while the region of negative curvature locally resembles an unstable inverted oscillator. A smooth potential such as the Volcano provides a natural interpolation between these two regimes.

In this locally unstable negative curvature region, A small displacement from a reference trajectory does not experience a restoring force; instead, the force drives the (classical) particle further away. Consequently, nearby classical trajectories separate in time, leading to local sensitivity to initial conditions. However, because the negative curvature region is spatially finite in smooth confining potentials, the instability persists only for a limited time window. The growth eventually saturates once the dynamics leaves the unstable region. 

This is commensurately reflected in the quantum mechanical case. In this locally unstable negative curvature region, the Heisenberg
operator commutator develops hyperbolic time dependence. Consequently, quantum states with significant amplitude in this region can inherit the local exponential instability of the underlying classical dynamics.

However, there is some important distinction here. For a confining potential, the vanishing velocity of the particle significantly enhances the dwell-time density near a regular turning point. In quantum mechanics, the wavefunction does not peak at the classical turning point. The potential is nearly flat in the sense that the naive WKB approximation breaks down and the correct local approximation is linear, not the quadratic one. With this linear potential, the time-independent Schrödinger equation becomes an Airy equation and the outermost antinode of the wavefunction peaks slightly inside the classically allowed region.

Additionally, quantum effects can impose a further temporal restriction through the Ehrenfest time \cite{Emerson2001, Shepelyansky2020}, which characterizes the timescale over which quantum dynamics remains well approximated by its classical or semiclassical counterpart. Beyond this timescale, quantum interference and wave-packet spreading can invalidate the semiclassical description, thereby limiting the regime over which exponential sensitivity governed by classical instability can persist.

Let us discuss this effect in detail for the Volcano Potential, where instability can be analyzed via its curvature,
\begin{align}
V''(x) = &-a_1 \left[ 4\nu^2 \cosh^{2\nu}x - 2\nu(2\nu - 1)\cosh^{2\nu - 2}x \right]\nonumber\\
&- a_2 \left[ 4\,\text{sech}^2 x - 6\,\text{sech}^4 x \right]
\end{align}

which can take various forms depending on the value of $\nu, a_1,a_2$\footnote{For a brief discussion of the case of varying $a_{1,2}$, keeping $\nu$ fixed, readers are directed to Appendix.\eqref{Appendix_B}.}. Since we focus on potentials with a attractive well (with or wothout a continuum), the low-lying eigenstates are concentrated near the central minimum and predominantly sample the positive-curvature region. Their probability distributions are approximately Gaussian near the minimum and are associated with locally stable oscillatory dynamics. As the excitation energy increases, the wavefunctions extend farther into the potential and acquire increasing support in the negative curvature region. In what follows, we will quickly analyze how straddling negative curvature generates exponential sensitivity, first at the classical level and then in quantum mechanics. Finally, we will also incorporate the Airy correction near the turning points, refining the semiclassical estimate.

\subsection{Introducing Sensitivity Through Classical Dynamics}
To understand the origin of the local instability, we analyze the dynamics in the neighbourhood of the outer classical turning point $x_c$, implicitly defined by $E = V(x_c)$, Kinetic Energy = 0. The local geometry of the potential around $x_c$ controls the dominant classical dynamical behaviour. Expanding the potential in a Taylor series around $x_c$ (upto second order),
\begin{equation}V(x) \approx V(x_c) + V'(x_c)(x-x_c) + \frac12 V''(x_c)(x-x_c)^2 \end{equation}

we eliminate the linear term, by shifting the coordinate origin by defining $x_d = x_c - \frac{V'(x_c)}{V''(x_c)}$, so that,
\begin{equation}V(x) \approx V(x_c)-\frac{V'(x_c)^2}{2 V''(x_c)} +\frac{1}{2}V''(x_c)(x-x_d)^2. \end{equation}

Dropping irrelevant constants, the potential can be approximated quadratically $V(x) \approx \frac12 V''(x_c)(x-x_d)^2$. In this approximation, the dynamics reduces to either a harmonic oscillator or an inverted harmonic oscillator depending on the sign of $V''(x_c)$.
For pure classical motion we would have:
\begin{equation}\mu \ddot{x} = -\frac{dV}{dx} = - V''(x_c)(x-x_d)\end{equation}
Putting \(q = x - x_d\), the equation becomes:
\begin{equation} \mu \ddot{q} + V''(x_c) q = 0\end{equation}

For initial conditions $x(0) = x_0$ and $p(0) = p_0$, the solution now depends on the sign of $V''(x_c)$.

\subsubsection*{\textbf{Case 1: Positive Curvature (\texorpdfstring{$V''(x_c)>0$}{V''(x_c)>0}) }}
In this case the equation of motion becomes $\ddot{q} + \omega^2 q = 0$ where $\quad \omega = \sqrt {\frac{V''(x_c)}{\mu}}$, and its solution is,
\begin{equation}q(t)=(x_0 - x_d)\cos(\omega t)
+ \frac{p_0}{\mu \omega}\sin(\omega t) \end{equation}

since $q=x-x_d$:
\begin{equation}x(t)=x_d + (x_0 - x_d)\cos(\omega t) + \frac{p_0}{\sqrt{\mu V''(x_c)}}\sin(\omega t) \end{equation}

This motion is bounded and oscillatory, with no exponential divergence between nearby trajectories.

\subsubsection*{\textbf{Case 2: Negative Curvature   (\texorpdfstring{$V''(x_c)<0$}{V''(x_c)<0}) }}
For this case, we have: $\ddot{q} - \Omega^2 q = 0$ where  $\Omega = \sqrt{\frac{-V''(x_c)}{\mu}}$ and its solution
\begin{equation}x(t)=x_d + (x_0 - x_d)\cosh(\Omega t) + \frac{p_0}{\sqrt{-\mu V''(x_c)}}\sinh(\Omega t). \end{equation}

For large $t$, the dominant term grows as $\sim e^{\Omega t}$, implying classical sensitivity to initial conditions. Thus, the local growth rate which plays the role of a local (but only transiently defined) classical Lyapunov exponent is
\begin{equation} \lambda_c = \sqrt{\frac{|V''(x_c)|}{\mu}}. \end{equation}

This transient instability, a purely \emph{local} effect, persists only while the trajectory remains inside the negative curvature region.\footnote{A nice way to probe this would be to define $\frac{\partial x(t;x_0,p_0)}{\partial x_0}$ as the classical sensitivity parameter for a trajectory, for which the OTOC in the semiclassical limit can be written in terms of the Poisson brackets, i.e. $[x(t),p(0)]
\longrightarrow
i\hbar\{x(t),p(0)\}_{\rm P.B.}
=
i\hbar\frac{\partial x(t)}{\partial x_0},$ culminating in
\begin{equation}
     C(t)
  = \hbar^2\left( \frac{\partial x(t;x_0,p_0)}{\partial x_0}\right)^2.
\end{equation}
This sensitivity could be of a polynomial kind in time as well, especially when the classical trajectory escapes to infinity. A concrete example involving the runaway Volcano potential can be found in Appendix.\eqref{AppE}.}
In this regard, see FIG.\eqref{fig:classical_turnning_curvature} for the curvature associated to Volcano potential in the parameter space of our interest. Note that for $\nu=-2,-1,0$, negative
curvature occurs for all $x>x_L\approx0.658$, a semi-infinite interval on the real line;
for $\nu=1$, it is confined to the interval $0.658<x<2.63$; and for $\nu=2$, it narrows further to $0.665<x<1.709$ as the walls become steeper.

\begin{figure}
    \centering
    \includegraphics[width=1\linewidth]{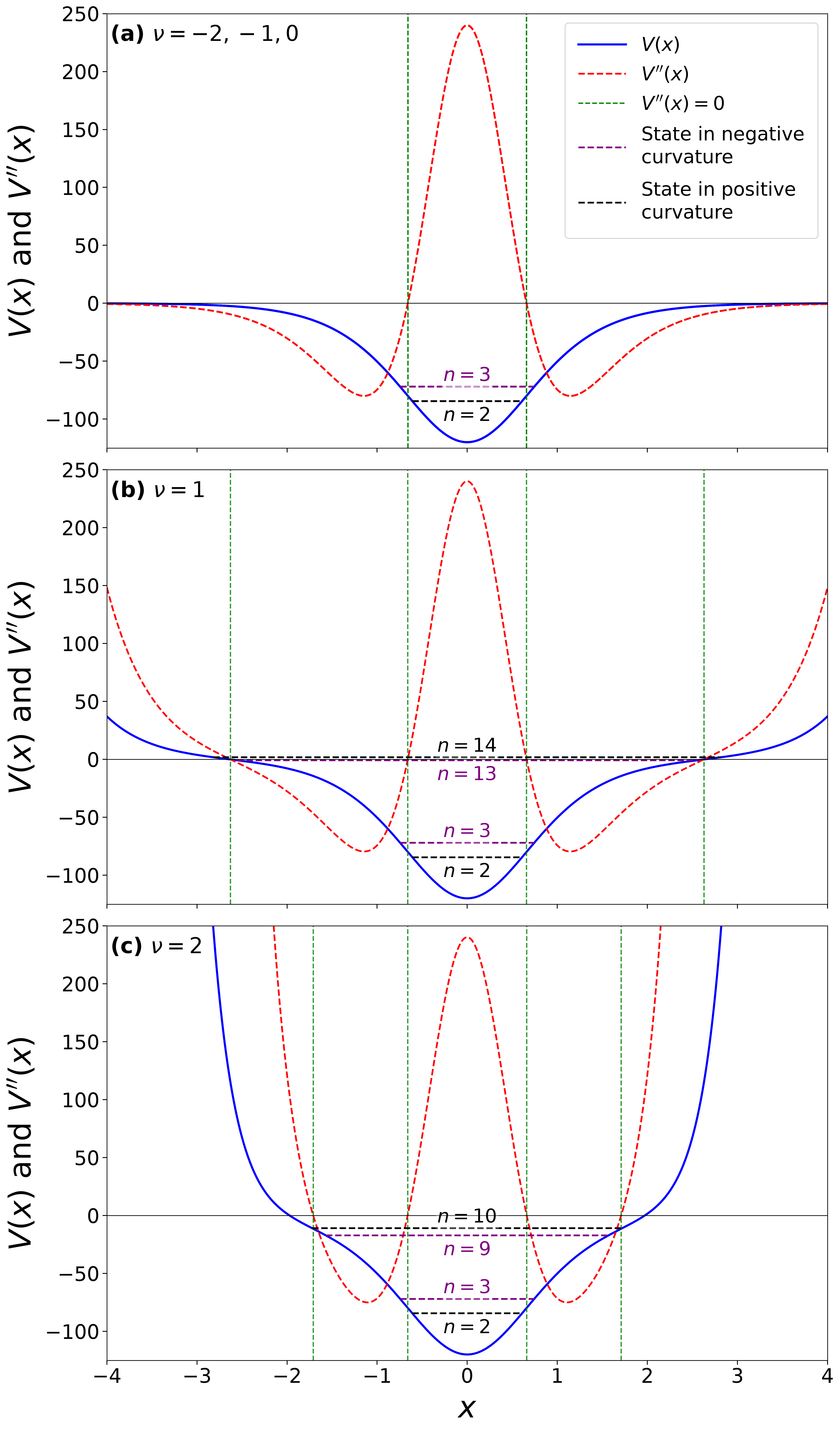}
    \caption{Potential (blue) and curvature (red dashed) curves for various values of $\nu$ with fixed $a_1 = -0.05$ and $a_2 = 120$. The horizontal dashed lines indicate the energies of the bound states, labeled by their state number ($n$). Purple and black lines denote states predominantly having probability density in the  region of negative and positive curvatures, respectively.}
    \label{fig:classical_turnning_curvature}
\end{figure}

Comparing classical turning points with the negative curvature windows shows that for
every $\nu$ considered in our work, the first state whose classical turning point
enters the negative curvature region is $n=3$, with the transition
occurring between $n=2$ and $n=3$. For $\nu\le0$, no further transition out
occurs because the region is semi-infinite, so all states $n\ge3$
remain inside it. For $\nu=1$, the last state inside the window is
$n=13$, with the further transition out into positive curvature region happening between $n=13$ and $n=14$. For
$\nu=2$, the perturbative shift from the $\cosh^4x$ term places the
last state inside the window at $n=9$, giving a transition out between
$n=9$ and $n=10$.

\subsection{Quantum Sensitivity}
A fully analogous structure appears in the quantum mechanical case. Using the same quadratic approximation around $x_c$, the Hamiltonian becomes

\begin{equation}\hat{H} = \frac{\hat{p}^2}{2\mu} + \frac12 V''(x_c)(\hat{x}-x_d)^2 \end{equation}

The approximated Hamiltonian is Quadratic, so Heisenberg equations of motion are linear and therefore admit the exact operator-valued solution:

\begin{equation}\frac{d\hat{A}}{dt} = \frac{i}{\hbar} \big[\hat{H}, \hat{A}\big] + \left( \frac{\partial \hat{A}}{\partial t} \right)_{H}\end{equation}
Leading to:
\begin{equation} \frac{d\hat{x}}{dt} = \frac{\hat{p}}{\mu} \quad\text{and} \quad \frac{d\hat{p}}{dt} = -V''(x_c)(\hat{x} - x_d)\end{equation}
Similarly,
\begin{equation} \frac{d^2\hat{x}}{dt^2} = \frac{1}{\mu}.\frac{d\hat{p}}{dt} 
\Rightarrow \quad \frac{d^2\hat{x}}{dt^2} + \frac{V''(x_c)}{\mu}(\hat{x} - x_d) = 0. \end{equation}
By solving this (for $V''(x_c)<0$), we get
\begin{equation}\hat{x}(t) = x_d + (\hat{x}(0) - x_d)\cosh(\tilde\Omega t) + \frac{\hat{p}(0)}{\sqrt{ - \mu V''(x_c)}}\sinh(\tilde\Omega t) \end{equation}
where $\tilde\Omega=\sqrt{\dfrac{-V''(x_c)}{\mu}}$. This is the operator analogue of the classical trajectory: the operator displacement grows exponentially with the same rate \(\tilde\Omega\).
 Using the canonical commutation relation $[\hat{x}(0),\hat{p}(0)] = i\hbar$, the required commutator evaluates to
\begin{equation}[\hat{x}(t),\hat{p}(0)] = i\hbar \cosh(\tilde\Omega t).\end{equation}

So, the OTOC
$C(t) = -\langle [\hat{x}(t),\hat{p}(0)]^2\rangle$,
then grows for $(\tilde\Omega t \gg 1)$ as
\begin{equation}C(t) = \hbar^2 \cosh^2(\tilde\Omega t) \simeq \frac{\hbar^2}{4}\,e^{2\tilde\Omega t},\end{equation}
i.e. at a rate of $2\sqrt{\dfrac{|V''(x_c)|}{\mu}}$. This sensitivity can be defined as semiclassical  sensitivity. Note that we define this sensitivity using $|V''|$ which seems to be including both positive and negative curvature regions, but the interpretation of transient growth is \textit{only} valid in $V''<0$. Otherwise, this is just a curvature frequency scale.

\subsection{Airy-corrected Semiclassical Approximation}

A more accurate estimate of the local instability evaluates the commutator not at the classical turning point $x_c$ but at the nearby point $x_m$ where the wave function reaches its maximum. Near the turning point the potential is better approximated linearly:
\begin{equation}V(x) \approx E + V'(x_c)(x-x_c).\end{equation}
Introducing the characteristic Airy scaling length
\begin{equation}\Delta = \left( \frac{\hbar^2}{2\mu |V'(x_c)|} \right)^{1/3}\end{equation}
and a dimensionless variable $z = (x-x_c)/\Delta$. Then  the Schr\"odinger equation reduces to the Airy equation \cite{Vallee2004}, 

\begin{equation}\frac{d^2\psi}{dz^2} - z\psi = 0 .\end{equation}

The solution of this equation $\psi_n(z)$ is given by the Airy function Ai(z), which oscillates for \(z<0\) and  attains its maximum at $z= -\eta_{A} = -1.01879... \approx -1.02 $, implying that the Airy estimate of the outermost turning-point maximum of the probability density lies slightly inside the classically allowed region, at\footnote{Unlike the volcano potential, in non-oscillatory cases, one has to be cautious about the validity of the Airy correction, as it may overestimate the magnitude of the correction. See the discussion later for the Braneworld case.}
\begin{equation}x_m = x_c - \eta_{A}\Delta . \label{eq:airy_correction} \end{equation}
This shift is a purely quantum effect: it reflects the exponential tail of the wavefunction penetrating the forbidden region, which displaces the maximum away from the classical turning point.
Therefore, the physically relevant curvature should be evaluated at $x_m$, not $x_c$. So, the improved estimate will be
\begin{equation}\lambda_{sc}^{(A)} = \sqrt{\frac{|V''(x_m)|}{\mu}} , \end{equation}
which again has the interpretation of sensitivity only when $V''<0$.
This improved value provides airy-corrected semiclassical version of the short-time hyperbolic growth rate as $2\lambda_{sc}^{(A)}$, especially for highly excited states.

\begin{figure}
    \centering
    \includegraphics[width=1\linewidth]{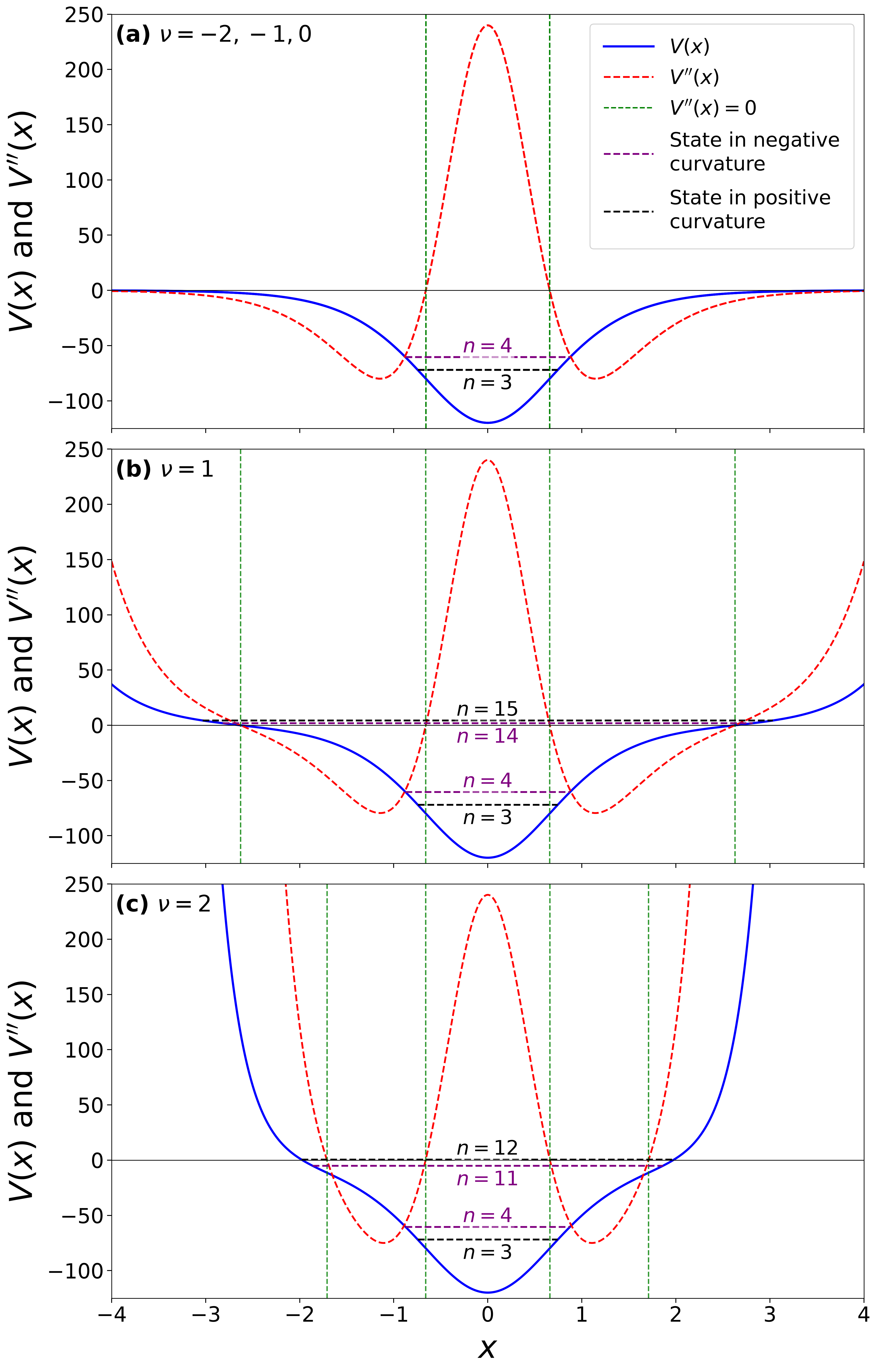}
    \caption{Potential (blue) and curvature (red dashed) curves for various values of $\nu$ with fixed $a_1 = -0.05$ and $a_2 = 120$. The horizontal dashed lines indicate the energies of the bound states, labeled by their state number ($n$). Purple and black lines denote states predominantly having probability density (\textbf{Airy-Corrected}) in the region of negative and positive curvatures, respectively. }
    \label{fig:Airy_corrected_curvature}
\end{figure}

As a concrete illustration, consider the set of potentials in FIG.~\eqref{fig:classical_turnning_curvature}. The classical turning point ($x_c$) analysis suggested that $n=3 $ is the first bound state entering the negative curvature region and hence should be the critical state. However, after including Airy correction (see FIG.~\eqref{fig:Airy_corrected_curvature}), the wavefunction maximum is pushed inward, and the $n=4$ becomes the first state whose outer turning point ($x_m$) lies in the negative curvature region. Therefore, $n=4$ acts as the actual Airy-corrected semiclassical estimate of the state that first exhibits the effects of local instability. Hence, for mixed curvature potentials, the Airy correction provides a more accurate semiclassical estimate of the crossover, therefore correctly identifying the quantum threshold between stable and unstable dynamics. Similar shifts appear in crossover cases for different values of $\nu$, which one can visualise from FIG.~\eqref{fig:Airy_corrected_curvature}.

\subsection{Extraction of the Early-time OTOC Growth Rate and Ehrenfest Time}
 For each eigenstate having its Airy-corrected turning point ($x_m$) in the negative curvature region, we numerically extract the quantum growth rate from the early-time regime of the OTOC. To quantify this growth and compare it with the semiclassical and Airy-corrected semiclassical sensitivities, we consider the logarithmic growth rate, $\frac{d}{dt}\ln{ C_n(t)}$.
 If the OTOC exhibits exponential growth, i.e. $C_n (t) \sim C_0\exp{\lambda_Q  t}$, the above quantity gives the corresponding transient growth rate $\lambda_Q$ in that regime. Thus, the quantum growth rate can be obtained from the slope of $\ln C_n(t)$. We determine this growth rate by performing a linear fit of $\ln C_n(t)$ over an appropriate early-time interval. the resulting slope represents the quantum growth rate over the chosen fitting interval and is denoted by $\lambda_Q^{(n)}.$  Rather than using the same fixed time interval for all eigenstates, one can choose the fitting window individually for each state using their corresponding Ehrenfest time ($t_E$), thus making sure that analysis is restricted to the regime where quantum and semiclassical dynamics correspondence is still valid.
 
 The Ehrenfest time can be estimated as \cite{Emerson2001,Berman1978, Zaslavsky:1981aw, Silvestrov2002},
\begin{equation}t_E^{(n) } =\frac{1}{\lambda_{sc}^{(A)}(n) } \ln\left(\frac{S_n}{\hbar}\right), \end{equation}
where $S_n$ is the classical action associated with the state $n$ and $\lambda_{sc}^{(A)}(n) $ is the corresponding airy-corrected semiclassical sensitivity. The action is given by $S_n = \oint p(x) dx,$
Where $p(x) = \sqrt{2\mu|E_n - V (x) |}$ is the momentum. For a bound state it is integrated between two classical turning points $-x_c$ and $x_c$.

\begin{figure}[!htbp]
    \centering
    \includegraphics[width=1\linewidth]{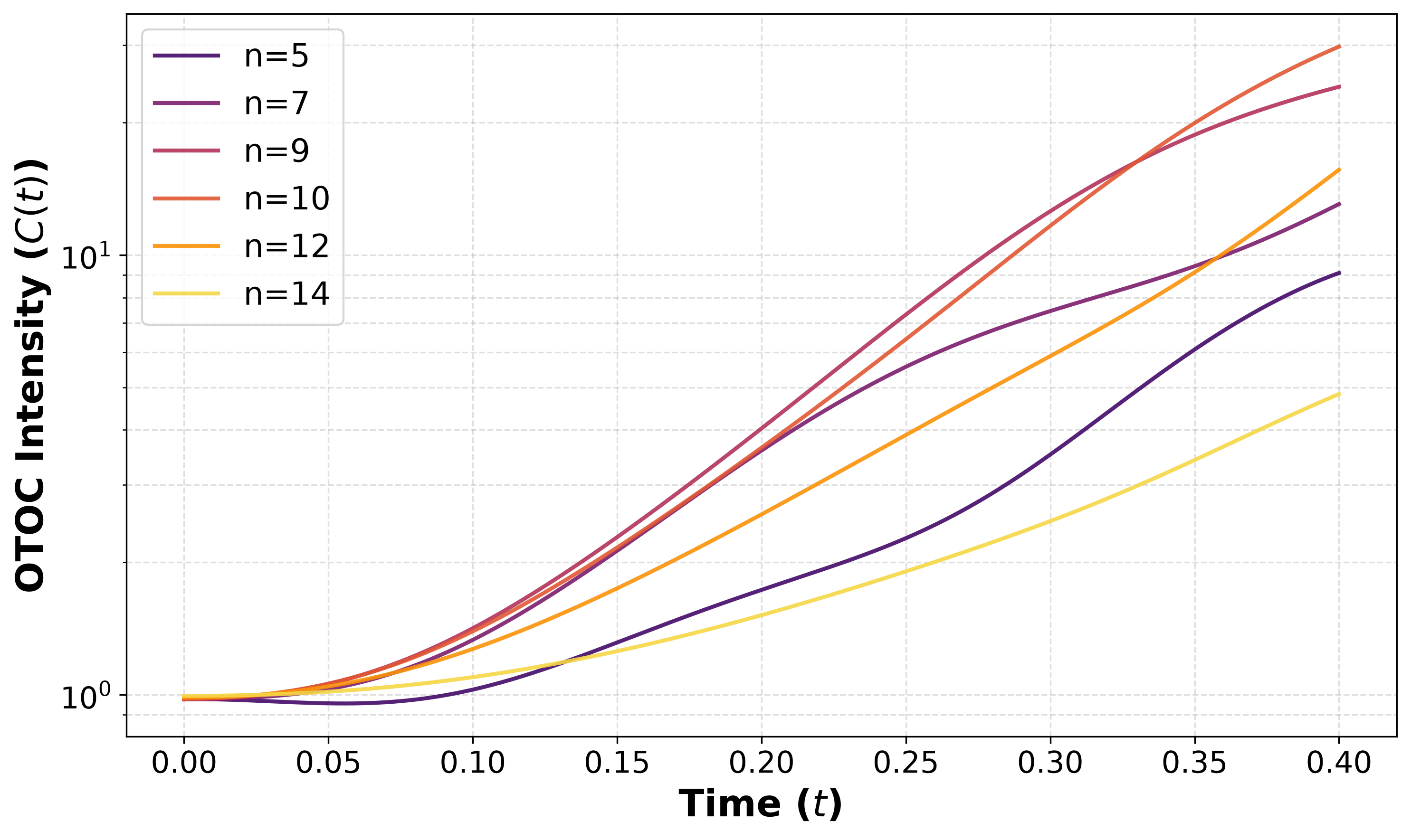}
    \caption{Transient exponential growth for bound states starting in the negative curvature region for the example case $\nu=-2$ (early time),fixed $a_1 = -0.05$ and $a_2 = 120$.}
    \label{fig:early_time_regime_otoc_gowth_for_nu_-2}
\end{figure}

\begin{figure*}[htbp]
    \centering
   
    \begin{subfigure}{0.485\textwidth}
    
        \includegraphics[width=1\linewidth]{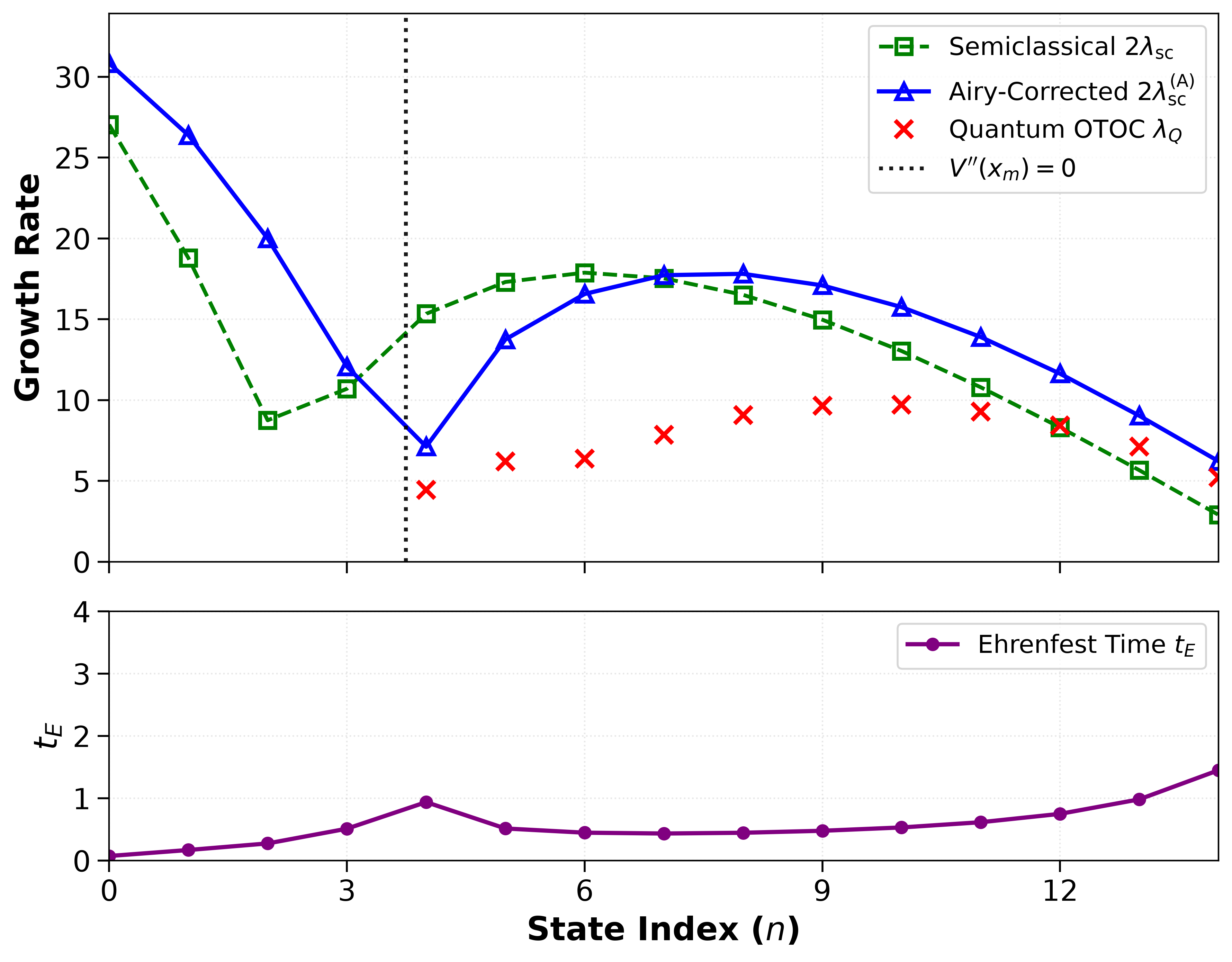}
              
        \caption{$\nu = -2$}
    \end{subfigure}\hfill
    \begin{subfigure}{0.485\textwidth}
        \includegraphics[width=1\linewidth]{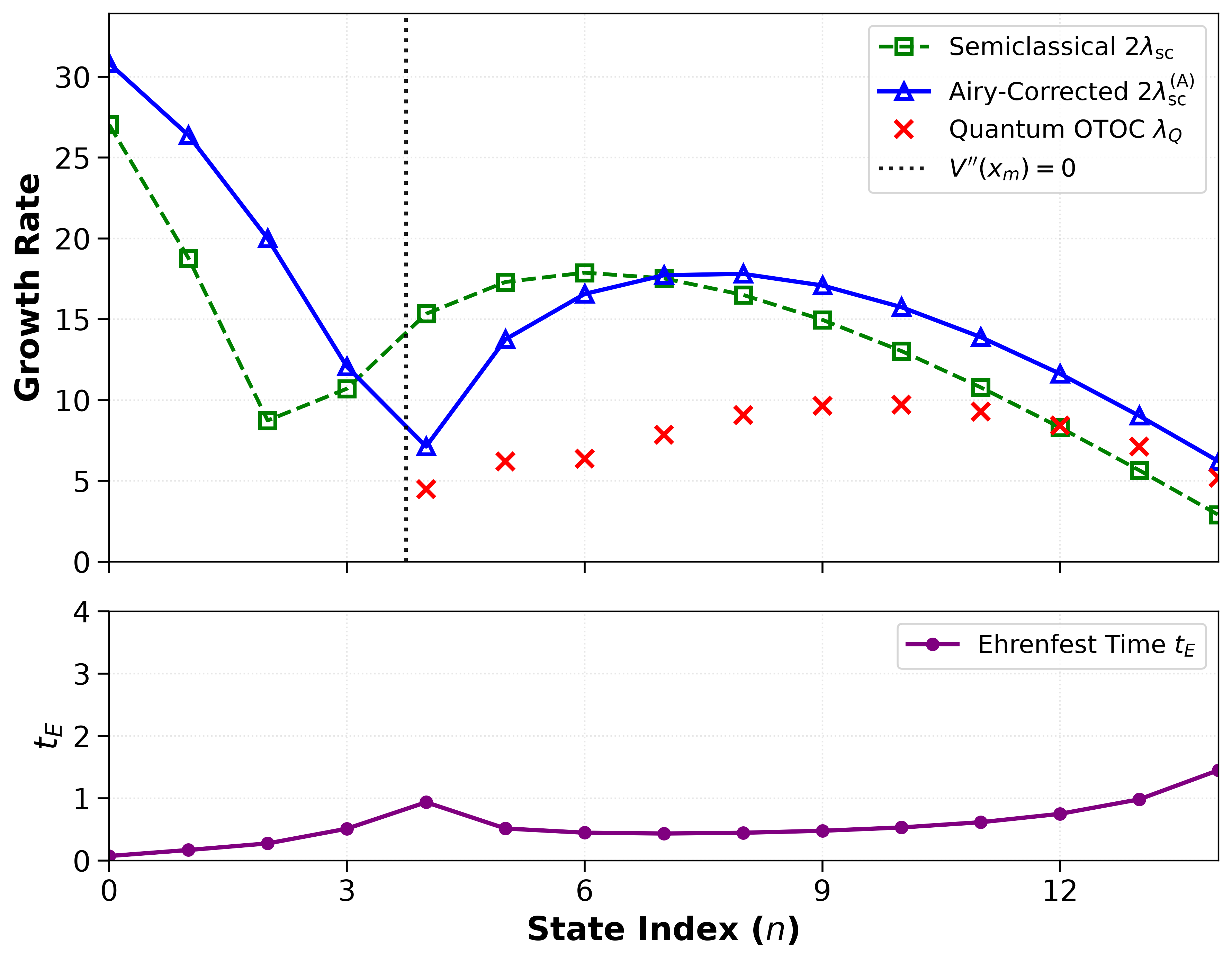}
                   
        \caption{$\nu = -1$}
    \end{subfigure}
    
    \vspace{0.3cm} 

    \begin{subfigure}{0.485\textwidth}
        \includegraphics[width=1\linewidth]{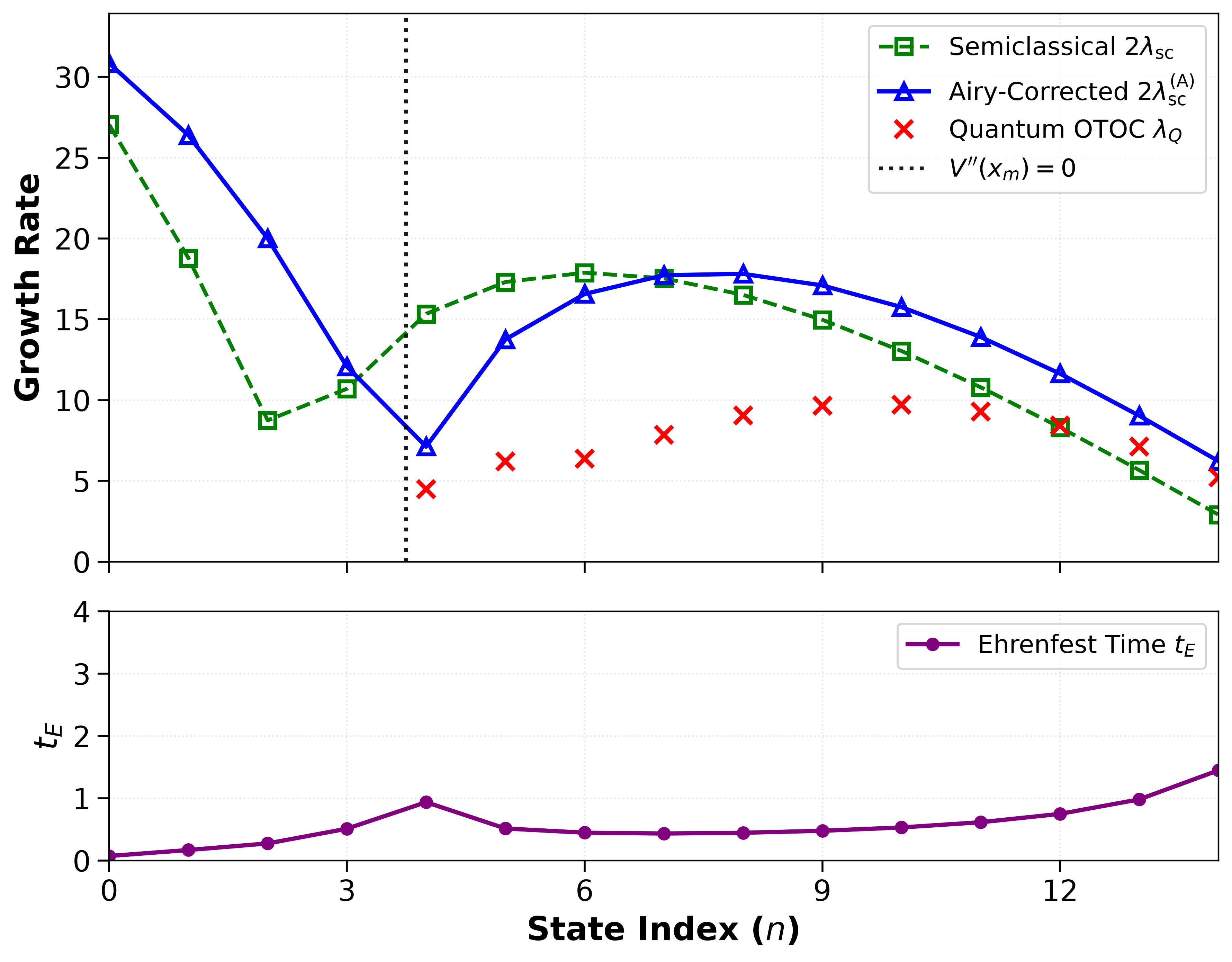}
               
        \caption{$\nu = 0$}
    \end{subfigure}\hfill
    \begin{subfigure}{0.485\textwidth}
        \includegraphics[width=1\linewidth]{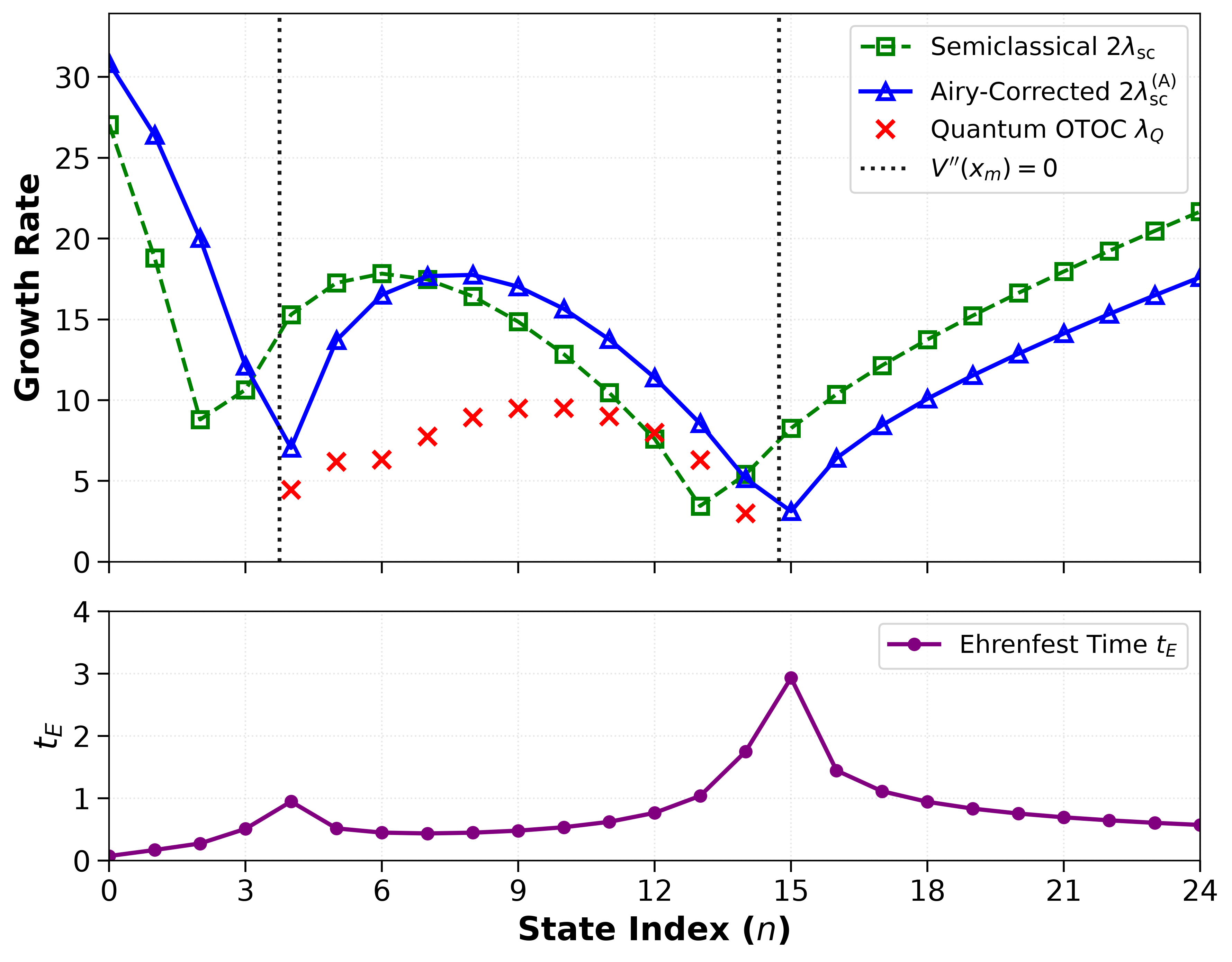}
          
        \caption{$\nu = 1$}
    \end{subfigure}
    
    \vspace{0.3cm}

    \begin{subfigure}{0.485\textwidth}
        
        \includegraphics[width=1\linewidth]{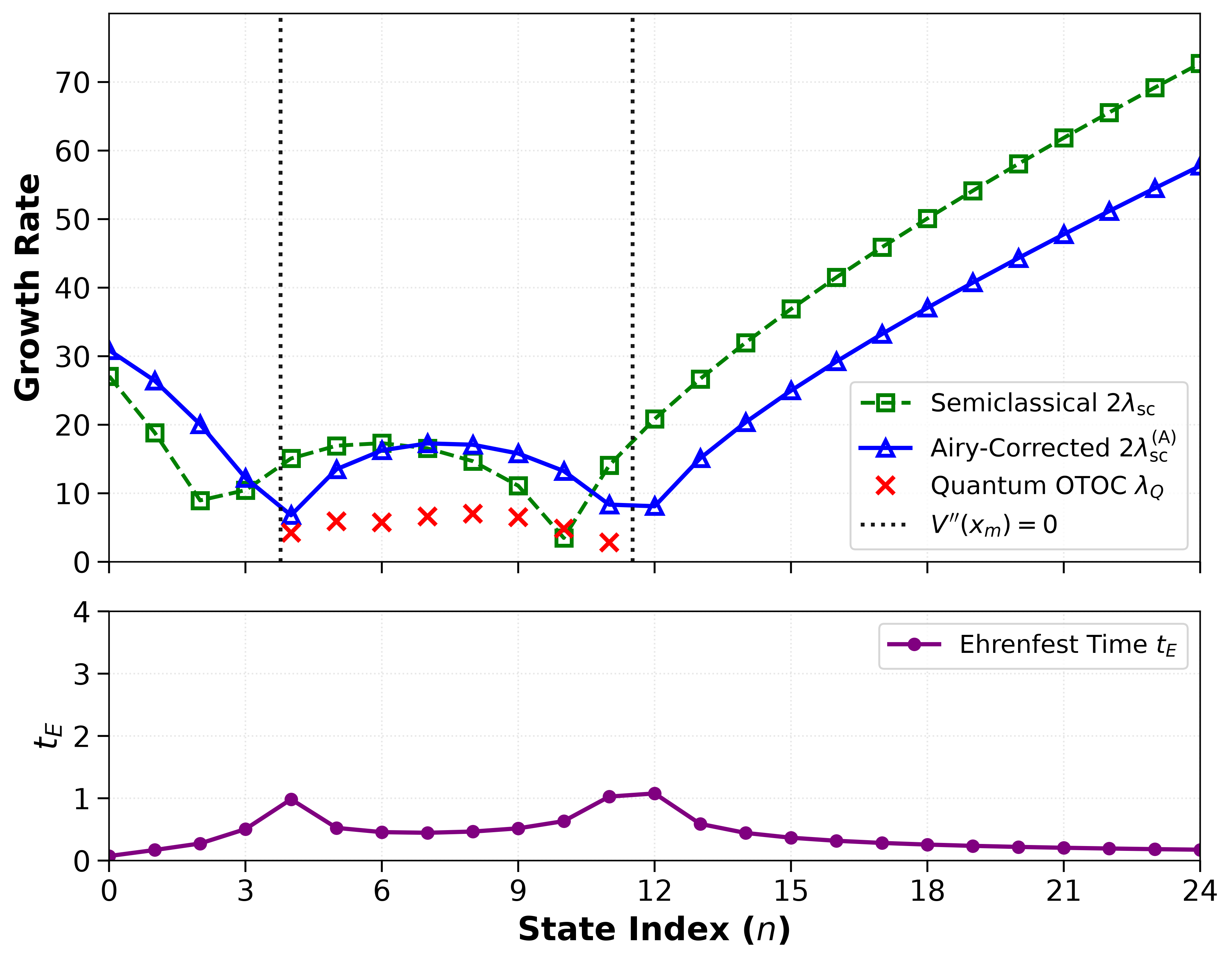}
        
        \caption{$\nu = 2$}
    \end{subfigure}

    \caption{Comparison of semiclassical sensitivity ($\lambda_{sc}$), Airy-corrected semiclassical sensitivity ($\lambda_{sc}^{(A)}$) and extracted OTOC exponential growth ($\lambda_{Q}$) across different $\nu$ values (all characterized by $a_1=-0.05$ and $a_2=120$). The vertical dotted lines signify the negative curvature region as sampled by the states. This is the region where we can read off the true extracted rate of local instability.}
    \label{fig:sensitivities_comparison}

\end{figure*}

This expression provides an of the time up to which the quantum dynamics can approximately follow the corresponding classical one. The extracted quantum OTOC growth rates in this window (which can be verified from the correspondence with the figures like FIG.~\eqref{fig:early_time_regime_otoc_gowth_for_nu_-2} and Table.\eqref{tab:nu_minus2_results}), are then compared with the corresponding semiclassical and Airy-corrected semiclassical sensitivities. The resulting state-by-state comparison for different values of $\nu $ is presented in FIG.~\eqref{fig:sensitivities_comparison}.

\begin{table}[htb]
    \centering
    \caption{Ehrenfest time $t_E$ and extracted quantum OTOC growth rate $\lambda_Q$ for selected eigenstates at $\nu=-2$ (see also FIG.~\eqref{fig:early_time_regime_otoc_gowth_for_nu_-2}) .}
    \label{tab:nu_minus2_results}
    \begin{tabular}{ccc}
        \toprule
        State index $n$ & $t_E$ & $\lambda_Q$ \\
        \midrule
        5  & 0.5149    &   6.2036 \\
        7  & 0.4344    &   7.8468 \\
        9  & 0.4781    &   9.6517 \\
        10 & 0.5316    &   9.7159 \\
        12 & 0.7497    &   8.4289 \\
        14 & 1.4501    &   5.2222 \\
        \bottomrule
    \end{tabular}
\end{table}

\begin{figure}[htbp]
    \centering
    \begin{subfigure}{\linewidth} 
        \includegraphics[width=1\linewidth]{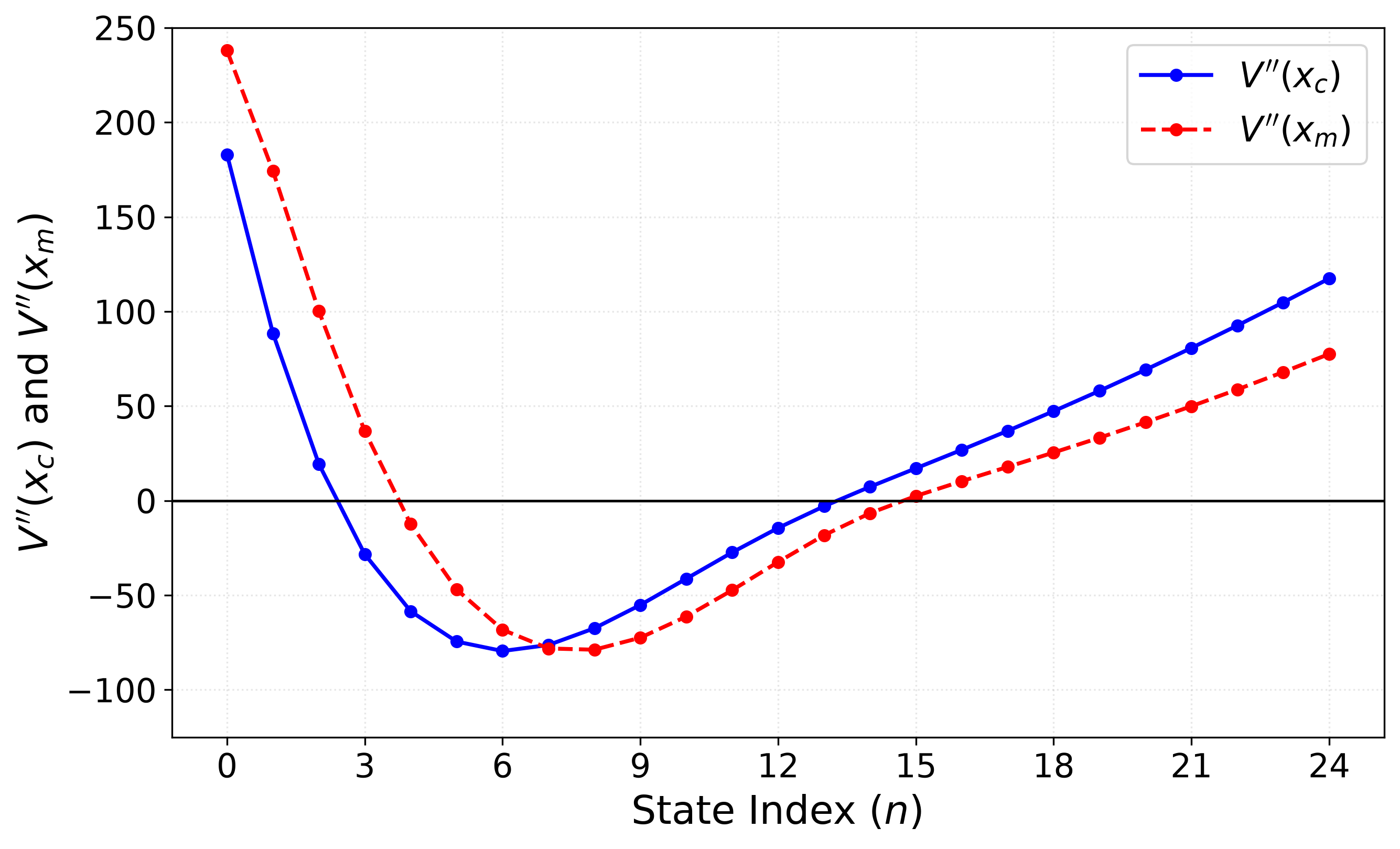}
        \caption{}
    \end{subfigure}

    \begin{subfigure}{\linewidth} 
        \includegraphics[width=1\linewidth]{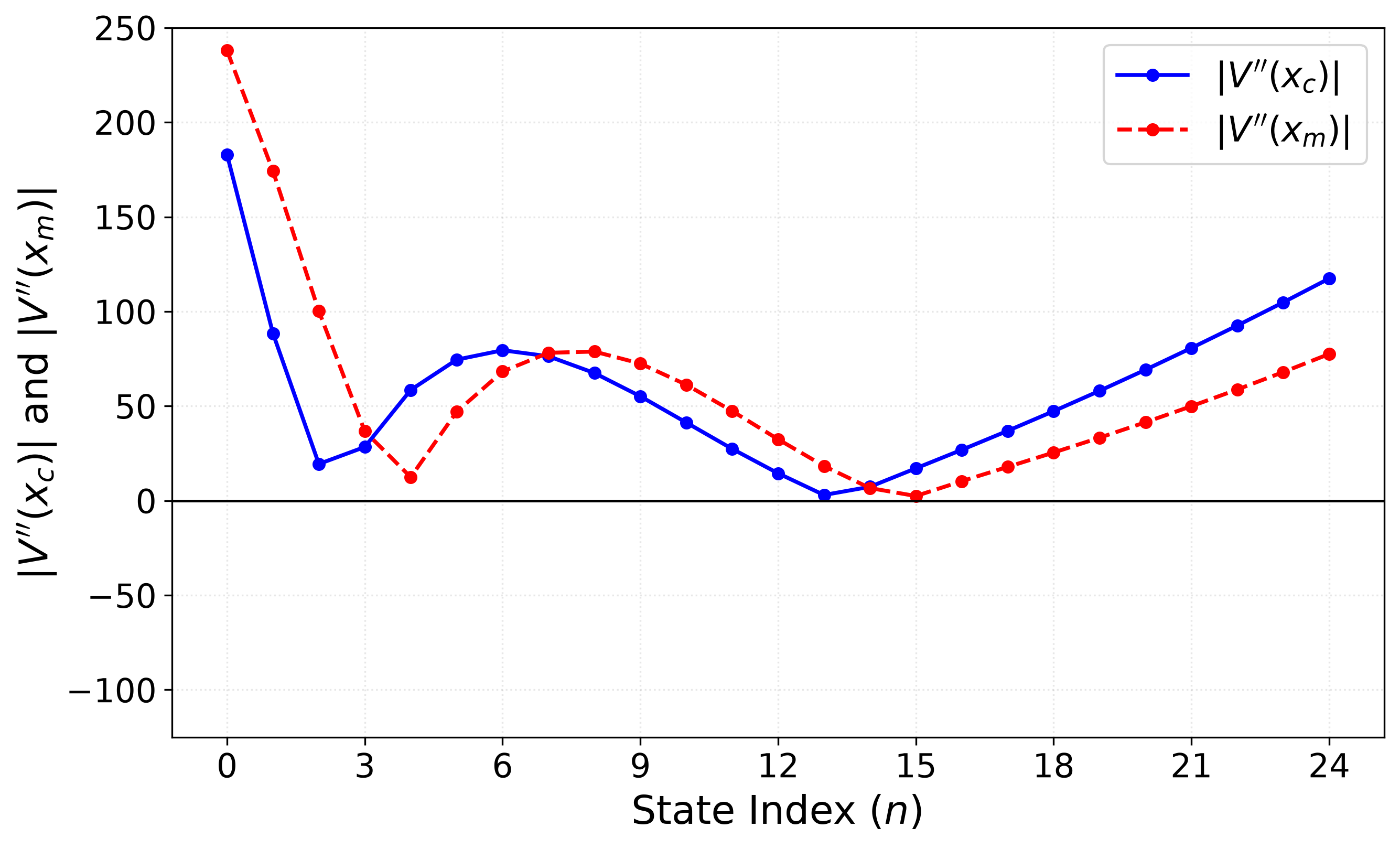}
        \caption{}
        
    \end{subfigure}
    \caption{Plotting signed values (a) \& absolute values (b) of Curvature at classical turning point ($x_c$) and maximum wavefunction amplitude point ($x_m$) against state index ($n$) for the $\nu=1$ case. } 
    \label{fig:Signed_and_absolute_value_of_Curvature_for_nu_1} 
\end{figure}

As discussed before, we take semiclassical and Airy-corrected semiclassical sensitivity only to make sense inside the negative curvature region of the potential where it depends on the absolute value of the curvature at classical turning point and its Airy-corrected version, respectively: $\lambda_{sc} \propto |V''(x_c)|^{1/2} , \quad \lambda_{sc}^{(A)} \propto |V''(x_m)|^{1/2}$.
Hence in FIG.~\eqref{fig:sensitivities_comparison}, the minima of $\lambda_{sc} $ and $\lambda_{sc}^{(A)}$ indicate the states for which the negativity of the curvature is minimum at $x_c$ and $x_m$. For example, in $\nu=1$ (and generally for $\nu >0$ cases), we have two curvature crossovers. FIG.~\eqref{fig:Signed_and_absolute_value_of_Curvature_for_nu_1} shows both the signed curvature and its absolute value as a function of the state index. The first state with negative curvature at $x_c$ is $n =3$, whereas the absolute curvature reaches its minimum at $n=2$, as also indicated by FIG.~\eqref{fig:sensitivities_comparison}. Similarly, at the second crossover, the last state with negative curvature at $x_m$ is $n=14$, while the absolute curvature reaches its minimum at $n=15$, which transitions out. Therefore, the minima of $\lambda_{sc} $ and $\lambda_{sc}^{(A)}$ should be interpreted as indicating the minimum magnitude of curvature, rather than directly identifying the critical state, since the two do not necessarily coincide. 

This distinction is also reflected in the behaviour of the Ehrenfest time (see FIG.~\eqref{fig:sensitivities_comparison}). The peaks in $t_E$ occur near the boundaries of the negative curvature region, where the local Airy-corrected semiclassical instability rate decreases and an initially localized wavepacket spreads more slowly. As a result, the quantum dynamics can remain correlated to the corresponding classical dynamics for a longer time.

\section{Special case: The P\"oschl--Teller Potential}\label{secPT}

We now specialize to the exactly solvable PT potential, which provides an ideal analytic benchmark for testing and demonstrating the interplay between semiclassical local instability and quantum corrections we have been discussing. This potential appears in a plethora of physical situations, hence the OTOC growth is also interesting in its own right. The PT potential is defined as
\begin{equation} 
V(x) = -V_0\,\operatorname{sech}^2(\alpha x), \qquad V_0>0,
\end{equation}
which is a finite well, and features a finite number of bound states, making all relevant quantities analytically tractable. Note that, for the generic volcano family we measure position in units of the
intrinsic length scale, so that $x$ was dimensionless. In the PT
benchmark we restore this scale explicitly through $\alpha$. In what follows, we try to understand some of the structures explored in generic Volcano potentials in previous sections, specialized for the PT case. Further, we would like to have an estimate for the OTOC in this case itself.

\subsection{Sensitivity}

To parametrize the spectrum, we introduce the positive constant $\Lambda$ via
\begin{equation}
\Lambda(\Lambda+1) = \frac{2\mu V_0}{\hbar^2\alpha^2},
\end{equation}
which allows us to write the potential strength as $V_0 = \frac{\hbar^2\alpha^2}{2\mu}\Lambda(\Lambda+1)$ and $\alpha\in \mathbb{R}$. The bound state energies, the regime we are interested in, are then given exactly by the textbook result
\begin{equation}
E_n = -\frac{\hbar^2\alpha^2}{2\mu}\,(\Lambda-n)^2,
\qquad n=0,1,\dots,\lfloor\Lambda\rfloor-1,
\end{equation}
where $\lfloor\Lambda\rfloor$ denotes the integer part of $\Lambda$.\footnote{There are then $N_B = \lceil\Lambda \rceil$ corresponding bound states whose wavefunctions may be written using Gegenbauer polynomials. A quick discussion maybe found in the very end of Appendix.\eqref{Appendix_D}.} For a given state $n$, the classical outer turning point $x_c(n)$ is determined by the condition $V(x_c)=E_n$. Substituting the explicit forms yields
\begin{equation}
-V_0\,\operatorname{sech}^2(\alpha x_c) = -\frac{\hbar^2\alpha^2}{2\mu}(\Lambda-n)^2.
\end{equation}
Using the definition of $\Lambda$, this simplifies to
\begin{equation}
\operatorname{sech}^2(\alpha x_c) = \frac{(\Lambda-n)^2}{\Lambda(\Lambda+1)} \equiv q_n,
\end{equation}
where we have introduced the shorthand $q_n$ for the value of $\operatorname{sech}^2$ at the turning point. This quantity lies between $0$ and $1$ and monotonically decreases as $n$ approaches $\Lambda$, signaling that the turning point moves outward toward the asymptotic region.

The local geometry of the potential is governed by its curvature. 
At the classical turning point, the curvature becomes
\begin{equation}
V''(x_c) = 2\alpha^2V_0\, q_n \left(3q_n - 2\right).
\end{equation}
The rate of instantaneous local instability ($\lambda_{sc}(n)$), as introduced before, measures the strength of the hyperbolic growth for a trajectory sitting at the turning point:
$\lambda_{sc}(n) = \sqrt{\frac{|V''(x_c)|}{\mu}}$.
Substituting the expression for $V''(x_c)$ and the definition of $V_0$, we obtain the compact form
\begin{equation}
\lambda_{sc}(n) = \sqrt{\frac{2\alpha^2 V_0}{\mu}}\,
\sqrt{\left| q_n \left(3q_n - 2\right) \right|}
.
\end{equation}
Expanding $q_n$ explicitly in terms of $\Lambda$ and $n$ gives the equivalent exact dependence
\begin{equation}
\lambda_{sc}(n) = \frac{\hbar\alpha^2}{\mu}\,(\Lambda-n)
\sqrt{\left| \frac{3(\Lambda-n)^2 - 2\Lambda(\Lambda+1)}{\Lambda(\Lambda+1)} \right|},
\end{equation}
which highlights that $\lambda_{sc}(n)$ vanishes linearly as $n \to \Lambda$ (near dissociation) and that it measures absolute sensitivity only when $q_n < 2/3$, i.e. when the turning point lies in the negative curvature region of potential.

Note further, the curvature changes sign at $x_L
=
\frac{1}{\alpha}
\operatorname{arcosh}
\sqrt{\frac32}
\simeq
\frac{0.6585}{\alpha}$, and the most negative curvature is $V''_{\min}
=
-\frac{2\alpha^2V_0}{3}.$  From here, we can get the first unstable state (without airy-correction) as fixed by:
\begin{equation}
    n_c^{\rm cl}
        =
        \left\lfloor
            \Lambda - \sqrt{\frac{2}{3}\Lambda(\Lambda+1)}
        \right\rfloor + 1
\end{equation}

However, as discussed before, a purely classical turning point analysis based on $x_c$ is insufficient for quantum states, which shifts the maximum inward. To capture this effect, we linearize the potential in the vicinity of $x_c$, and calculate the Airy correction. Consequently, Airy estimate of the outermost turning-point probability maximum is located at (see \eqref{eq:airy_correction})
\begin{equation}
x_m(n) = x_c(n) - \eta_A \Delta(n),
\end{equation}
where in the quantum shift, $\Delta(n)$ is the Airy scaling length
\begin{equation}
\Delta(n) = \left( \frac{\hbar^2}{2\mu |V'(x_c)|} \right)^{1/3}.
\end{equation}
This shift is a purely quantum length that grows as the potential becomes flatter near the turning point. Using the derivative of the PT potential at the turning point:
$|V'(x_c)| = 2\alpha V_0\, q_n \sqrt{1-q_n}$, the Airy scaling length is explicitly
\begin{equation}
\Delta(n) = \left( \frac{\hbar^2}{4\mu \alpha V_0\, q_n \sqrt{1-q_n}} \right)^{1/3}.
\end{equation}
It is instructive to write down the dimensionless Airy length:
\begin{equation}
\alpha\Delta(n)
=
\left[
\frac{
1
}{
2\Lambda(\Lambda+1)
q_n\sqrt{1-q_n}
}
\right]^{1/3}.
\label{eq:alpha-Delta}
\end{equation}
Immediately this leads to the Airy-corrected turning estimate:
\begin{equation}
q_m
\equiv
\sech^2(\alpha x_m)
=
\sech^2
\left[
\cosh^{-1}\left(q_n^{-1/2}\right)
-
\eta_A\alpha\Delta(n)
\right].
\label{eq:qm-exact-Airy}
\end{equation}
Since $x_m<x_c$ on the positive side of the well and
$\sech^2(\alpha x)$ decreases monotonically with $x>0$, one necessarily has $q_m>q_n$.

The correct semiclassical marker of the transient instability $\lambda_{sc}^{(A)}(n)$ is then defined by evaluating the curvature at the actual peak of the outermost antinode rather than at the classical turning point: i.e. 
\begin{equation}
\lambda_{sc}^{(A)}(n) = \sqrt{\frac{|V''(x_m)|}{\mu}},~~~ V''(x_m)=2\alpha^2V_0q_m(3q_m-2)
\end{equation}
Since $x_m = x_c - \eta_A \Delta$, considering the small separation between the two turning estimates, one could find an analytic expansion in orders of the dimensionless Airy length. 

Especially, far from dissociation, the Airy shift is small compared to the distance to the caustic, so the quantum peak sits essentially at the classical turning point.
Near dissociation, where $\Lambda-n \equiv \kappa_n \ll 1$ is very small, we have
\begin{equation}
q_n \sim \kappa_n^2/[\Lambda(\Lambda+1)],\quad |V'(x_c)| \sim \kappa_n^2.
\end{equation}
Consequently, the Airy shift scales as $\Delta(n) \sim \kappa_n^{-2/3}$, which diverges. \footnote{See Appendix \eqref{appcritical} for more details and subtleties on the critical scaling structure.} This means that for high-lying states, the peak could move significantly inward relative to $x_c$, where the curvature is typically larger in magnitude. This leads to the inequalities
\begin{equation}
\lambda_{sc}^{(A)}(n) > \lambda_{sc}(n), \quad
\lambda_{sc},\lambda_{sc}^{(A)}(n) \to 0 \quad \text{as} \quad n \to \Lambda.
\end{equation}
Both exponents vanish at dissociation because the potential becomes asymptotically flat, but the Airy-corrected semiclassical exponent remains larger for all intermediate states.

In the Volcano potential well bound states, which are structurally similar to PT well, this explains why the semiclassically predicted critical state $n=3$ is stabilized by the Airy shift—the peak $x_m(n=3)$ remains in positive curvature—while $n=4$ has its peak just inside negative curvature region. Hence, this demonstrates that the local instability is activated only when the quantum wavefunction's antinode crosses the curvature boundary.

\subsection{Towards an Analytic OTOC}
Although the PT potential is exactly solvable, finding full OTOCs analytically is much harder as it involves matrix elements of \(x\) and \(p\) between bound and continuum sectors as well. However, this is an instructive exercise in assessing how OTOCs for non-trivial symmetric potentials work. In this section, we want to get a working idea about how the OTOCs in this case may look like, even under some approximations. To start with, our approximation in this work has been to be confined within bound states. Then we can consider the set of such states in the finite-dimensional bound state subspace
\begin{equation}
\mathcal{H}_B
=
\operatorname{span}
\left\{
|n\rangle:\ n<\Lambda
\right\}
\end{equation}
Define the
spectral projector
\begin{equation}
P_B
=
\sum_{n<\Lambda}
|n\rangle\langle n| \implies 
[H,P_B]=0.
\label{eq:H-P-commute}
\end{equation}
Consequently, $\mathcal{H}_B$ is an invariant subspace of the exact Hamiltonian evolution. Defining the projected Hamiltonian
\begin{equation}
H_B=P_BHP_B,
\end{equation}
we have, on $\mathcal{H}_B$,
\begin{equation}
e^{-iH_Bt/\hbar}
=
P_Be^{-iHt/\hbar}P_B.
\label{eq:restricted-evolution}
\end{equation}
The observables are projected according to\footnote{The reader could note here that projection onto the finite-dimensional bound state subspace modifies
the canonical algebra: $[x_B,p_B]\neq i\hbar P_B$, since the trace of
any finite-dimensional commutator vanishes, $\Tr[x_B,p_B]=0$, whereas
$\Tr(i\hbar P_B)=i\hbar\dim\mathcal{H}_B\neq0$. This does not signal an
inconsistency: $x_B$ and $p_B$ are well-defined projected observables,
but they no longer form an exact canonical pair. Canonical identities
that rely on $[x,p]=i\hbar$ must therefore be replaced by their
projected counterparts, with deviations measuring the effect of
discarding transitions outside the bound state subspace.}
\begin{equation}
x_B=P_BxP_B,
\qquad
p_B=P_BpP_B.
\label{eq:projected-operators}
\end{equation}
It is important to distinguish this construction from the full
canonical OTOC. The Hamiltonian evolution on $\mathcal{H}_B$ is the
exact restriction of the full evolution, but the operators $x_B$ and
$p_B$ omit matrix elements through the continuum. Thus the quantity we are interested in
is the projected bound-sector OTOC; it is not identical to the full OTOC constructed from $x$ and $p$ on the complete Hilbert space.

For bound states, the matrix elements like $x_{mn}
\equiv
\langle m|x|n\rangle; m,n<\Lambda$, can be calculated by implementing the parity selection rules, $x_{mn}=0$ if 
$m+n$ is even. Thus $x$ (parity odd) connects states of opposite parity. In general,
\begin{equation}
|m-n|=1,3,5,\ldots
\end{equation}
are allowed whenever the corresponding states exist. In particular, there is no exact harmonic-oscillator-like nearest-neighbour selection rule for the position operator in a general potential. One can then show, between exact energy eigenstates, momentum matrix elements $p_{mn}= \frac{i\mu}{\hbar}
(E_m-E_n)x_{mn}$ also obey the same parity
selection rule.

For a deep well and fixed low-lying $n$, the potential is
approximately harmonic near its minimum, with a harmonic frequency $\omega_0 = \alpha\sqrt{\frac{2V_0}{\mu}}.$ Before evaluating the integral for matrix elements like $x_{n,n+1}$, it is useful to establish its scaling
with the inverse length scale $\alpha$. Introduce the dimensionless $y=\alpha x$, then the normalized wavefunction in the original coordinate is $\psi_n(x) = \sqrt{\alpha}\,\phi_n(\alpha x)$. Hence the exact matrix element necessarily has the form:
\begin{equation}
x_{mn}
=
\frac{1}{\alpha}
f_{mn}(\Lambda),
\label{eq:PT-xmn-scaling}
\end{equation}
where
\begin{equation}
f_{mn}(\Lambda)
\equiv
\int_{-\infty}^{\infty}
\phi_m^*(y)\,
y\,
\phi_n(y)\,dy
\end{equation}
is dimensionless. The exact bound state wavefunctions (with $z=\tanh y$) can be written as \cite{Poschl:1933zz,CooperKhareSukhatme1995}
\begin{equation}
\psi_n(z)
=
\mathcal{N}_n
(1-z^2)^{\kappa_n/2}
C_n^{\kappa_n+\frac12}(z),
\qquad
\kappa_n=\Lambda-n.
\label{eq:PT-z-wavefunction}
\end{equation}
It is convenient to separate the dimensional normalization factor:
\begin{equation}
\mathcal{N}_n
=
\sqrt{\alpha}\,
\mathcal{M}_n(\Lambda),
\label{eq:PT-N-scaling}
\end{equation}
where $\mathcal{M}_n$ is dimensionless, and there is simply no other dimensional length left in the eigenfunctions. The exact integrals for the matrix element can be performed with some rigour, and they analytically work out as combinations of Beta and Gamma function, the full form of which is not very illuminating. We omit the full computation here, and the reader can refer to Appendix.\eqref{appmatrixpt} for the details. The total structure, however, stays as \eqref{eq:PT-xmn-scaling}. One can always check that it reduces to the nearest neighbour harmonic oscillator element when we take the deep well limit $\Lambda\gg1$ with fixed low-lying $n$.

Now for a bound eigenstate $|n\rangle$, we define the projected OTOC:
\begin{equation}
C_n^{(B)}(t)
=
\langle n|
[x_B(t),p_B]^\dagger
[x_B(t),p_B]
|n\rangle,
\label{eq:OTOC-def}
\end{equation}
Since the bound states diagonalize $H_B$,
\begin{equation}
x_B(t)
=
\sum_{m,k<\Lambda}
e^{i(E_m-E_k)t/\hbar}
x_{mk}
|m\rangle\langle k|.
\label{eq:x-t-expansion}
\end{equation}
Define and expand the amplitude:
\begin{widetext}
    
\begin{equation}
A_{mn}^{(B)}(t)
\equiv
\langle m|
[x_B(t),p_B]
|n\rangle
=
\frac{i\mu}{\hbar}
\sum_{k<\Lambda}
x_{mk}x_{kn}
\left[
(E_k-E_n)e^{i(E_m-E_k)t/\hbar}
-
(E_m-E_k)e^{i(E_k-E_n)t/\hbar}
\right].
\label{eq:A-x-only}
\end{equation}
\end{widetext}
The exact projected OTOC is therefore sum over all bound-bound OTOC elements, on which we also have to impose the parity constraint as it contains products of two parity-odd operators, to get
\begin{equation}
C_n^{(B)}(t)
=
\sum_{\substack{m<\Lambda\\m+n\ {\rm even}}}
\left|
A_{mn}^{(B)}(t)
\right|^2.
\label{eq:OTOC-parity}
\end{equation}
Because $\mathcal{H}_B$ contains only finitely many real energy
eigenvalues, \eqref{eq:OTOC-parity} is a finite sum of
oscillatory terms involving bound state Bohr frequencies, $\omega_{mn} = \frac{E_m-E_n}{\hbar}$, and thus remains bounded for all time. Since different oscillation frequencies can be different, the interference could produce quasiperiodicity, beats etc. Note that what we are going to calculate will be purely oscillatory and bounded. Hence the transient
asymptotic exponential growth over a restricted time interval can only be established from the
actual matrix elements. In
particular, an apparent exponential fit over a short interval should
be distinguished from ordinary short-time Taylor growth.
\footnote{Put more directly, this finite Fourier sum can approximate a rapidly rising exponential envelope over a finite interval before interference causes turnover, saturation-like behaviour, or recurrence, but can never support sustained hyperbolic growth.}

For a fixed bound state $|n\rangle$, we define the nearest neighbour (NN) matrix elements (which are not enough for the full $C_n^{(B)}(t)$):
\begin{equation}
x_+
\equiv
x_{n,n+1},
\qquad
x_-
\equiv
x_{n,n-1},
\label{eq:PT-xpm}
\end{equation}
and
\begin{equation}
x_+^{(2)}
\equiv
x_{n+1,n+2},
\qquad
x_-^{(2)}
\equiv
x_{n-1,n-2}.
\label{eq:PT-xpm2}
\end{equation}
Note again the generic matrix elements can be found in Appendix \eqref{appmatrixpt}, and in principle, can be used to calculate all bound-bound transitions beyond NN.
For real bound state wavefunctions the position matrix elements may
be chosen real, so that $x_{mn}=x_{nm}$. For transitions above and below $n$, introduce the positive
frequencies: 
\begin{align}
\omega_+
&\equiv
\frac{E_{n+1}-E_n}{\hbar},
~
\omega_-
\equiv
\frac{E_n-E_{n-1}}{\hbar},\nonumber \\
\omega_+^{(2)}
&\equiv
\frac{E_{n+2}-E_{n+1}}{\hbar},~
\omega_-^{(2)}
\equiv
\frac{E_{n-1}-E_{n-2}}{\hbar}.
\label{eq:PT-omega-pm2}
\end{align}
All of which we can calculate in terms of $\Lambda$ and $n$ using the exact spectrum, provided all these bound states exist and have energies in the expected ordering. We can further calculate momentum elements $p_{mn}
=
\frac{i\mu}{\hbar}
(E_m-E_n)x_{mn}$, which is an identity of the full, unprojected canonical theory, and hence unchanged. Under the NN approximation, the product of two
position or momentum transitions changes the level number by
$0$ or $\pm2$ (with the
same parity). Therefore the only possible final states are
\begin{equation}
m=n,\qquad n+2,\qquad n-2,
\end{equation}
provided that they belong to the bound state spectrum. Hence the truncated OTOC is:
\begin{equation}
C_{n,\mathrm{NN}}^{(B)}(t)
=
|A_{nn}(t)|^2
+
|A_{n+2,n}(t)|^2
+
|A_{n-2,n}(t)|^2.
\label{eq:PT-NN-three-channels}
\end{equation}

If one goes beyond the NN approximation, but stays in the bound state manifold, the above sum will receive contributions from more transition amplitudes. But we keep ourselves restricted to the above approximation, since we did not assume a fixed number of bound states yet. 
Now, naturally, for diagonal channel $m=n$ and provided $|k-n|=1$, the intermediate states are $k=n\pm1$ as parity permits only all opposite-parity intermediate states. And in our truncated case, these are just $n\to n+1$ and $n\to n-1$ transitions. For $m=n+2$, the only intermediate state is $k=n+1$, including $n+2 \to n+1$ transitions, and For $m=n-2$, the only intermediate state is $k=n-1$, including $n-2 \to n-1$ transitions in their respective amplitudes. Combining the three channels, the nearest-neighbour projected
bound state OTOC is
\begin{widetext}
\begin{equation}
\begin{aligned}
C_{n,\mathrm{NN}}^{(B)}(t)
&=
4\mu^2
\left[
\omega_+x_+^2\cos(\omega_+t)
-
\omega_-x_-^2\cos(\omega_-t)
\right]^2
\\[4pt]
&\quad+
\mu^2x_+^2
\left(x_+^{(2)}\right)^2
\Bigl[
\omega_+^2
+
\left(\omega_+^{(2)}\right)^2
-
2\omega_+\omega_+^{(2)}
\cos\left(
[\omega_+-\omega_+^{(2)}]t
\right)
\Bigr]
\\[4pt]
&\quad+
\mu^2x_-^2
\left(x_-^{(2)}\right)^2
\Bigl[
\omega_-^2
+
\left(\omega_-^{(2)}\right)^2
-
2\omega_-\omega_-^{(2)}
\cos\left(
[\omega_--\omega_-^{(2)}]t
\right)
\Bigr].
\end{aligned}
\label{eq:PT-NN-OTOC-final}
\end{equation}
\end{widetext}

This is term-by-term oscillatory. For the exact PT spectrum, the neighbouring level
spacings satisfy
\begin{equation}
\omega_+-\omega_+^{(2)}
=
\frac{\hbar\alpha^2}{\mu},~~
\omega_-^{(2)}-\omega_-
=
\frac{\hbar\alpha^2}{\mu},
\end{equation}
which gives the beat frequency. Note again that this is a very constrained version of the total OTOC, which, even in the bound state sector would have to include all possible transitions. 

A useful consistency check is obtained from the deep-well, low-energy limit of the PT potential. In this limit, the wavefunctions (atleast $n+2$ to $n-2$) are concentrated near $x=0$, where
the quadratic approximation  is
valid. For fixed $n$ and $\Lambda\gg1$ one can show,
\begin{equation}
\omega_+,
\omega_-,
\omega_+^{(2)},
\omega_-^{(2)}
=
\frac{\hbar\alpha^2}{\mu}\Lambda
\left[
1+\mathcal{O}\left(\frac{1}{\Lambda}\right)
\right].
\end{equation}
Meanwhile, the harmonic well approximation for PT around $x=0$ gives the frequency $\omega_0=
\frac{\hbar\alpha^2}{\mu}
\sqrt{\Lambda(\Lambda+1)}$, whose large $\Lambda$ asymptotics is exactly as above. Thus all neighbouring transition frequencies become equal relative to
this common large frequency scale. The matrix amplitudes in this limit become
\begin{equation}
    A_{nn}(t)
\rightarrow
i\hbar\cos(\omega_0t),~\{|A_{n+2,n}(t)|^2
,~|A_{n-2,n}(t)|^2\}
\rightarrow
0.
\end{equation}
Thus the full bound OTOC $C_{n,\mathrm{NN}}^{(B)}(t)
\longrightarrow
\hbar^2\cos^2(\omega_0t)$, which is the exact harmonic-oscillator result \cite{Hashimoto:2017oit}, which provides a strong consistency check.

\subsection{Numerical vs analytical bound sector OTOC}

\begin{figure}[htbp]
    \centering
    \includegraphics[width=1\linewidth]{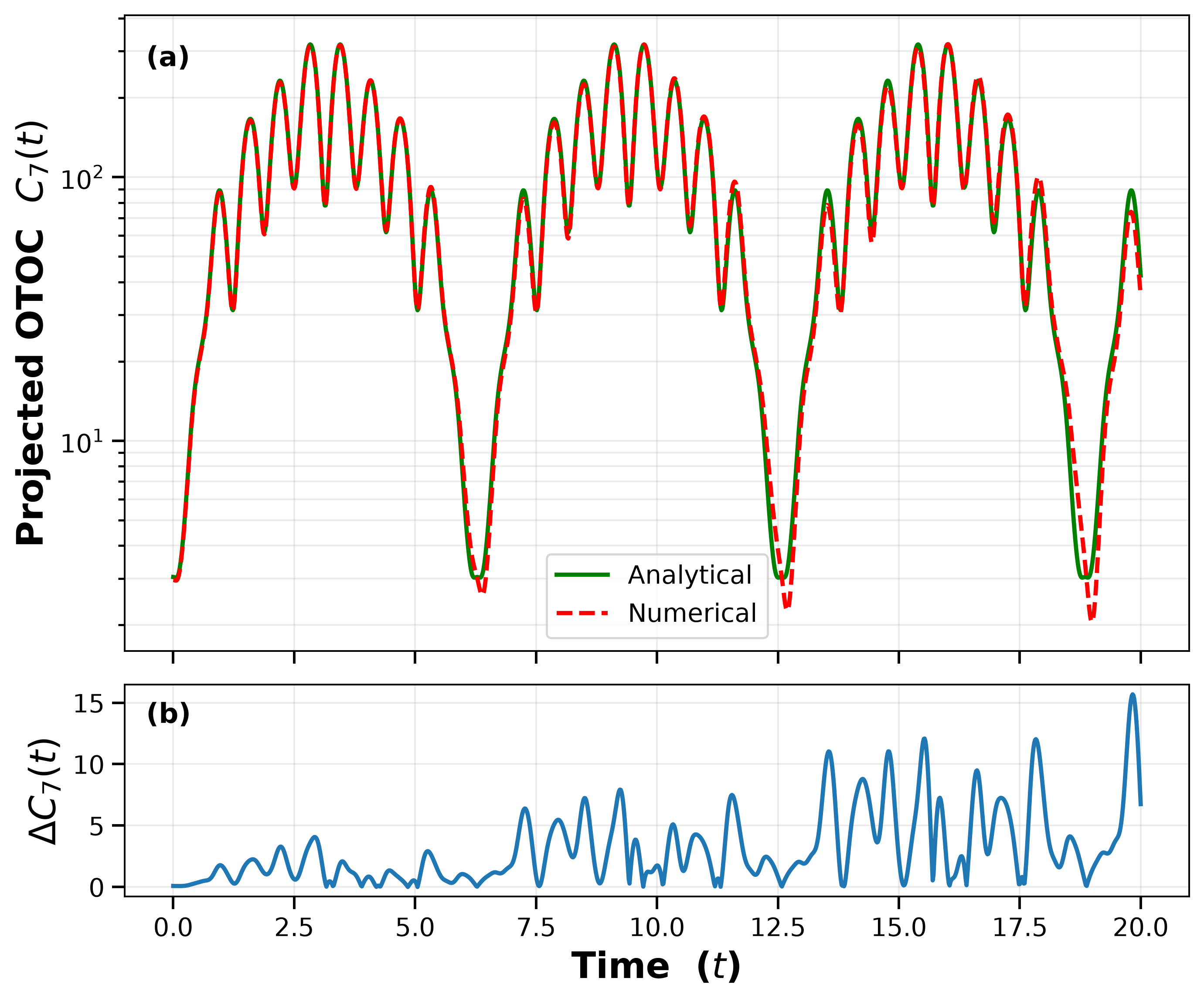}
    \caption{(a) Comparison of the projected $n=7$ OTOC obtained from the analytical Cauchy-product position matrix elements and from an independent numerical calculation. Both calculations are restricted to the bound state sector.
    (b) Absolute difference between the analytical and numerical OTOCs.}
    \label{fig:analytical_vs_numerics_otoc}
\end{figure}

To illustrate the analytical calculations of the projected OTOC, we compare the result obtained from the analytical position matrix elements (from Appendix.\eqref{appmatrixpt}) with an independent calculation using the discretized Hamiltonian. The comparison is performed for the $n=7$ bound state of the P\"oschl-Teller Potential (with the parameters as $\Lambda =12, \alpha=1, \hbar=1, \mu=1$) by summing over transition channels from all other bound states. In FIG.~\eqref{fig:analytical_vs_numerics_otoc}(a), the analytical result, obtained using the Cauchy-product expression for the position matrix elements, is compared with the corresponding numerical result. The two curves follow each other closely over the entire time interval. However, small differences become more and more pronounced at later times, due to accumulated phase error from numerical energies.
The absolute difference between analytical and numerical results,
\[\Delta C_7(t) = \left| C_7^{\mathrm{Analytical}}(t)
- C_7^{\mathrm{Numerical}}(t) \right|.\]
is plotted in FIG.~\eqref{fig:analytical_vs_numerics_otoc}(b).

\begin{figure}[htbp]
    \centering
\includegraphics[width=1\linewidth]{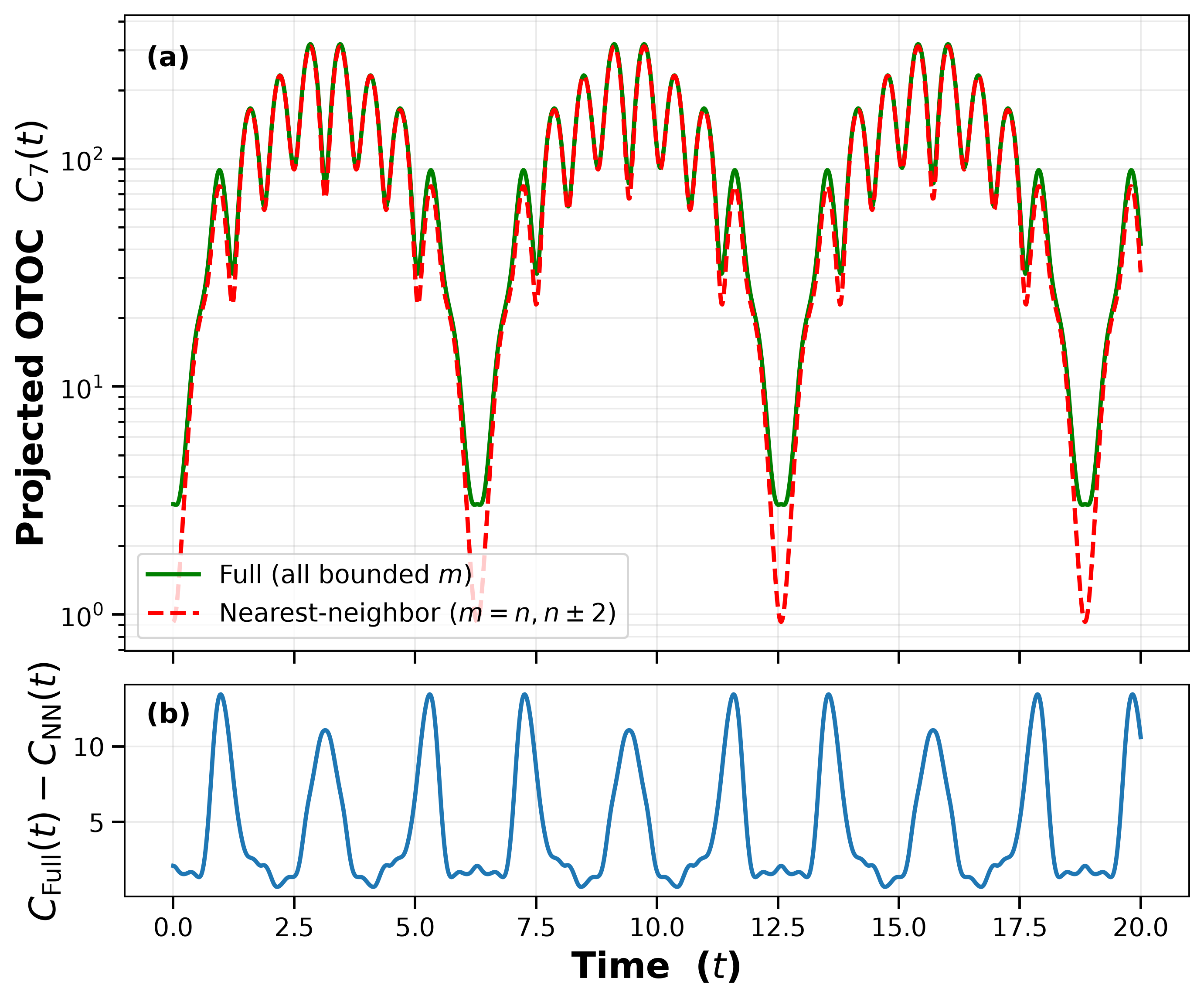}
    \caption{(a) Comparison of the full bound state projected OTOC with the nearest-neighbour approximated OTOC. Here, both are based on analytical approach.
    (b) Contribution omitted by the nearest-neighbour truncation.}
    \label{fig:C_full_vs_C_NN}
\end{figure}

For completeness, we next examine the effect of restricting the analytical sum over intermediate states to NN ones, as discussed in the last section in accordance to parity. FIG.~\eqref{fig:C_full_vs_C_NN} compares the full bound state result with the NN approximation, in which only final commutator channels $A_{mn}$ satisfying $(m-n )= 0, \pm 2$ are retained. As seen in the FIG.~\eqref{fig:C_full_vs_C_NN}(a), the nearest-neighbour result follows the same overall oscillatory pattern as the full result. The location of the main maxima and minima are similar, while the amplitudes show visible differences. The difference between these two values are plotted in FIG.~\eqref{fig:C_full_vs_C_NN}(b), where the omitted contribution is non-zero but much smaller than the full bound state projected OTOC. Therefore, the NN restriction provides a good enough approximation in this restricted manifold.

Note that the individual contributions from the channels are
\[|A_{m7}(t)|^2 \quad ; \quad\text{where} \quad A_{m7}(t) = \langle m|[x(t),p]|7\rangle .\]
So, that the full bound state projected OTOC can be written as

\[C_7(t)=\sum_m |A_{m7}(t)|^2 ,\]
where $m$ should be a bound state for the same potential. The Individual Channel-by-channel contribution in $C_7(t)$ is shown in the Table.~\eqref{tab:channel_contributions}. The Channel-by-channel analysis further shows that the dominant contributions to the projected OTOC comes from the nearest-neighbour channels ($m=n,n\pm2$ ) themselves, while the contribution from the remaining channels are comparatively smaller.

\begin{table}[H]
\centering
\caption{Channel-by-channel contributions}

\label{tab:channel_contributions}
\begin{tabular}{ccc}
\toprule
$m$ & $\left|A_{m7}(t)\right|^2$ & Channel \\
\midrule
 0   &  0              &  Beyond NN \\
 1   &  0.00070408226  &  Beyond NN \\
 2   &  0              &  Beyond NN \\
 3   &  0.74449419     &  Beyond NN \\
 4   &  0              &  Beyond NN \\
 5   &  16.354901      &  NN \\
 6   &  0              &  NN \\
 7   &  45.749637      &  NN \\
 8   &  0              &  NN \\
 9   &  35.707724      &  NN \\
 10  &  0              &  Beyond NN \\
 11  &  3.1730062      &  Beyond NN \\
\bottomrule
\end{tabular}
\end{table}

\section{Comments on Braneworld Volcano Potentials}\label{secbrane}
As discussed in the introduction, a notable situation where Volcano potentials appear is in the study of field localization in Braneworld scenarios. The bulk graviton perturbations, for example, in this case can often be reduced to a one-dimensional Schrödinger-like equation for the extra-dimensional profile of the corresponding modes. They often consist of a potential well near the location of the brane along the bulk dimensions, together with asymptotically lower barriers or tail; exactly like the Volcanic ones we have been discussing. The bound states of this effective quantum-mechanical problem then correspond to modes localized around the brane, while the continuum describes modes that can propagate away from it, i.e. leak into the bulk \cite{Maartens:2003tw}. Consequently, one would expect the curvature of the effective potential will provide a useful local characterization of the dynamics experienced by different modes. This is where our study of OTOCs could become important. 

For example, consider a $(4+1)$-dimensional warped metric,
\begin{equation}
ds^2
=
e^{2A(z)}
\left(
\eta_{\mu\nu}\,dx^\mu dx^\nu
+dz^2
\right),
\qquad
\mu,\nu=0,...3,
\end{equation}

where the conformal warp factor $e^{2A(z)}$ depends on the extra-dimensional coordinate $z$. After separating the dependence on the brane coordinates from that on the extra dimension, a bulk fluctuation can typically be written in the form $\Phi(x,z)
=
e^{-\beta A(z)}
\psi(z)\phi(x)$.
The resulting equation for the $z$ dependend mode takes the Schrödinger form with an effective potential:
\begin{equation}
\left[
-\frac{\hbar^2}{2\mu}\frac{d^2}{dz^2}
+V_{\rm eff}(z)
\right]\psi_n(z)
=
E_n\psi_n(z)
\end{equation}
where the energy eigenvalue is $E_n = \dfrac{m_n^2}{2}$ in our convention.
For many commonly studied Braneworld backgrounds, the effective potential the field feels is determined entirely by the warp factor, as the exponential factor suppresses the fields in the bulk. A frequently occurring structure is comparable to our \eqref{volca}
where the constants $a_1$ and $a_2$ depend on the bulk field and their coupling to the geometry. For spin-$2$ fields, the lowest normalizable mode $\psi_0$ is massless, and is localized around the brane, identified with the massless four-dimensional graviton.

A typical thick-brane potential for a bulk graviton can be seen in \cite{Koley2008}, which will take following form with our notations
\begin{equation}\label{eq:brane_potential}
 V_{\rm eff}(z) = -\frac{3}{4}.\frac{b^2}{(1 + b^2 z^2)} + \frac{21}{8} .\frac{b^4z^2}{(1+b^2z^2)^2}.\end{equation}
 
Here $b$ is a parameter related to the thickness of the brane, and $z$ is the conformal coordinate of the extra bulk direction. Note that the potential is asymptotically flat, going to $0^+$ at $z\to \pm \infty$. It represents a finite localized well at the origin with $V_{\rm eff}(0)=-\frac{3}{4}b^2<0$.

Now, if we perform a transformation with $bz = \sinh(x)$, we get

\begin{align}
V(x)  = \frac{15 b^2}{8}\text{sech}^2x - \frac{21 b^2}{8} \text{sech}^4 x  
\end{align}

By comparing this to standard form of Volcano Potential \eqref{volca}, where the relation is only via the shape, we get
\begin{equation} \nu=-2, \quad a_1 = \frac{21 b^2}{8} , \quad a_2 = - \frac{15b^2}{8}.\end{equation}

The $b$ here sets the overall energy scale. 
Note that we should view this transformed model as a spectral toy model motivated by the braneworld potential rather than as the actual bound graviton problem, since the coordinate transformation also changes the kinetic term. 

We stick to the $z$ coordinate and the form in \eqref{eq:brane_potential}, where the curvature for the thick-brane potential is,

\[ V''_{\rm eff}(z) = \frac{9 b^4}{4} \left[\frac{ 5b^4 z^4 -20 b^2 z^2+ 3 }{ (1+b^2z^2)^4 } \right]\]

Here $V''_{\rm}<0$  gives two symmetrical regions of negative curvature for this potential:

\[-\frac{1.9606}{b} < z < - \frac{0.3951}{b} ;~~ \frac{0.3951}{b} < z < \frac{1.9606}{b} . \]

The brane potential has only one bound state, which can be identified with  zero-energy graviton bound state with following normalized eigenfunction and energy\footnote{Note that it is a threshold bound state, but normalizable as $\int^\infty |\psi_0|^2 dz<\infty$.},
\[\psi_0(z) =  \sqrt{\frac b2} (1+b^2z^2)^{-3/4} , \qquad E_0 =0.\]
Hence, the classical turning point $z_c$ can be defined as $V_{\rm eff}(z_c) = E_0 = 0$, which gives
\[z_c = \pm \frac{1}{b} \sqrt{\frac{2}{5}} = \pm \frac{0.63246}{b} .\]

Therefore, for every value of $b$, both classical turning points of the zero-energy graviton bound state lie within the negative curvature region of the brane potential.

\begin{figure}[H]
    \centering
    \includegraphics[width=1\linewidth]{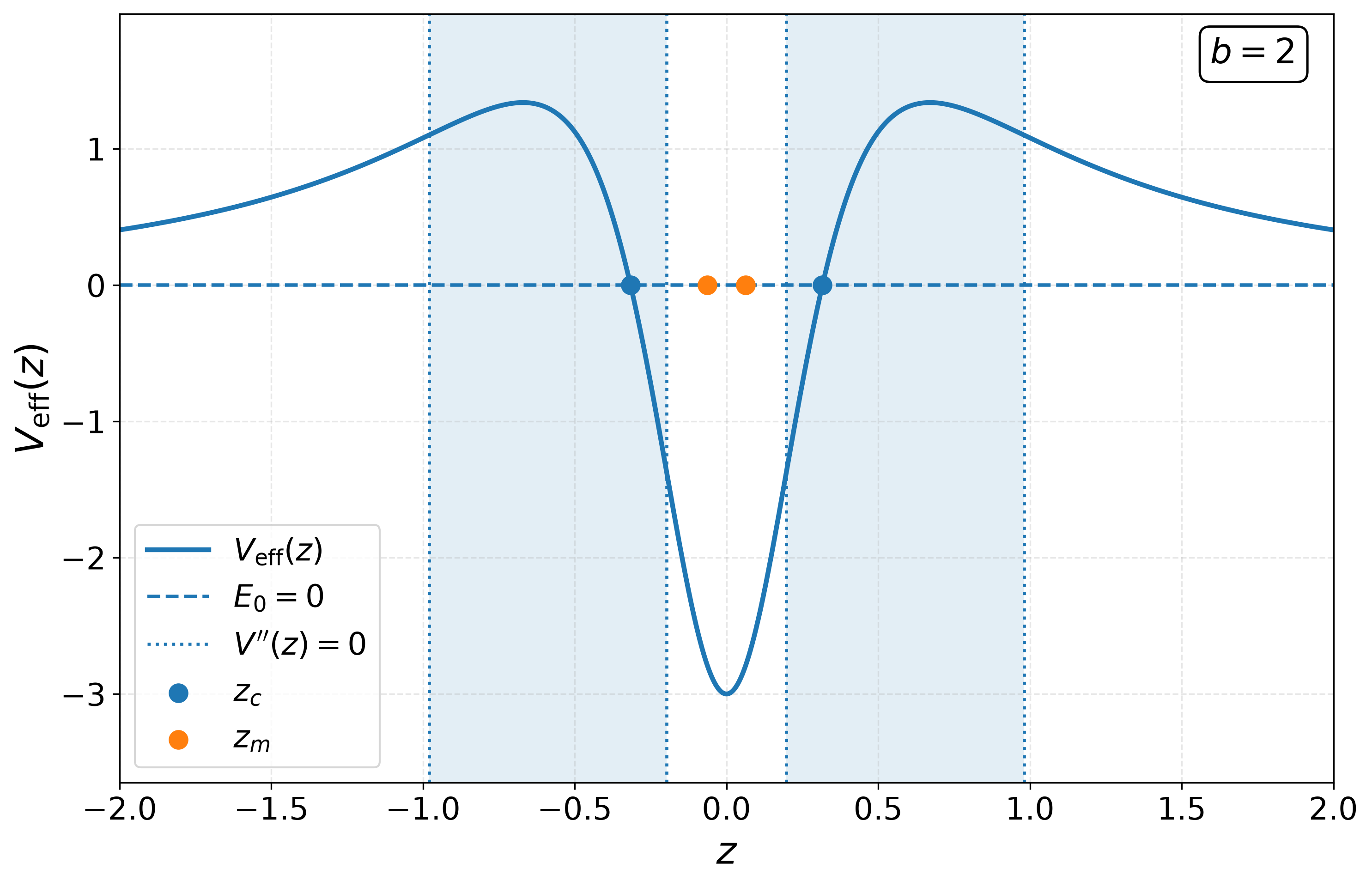}
    \caption{Thick Brane potential $V_{\rm eff}(z)$, highlighting the negative curvature regions, the zero-energy bound state $E_0=0$, and the corresponding classical turning point $z_c$ and Airy-corrected point $z_m$. }
    \label{fig:brane_potential_b_2}
\end{figure}

Now if we perform naive Airy-correction using Eq.~\eqref{eq:airy_correction}, which gives (with $\hbar=\mu=1$)
\[z_m = \mp \frac{0.12727}{b},\]
the story changes significantly.
The Airy-corrected outermost turning points ($z_m$) of the the zero-energy graviton bound state seems to lie well within the positive-curvature region of the braneworld potential (see FIG.\eqref{fig:brane_potential_b_2} for example), unlike classical turning points. Therefore, we can't expect even transient instabilities in the corresponding OTOCs.

But there is something more than what meets the eye in this case. The zero mode solution here is nodeless and non-oscillatory. In fact the probability density decreases monotonically away from the brane and has only one maximum. So, even though the Airy local solution is valid near $z_c$, there are no oscillations whose outermost maximum can be morphed into the first Airy maximum. Hence, a single Airy-corrected
sampling point can be misleading. Further, one can check that the exact zero mode has a substantial support in the negative curvature region and a substantial tail beyond the classical turning points.  A useful quantum diagnostic of the curvature sampled by the bound state is then provided by the expectation value of the potential curvature,

\[\langle V''_{\rm eff}\rangle =  \int_{-\infty}^{\infty} {\psi^*(z)  V''_{\rm eff}(z) \psi(z)~dz }\]

For normalized zero-energy graviton state and curvature of brane potential, this takes following form

\[ \langle V''_{\rm eff}\rangle_0 = \frac{9b^{5}}{8} \int_{-\infty}^{\infty} \frac{5b^{4}z^{4}-20b^{2}z^{2}+3} {\left(1+b^{2}z^{2} \right)^{11/2}} \,dz.\]

For $u=bz, dz=du/b$ this becomes
\[ \langle V''_{\rm eff}\rangle_0= \frac{9b^{4}}{8} \int_{-\infty}^{\infty} \frac{5u^{4}-20u^{2}+3} {\left(1+u^{2}\right)^{11/2}} \,du.\]

Now, using standard integral formulae, this can be evaluated in terms of Beta functions, as for $m > k + \frac{1}{2}$:
\begin{align*}
    \int_{-\infty}^{\infty} \frac{u^{2k}}{\left(1+u^{2}\right)^{m}} \,du 
    & = B \left( k + \frac{1}{2}, m-k - \frac{1}{2} \right) \\
    & = \frac{\Gamma\!\left(k+\tfrac12\right)\Gamma\!\left(m-k-\tfrac12\right)} {\Gamma(m)},
\end{align*}

and the required expectation value evaluates to,
\[\langle V''_{\rm eff}\rangle_0  = \frac{26}{35} \, b^4 >0.\]

Thus, although the wavefunction has extended regions of support in the negative curvature region, the global signed curvature expectation value of the zero mode is still positive. This suggest that the OTOC associated with the zero mode should reflect the stabilizing character of the positive quantum-averaged curvature\footnote{One can explicitly check at early times, $[x(t),p]=
i\hbar
\left[
1-\frac{V''(x)}{2\mu}t^2+\cdots
\right]$, so roughly $C_n(t)\simeq\hbar^2\left[
1-\frac{\langle V''\rangle_n}{\mu}t^2+\cdots
\right]$. In this case at genuinely short times, the full auxiliary OTOC initially decreases rather than grows due to the sign of $\langle V''\rangle_n$.}. This can be compared to our other cases, like the PT potential case, for high-lying states, the outer turning region can dominate semiclassical sensitivity, making a local Airy criterion useful. For nodeless or broadly distributed states like the graviton zero mode, it is not a reasonable local diagnostic anymore.

With only one bound state, the projected bound-only OTOC become identically zero as the projection operator $P_B = |0\rangle\langle0|$  gives zero position and momentum matrices. The nontrivial contributions to OTOC comes entirely from the continuum. But of course, we need to be cautious in what we mean by OTOC in this case, as the matrix elements belong to the auxiliary
one-dimensional Schr\"odinger problem governing the modes. Its evolution parameter may not be identified
directly with the physical Lorentzian time of the bulk theory. So an interpretation of this OTOC may say little about the dynamics of the zero-mode graviton itself.

\section{Discussions and Conclusions}\label{sec6}
Motivated by their appearance in various branches of theoretical physics, in this work we examined the dynamics of out-of-time-ordered correlators in Volcano potentials. For mixed curvature potentials, the physical lesson is quite clear: exponential OTOC growth could just be a signature of finite window hyperbolic growth due to local inverted oscillator dynamics, not quantum chaos. This notion is further refined when discussing classical and quantum regimes: a
criterion based solely on the classical turning point can identify
the onset of negative curvature too early, as the outermost node of quantum wavefunction actually peaks before reaching the classical turning point, and connects oscillatory allowed-region behaviour to an exponentially decaying forbidden tail. Hence we used Airy-corrected estimates, wherever such estimates are controlled, in the quantum case and compared them explicitly to the one without correction. For the PT potential case, we could do this analytically to find the sensitivity and crossover points in both regimes. The same problem also allowed us to clarify the pure
bound state contribution to the quantum OTOC. Projecting onto the finite bound state spectral subspace gives a well-defined
finite-dimensional analytic benchmark for the OTOC itself, which we could then compare with numerics.

One has to note that our parameter space in this case is quite special, since we focus mostly on the sensitivity of the OTOC starting from the bound initial states. We had also chosen this regime to make sure we have enough bound states to ensure non-trivial contribution from them. We have made the logic of this clear at various places in the text, and further discussed other parameter spaces where one might find bound states. However, despite this choice, the correlation between curvature and sensitivity persists across the parameter sets studied. This makes physically interesting applications, like the thick braneworld potentials discussed in Sec.\eqref{secbrane}, with one isolated bound system, quite alluring to follow-up on.

There are a few different ways we would like to extend our current study, since it gives a very nice playground for working with local instabilities in quantum potentials. First, it would be useful to test the relation between local
negative curvature and transient OTOC growth in other one-dimensional
mixed-curvature systems, and if possible formalize unique analytic crossover criterion. But since one-dimensional systems are integrable, more substantial extension would be to consider systems with two or more degrees
of freedom, where genuine Hamiltonian chaos becomes possible. One then needs to ask, can OTOC distinguish between a local change in curvature and a chaotic phase space at the state-resolved level? This may be an even more pertinent question to ask in the case of stadium billiard problems with mixed curvature of the boundary \cite{Das:2025tuc}, with focusing and defocusing walls, where true quantum chaos can occur. In these situations chaotic regimes coexist with regular trajectories, and it would be an interesting complementary exercise to see how OTOCs behave while transitioning between these boundaries.

Finally, it would be interesting to investigate related
PT-type spectral problems arising in AdS$_2$ and their
possible connection to JT gravity \cite{Almheiri:2014cka, Maldacena:2016hyu, Jensen:2016pah}. Scalar and higher-spin wave equations in global AdS backgrounds can often be reduced, after separation of variables, to finite-interval Schr\"odinger problems
with PT-type potentials, with the AdS
boundary providing the corresponding confining boundary conditions. An interesting question would be whether this spectral problem, probed through OTOCs, can provide a useful
quantum-mechanical representation of some aspects of the JT physics. 

We would like to come back to these questions in future communications.

\section*{Acknowledgements}
ABan acknowledges Mr. Keerthan Shriram Karvaje for his interest during the early stages of this work. 

ABan was supported in part by an OPERA grant and a seed grant NFSG/PIL/2023/P3816 from BITS-Pilani, and further an early career research grant ANRF/ECRG/2024/002604/PMS from ANRF India. He also acknowledges financial support from the Asia Pacific Center for Theoretical Physics (APCTP) via an Associate Fellowship.

RJ gratefully acknowledges the University Grants Commission (UGC), Government of India, for financial support through the Junior Research Fellowship (JRF) (NTA Ref. No.: 231610192580).

RS would like to thank BITS Pilani for institutional encouragement of undergraduate research.

Some of the numerics used in this manuscript have been augmented by the use of a LLM. 

\vspace{0.5cm}

\appendix

\section{Exact bound state solutions}\label{Appendix_D}
\subsection{Special parameter space}
For the general volcano potential (which goes to negative infinity asymptotically), the natural quantum mechanical spectrum does not consist of true bound states. However, there are some analytically tractable regions with isolated bound states.
Throughout this appendix we set $\hbar=\mu=1$, so that the kinetic
term in the Schr\"odinger equation is $-\tfrac{1}{2}\psi''$, which is different from the convention of \cite{Koley:2006ku}. The
generic Schr\"odinger equation for the Volcano potential then reads
\begin{equation}
  \psi''(x)
  +
  2\left[
  E
  +
  a_1\cosh^{2\nu}x
  +
  a_2\sech^2x
  \right]\psi(x)
  =
  0.
  \label{eq:schrodinger}
\end{equation}

The remarkable feature of the construction in \cite{Koley:2006ku} is that a special relation among the parameters permits an exact solution:  $ a_2 =  \frac{\nu}{8}\left(\nu+2\right),
  ~ E= -\frac{\nu^2}{8}$.
One can see how to get this, by considering
\begin{equation}
  \psi(x)
  =
  (\cosh x)^{-\nu/2}F(x).
  \label{eq:ansatz}
\end{equation}
Define
\begin{equation}
  u(x)
  =
  (\cosh x)^{-\nu/2} \implies
  \frac{u'}{u}
  =
  -\frac{\nu}{2}\tanh x,
\end{equation}
and
\begin{align}
  \frac{u''}{u}
  &=
  -\frac{\nu}{2}\sech^2x
  +
  \frac{\nu^2}{4}\tanh^2x
  \\
  &=
  \frac{\nu^2}{4}
  -
  \frac{\nu(\nu+2)}{4}\sech^2x.
\end{align}
Substitution into \eqref{eq:schrodinger} yields:
\begin{widetext}
\begin{equation}
  F''
  -
  \nu\tanh x\,F'
  +
  \left[
    2a_1\cosh^{2\nu}x
    +
    2E+\frac{\nu^2}{4}
    +
    \left(
      2a_2-\frac{\nu(\nu+2)}{4}
    \right)\sech^2x
  \right]F
  =
  0.
  \label{eq:F-x}
\end{equation}
\end{widetext}
At $E=-\nu^2/8$ and $a_2=\nu(\nu+2)/8$, this simplifies and reduces to
\begin{equation}
  F''
  -
  \nu\tanh x\,F'
  +
  2a_1\cosh^{2\nu}x\,F
  =
  0.
  \label{eq:F-special}
\end{equation}
Note that this is a very special point in the parameter space.
This is now in canonical form, and we can introduce the new coordinate
\begin{equation}
  y(x)
  =
  \int^x(\cosh t)^\nu\,d t,
  \label{eq:liouville-coordinate}
\end{equation}
which reduces this to the normal form
\begin{equation}
  F_{yy}+2a_1F=0.
  \label{eq:free-y}
\end{equation}
Thus the exact solution is
\begin{equation}
  F(y)
  =
  C_1\cos(\sqrt{2a_1}\,y)
  +
  C_2\sin(\sqrt{2a_1}\,y),
\end{equation}
which again was only possible due to a special choice of parameters.
Returning to $x$, the full solution looks like:
\begin{widetext}
\begin{equation}
  \psi(x)
  =
  \frac{
    C_1\cos\left[
      \sqrt{2a_1}
      \displaystyle\int^x(\cosh t)^\nu\dd t
    \right]
    +
    C_2\sin\left[
      \sqrt{2a_1}
      \displaystyle\int^x(\cosh t)^\nu\dd t
    \right]
  }{
    (\cosh x)^{\nu/2}
  }
  \label{eq:KK-exact}
\end{equation}
\end{widetext}
This solution can now be appropriately normalized.
Note that the integral for the new coordinate in \eqref{eq:liouville-coordinate} can be written as a hypergeometric function
\begin{equation}
  y(x)
  =
  \sinh x\,
  {}_2F_1
  \left(
    \frac12,
    \frac{1-\nu}{2};
    \frac32;
    -\sinh^2x
  \right),
  \label{eq:y-hypergeom}
\end{equation}
up to an additive constant. For example,
\begin{align}
  \nu=1:
  &\qquad
  y=\sinh x, \nonumber
  \\
  \nu=2:
  &\qquad
  y=\frac{x}{2}+\frac{\sinh2x}{4},\nonumber
  \\
  \nu=3:
  &\qquad
  y=\sinh x+\frac13\sinh^3x.
\end{align}
But note that for fixed $\nu$, this is just one set of bound states with a distinguished exactly solvable energy. The square integrability of both independent asymptotic solutions is
also the reason that some care is required in interpreting
\eqref{eq:KK-exact} as a quantum-mechanical bound state. For
$a_1>0$ and $\nu>0$, the potential falls to $-\infty$ sufficiently
rapidly that the endpoints are of limit-circle type. Normalizability
alone therefore does not select a unique solution or a unique
self-adjoint Hamiltonian. A choice of self-adjoint boundary conditions
at $x=\pm\infty$ is additionally required. The special relation for eigenvalues should therefore be interpreted as identifying
an \emph{exactly solvable energy and corresponding pair of
square-integrable solutions}; whether this energy belongs to the
spectrum as an eigenvalue depends on the chosen self-adjoint
realization of the Hamiltonian.

\subsection{Solvability at $\nu=-2$}The potential here becomes
\begin{equation}
V(x)=-a_1\sech^4x-a_2\sech^2x.
\end{equation}
Throughout we take $a_1>0,a_2>0$ as in the family of FIG.\eqref{fig:potential_and_curvature_a1_positive}.
With this sign choice, $V(x)<0$ for all $x$, $V(0)=-a_1-a_2$, and
\begin{equation}
V(x)\to 0^-\qquad\text{as }|x|\to\infty.
\end{equation}
This is a \emph{finite attractive well}, in contrast to the runaway case $a_1>0,\nu>0$, where $V\to-\infty$. But the situation is not very different from the bound wells we have discussed in the main text. We again set $\hbar=\mu=1$, so that the kinetic term in the Schr\"odinger equation is $-\tfrac{1}{2}\psi''$. Since $V(x)\to 0^-$ as $|x|\to\infty$, any bound state must have $E<0$ $(E=-\kappa^2/2$ with $\kappa>0)$. Set $z=\tanh x$ to get
\begin{equation}
  \frac{\dd}{\dd x}
  =
  (1-z^2)\frac{d}{dz}.
\end{equation}
Consequently,
The Schr\"odinger equation becomes
\begin{equation}
\psi_{zz}-\frac{2z}{1-z^2}\psi_z
+\left[-\frac{\kappa^2}{(1-z^2)^2}
+2a_1+\frac{2a_2}{1-z^2}\right]\psi=0,
\label{eq:psi-eq}
\end{equation}
which has regular singular points at $z=\pm1$ and an irregular singular point at $z=\infty$; it is a confluent Heun equation. Accordingly we make the ansatz
\begin{equation}
\psi(z)=(1-z^2)^{\kappa/2}\mathcal{P}(z)
\label{eq:ansatzheun}
\end{equation}
Substituting \eqref{eq:ansatzheun} into \eqref{eq:psi-eq} and dividing by $(1-z^2)^{\kappa/2}$, one obtains the following equation for the undetermined $\mathcal{P}$:
\begin{equation}
(1-z^2)\mathcal{P}''-2(\kappa+1)z\mathcal{P}'
+\left[B-2a_1z^2\right]\mathcal{P}=0,
\label{eq:Peq}
\end{equation}
where
\begin{equation}
B\equiv 2a_1+2a_2-\kappa(\kappa+1).
\label{eq:B}
\end{equation}
For generic $a_1,a_2,\kappa$, the solution of \eqref{eq:Peq} is a Heun function, and it is \emph{not} a polynomial. We now ask when \eqref{eq:Peq} admits a \emph{polynomial} solution $\mathcal{P}_n(z)$ of degree $n$, i.e. whether there is a quasi-exactly-solvable (QES) sector. Write
\begin{equation}
\mathcal{P}_n(z)=\sum_{j=0}^{n}c_jz^j.
\end{equation}
Substituting into \eqref{eq:Peq} and collecting powers of $z^m$ gives the three-term recurrence
\begin{widetext}
\begin{equation}
(m+2)(m+1)\,c_{m+2}
+\left[B-m^2-(2\kappa+1)m\right]c_m
-2a_1\,c_{m-2}=0,
\label{eq:recurrence}
\end{equation}
\end{widetext}
valid for $m\ge0$, with
\begin{equation}
c_{-1}=c_{-2}=0.
\end{equation}
The recurrence couples $c_{m+2}$, $c_m$, and $c_{m-2}$. Consequently the even and odd coefficients decouple: the polynomial has definite parity.

However, for $a_1\neq0$ this recurrence cannot terminate. To see this
directly, suppose that there were a polynomial solution of degree $n$,
\begin{equation}
\mathcal{P}_n(z)=c_nz^n+\cdots,
\qquad
c_n\neq0.
\end{equation}
In \eqref{eq:Peq}, all derivative terms have degree at most $n$,
whereas the term $-2a_1z^2\mathcal{P}_n(z)$
contains the contribution
\begin{equation}
-2a_1c_nz^{n+2}.
\end{equation}
There is no other contribution of degree $n+2$ that can cancel this
term. Therefore a polynomial solution would require
\begin{equation}
a_1c_n=0.
\end{equation}
Since $a_1>0$ and $c_n\neq0$, this is impossible.
The same obstruction follows directly from the recurrence relation above.
For a polynomial of degree $n$ one has $c_{n+2}=c_{n+4}=0$.
Setting $m=n+2$ in \eqref{eq:recurrence} then gives
$-2a_1c_n=0$,
which again contradicts $a_1>0$ and $c_n\neq0$. The same obstruction holds for $a_1<0$; the sign of $a_1$ is irrelevant to the termination argument.

The ansatz \eqref{eq:ansatzheun} does \emph{not} generate a finite
polynomial sector analogue to QES. The bound state wavefunctions instead have the
form
\begin{equation}
\psi_n(x)=(\sech x)^{\kappa_n}\mathcal{P}_n(\tanh x),
\end{equation}
with $\mathcal{P}_n$ in general, a non-terminating solution of the
confluent Heun equation with parity $(-1)^n$, and $\kappa_n$ is related to the energy at level $n$ by $E_n=-\kappa_n^2/2$. The eigenvalue condition is that $\mathcal{P}_n(z)$ be regular at $z=\pm1$; this selects the discrete values $\kappa_n$. The normalization integral in $z$ variable is
\begin{equation}
\int_{-\infty}^{\infty}|\psi_n(x)|^2\dd x
=\int_{-1}^{1}(1-z^2)^{\kappa_n-1}|\mathcal{P}_n(z)|^2\dd z.
\end{equation}
For $\mathcal{P}_n(z)=\sum_jc_jz^j$, this becomes
\begin{equation}
\sum_{j,k}c_jc_k\int_{-1}^{1}(1-z^2)^{\kappa_n-1}z^{j+k}\dd z.
\end{equation}
The integrals vanish for $j+k$ odd, and for $j+k=2m$ even they are Beta functions \cite{GradshteynRyzhik2014}:
\begin{equation}
    \int_{-1}^{1}(1-z^2)^{\kappa-1}z^{2m}\dd z
=B\!\left(\kappa,m+\tfrac12\right)
=\frac{\Gamma(\kappa)\Gamma(m+\frac12)}{\Gamma(\kappa+m+\frac12)}
\end{equation}
The polynomial obstruction disappears in the special limit $a_1=0$.
In this case \eqref{eq:Peq} reduces to
\begin{equation}
(1-z^2)\mathcal{P}''
-2(\kappa+1)z\mathcal{P}'
+B\mathcal{P}=0,
\end{equation}
with
\begin{equation}
B=2a_2-\kappa(\kappa+1).
\end{equation}
For a polynomial of degree $n$, the coefficient of the highest power
requires
\begin{equation}
B=n(n+2\kappa+1),
\end{equation}
and hence $2a_2=(n+\kappa)(n+\kappa+1)$.
Defining
\begin{equation}
\Lambda
=
\frac{\sqrt{1+8a_2}-1}{2},
\qquad
2a_2=\Lambda(\Lambda+1),
\end{equation}
one obtains
\begin{equation}
\kappa_n=\Lambda-n,
\qquad
E_n=-\frac{1}{2}(\Lambda-n)^2,
\end{equation}
with the normalizability condition
$\kappa_n=\Lambda-n>0$.
The corresponding polynomial is a Gegenbauer polynomial,
\begin{equation}
\mathcal{P}_n(z)\propto
C_n^{\kappa_n+\frac12}(z),
\end{equation}
so that
\begin{equation}
\psi_n(x)
\propto
(\sech x)^{\kappa_n}
C_n^{\kappa_n+\frac12}(\tanh x).
\end{equation}
This is the exactly solvable attractive PT well limit. With the present convention $\hbar=\mu=1$ and $\alpha=1$, the parameter defined here coincides with the $\Lambda$ of the main text , since $\Lambda(\Lambda+1)=2\mu V_0/(\hbar^2\alpha^2)$ reduces to $\Lambda(\Lambda+1)=2a_2$ for a $\sech^2$ well of depth $V_0=a_2$.

\section{Potential and Curvature for various \texorpdfstring{$\nu$}{nu} }\label{AppA}
Let us quickly, for completeness, discuss some parameter spaces of the potential
\begin{equation}V(x) = -\left(a_1 \cosh^{2\nu} x + a_2 \text{sech}^2 x \right)\end{equation}
which falls outside of the regime considered in the main text.

\subsubsection*{\textbf{(I) $|a_1|$ and $a_2$ are comparable $[a_1 = -100 , a_2 = 120]$}}

\begin{figure}[H]
    \centering
    \includegraphics[width=0.99\linewidth]{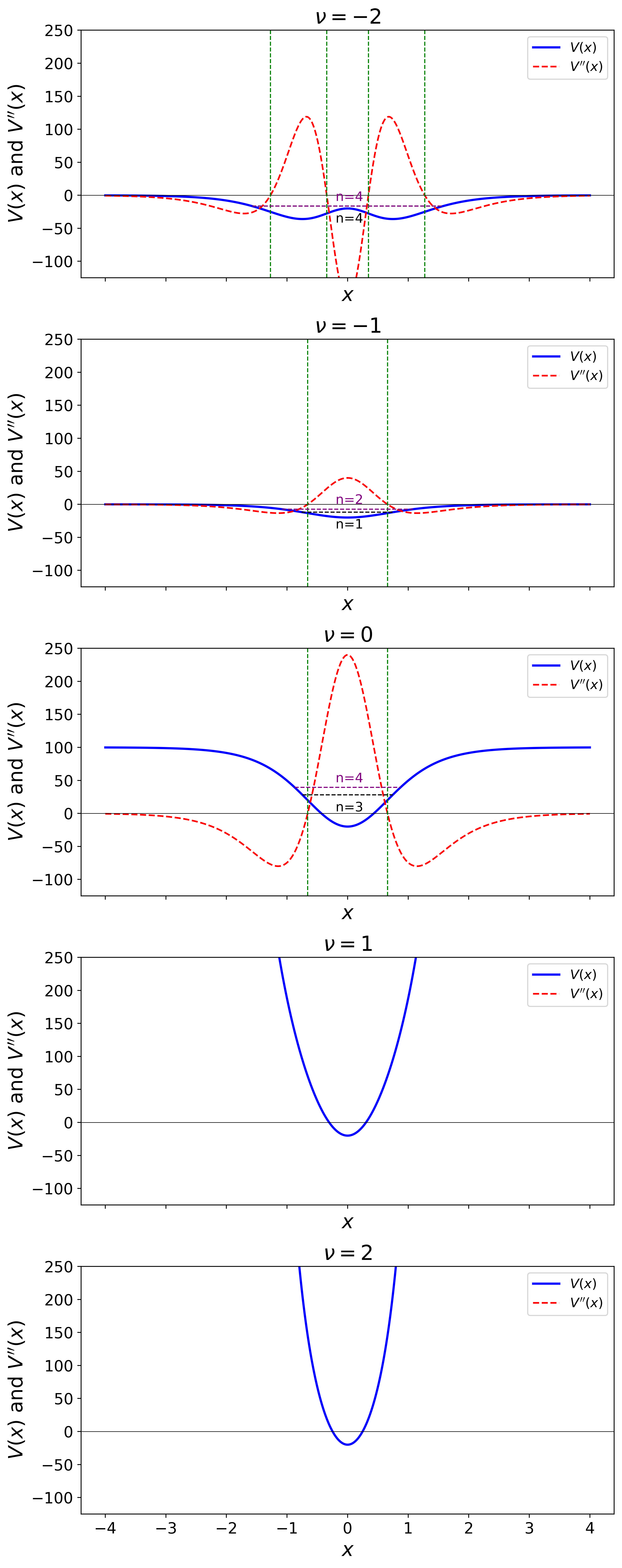}
    \vspace{-2mm}
    \caption{Potential (blue) and Curvature (red dashed) for $a_1 = -100 , a_2 = 120$.}
    \label{fig:potential_and_curvature_a1_comp_a2}
\end{figure}

Note that negative curvature regions don't exist for $\nu>0$. See FIG.\eqref{fig:potential_and_curvature_a1_comp_a2} for some representation.

\subsubsection*{\textbf{(II) $|a_1|>> a_2$  and  $a_1 = -100 , a_2 = 0.05$}}
\begin{figure}[H]
    \centering
    \includegraphics[width=0.99\linewidth]{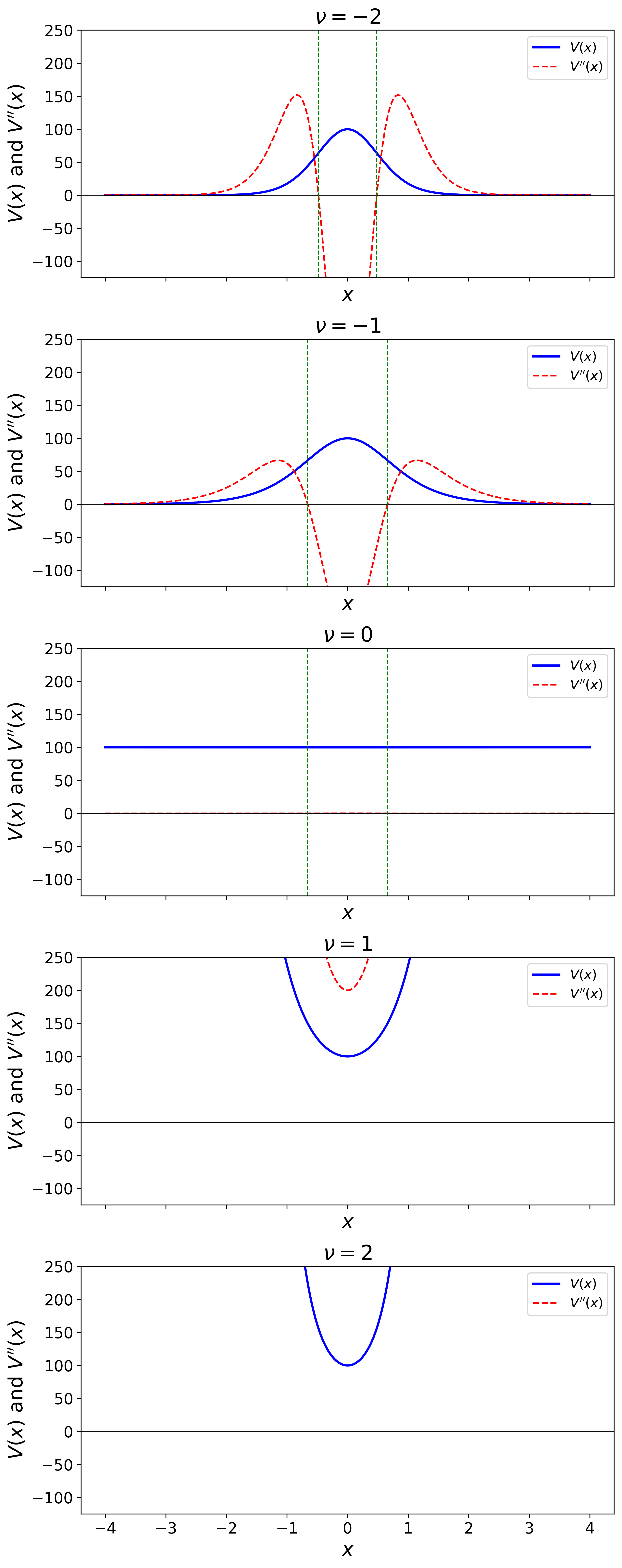}
    \vspace{-2mm}
    \caption{Potential (blue) and Curvature (red dashed) for $a_1 = -100 , a_2 = 0.05$.}
    \label{fig:potential_and_curvature_-a1_bigger_a2}
\end{figure}

Bound states here exist only if $\nu > 0$, which have no negative curvature region. See FIG.\eqref{fig:potential_and_curvature_-a1_bigger_a2} for some representation.

\subsubsection*{\textbf{(III) $a_1 = +ve$ and $a_1$ $>>$ $a_2$  $[a_1 = 100 , a_2 = 0.05]$}}
\begin{figure}[H]
    \centering
    \includegraphics[width=\linewidth]{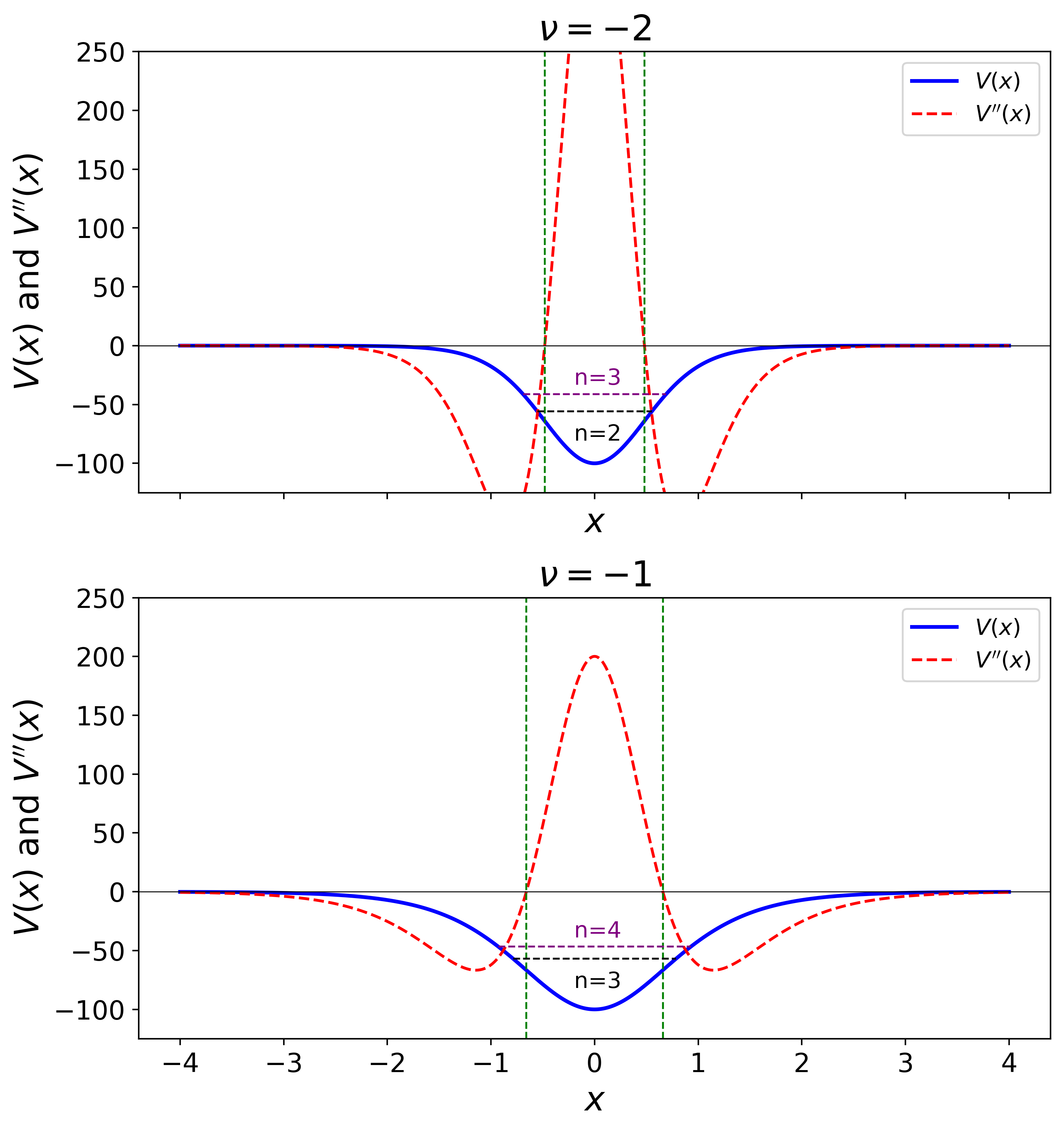}
    \caption{Potential (blue) and Curvature (red dashed) for $a_1 = 100 , a_2 = 0.05$}
    \label{fig:potential_and_curvature_a1_positive}
\end{figure}

Bound states in this case exist only if $\nu < 0$.See FIG.\eqref{fig:potential_and_curvature_a1_positive} for some representation.

\begin{table*}[htbp]
\centering
\caption{Summary of the properties of different parameter spaces and their suitability for the present analysis.}
\renewcommand{\arraystretch}{1.4}
\setlength{\tabcolsep}{5pt}

\begin{tabular}{|c|c|c|c|c|}
\hline
\makecell{\textbf{Parameter space}\\\textbf{$(a_1,a_2)$}}
&
\textbf{Bound states}
&
\makecell{\textbf{Bound states in}\\ \textbf{negative curvature region}}
&
\textbf{Remark}
&
\textbf{Figure} \\
\hline

\makecell{$a_1>0$\\$(100,0.05)$}
&
\makecell{Exist only for\\$\nu<0$}
&
\makecell{Not present for\\the complete set of\\$\nu=\{-2,-1,0,1,2\}$}
&
\makecell{Non-Hermitian for $\nu>0$;\\
therefore unsuitable for the\\
present analysis}
&
FIG.~\eqref{fig:potential_and_curvature_a1_positive}
\\
\hline

\makecell{$a_1<0,\ |a_1|\gg a_2$\\$(-100,0.05)$}
&
\makecell{Exist only for\\$\nu>0$}
&
\makecell{negative curvature region\\
exists only for $\nu<0$}
&
\makecell{Bound states and negative\\
curvature do not coexist;\\
therefore unsuitable}
&
FIG.~\eqref{fig:potential_and_curvature_-a1_bigger_a2}
\\
\hline

\makecell{$a_1<0,\ |a_1|\sim a_2$\\$(-100,120)$}
&
\makecell{Exist for all\\
$\nu=\{-2,-1,0,1,2\}$}
&
\makecell{Absent for $\nu>0$;\\
present for $\nu<0$}
&
\makecell{Suitable only for\\
$\nu<0$; unsuitable for\\
the full $\nu$ range}
&
FIG.~\eqref{fig:potential_and_curvature_a1_comp_a2}
\\
\hline

\makecell{$a_1<0,\ |a_1|\ll a_2$\\$(-0.05,120)$}
&
\makecell{Exist for all\\
$\nu=\{-2,-1,0,1,2\}$}
&
\makecell{Present for the\\
considered values of $\nu$;\\
bound states lie in this region}
&
\makecell{\textbf{Suitable parameter space}\\
for the present study}
&
FIG.~\eqref{fig:classical_turnning_curvature},~\eqref{fig:Airy_corrected_curvature}
\\
\hline

\end{tabular}
\end{table*}

\section{Numerical robustness and Box-Size dependence of the Continuum contribution}\label{AppConv}
In our numerical OTOC calculation, we discretize the Schr\"odinger equation on a finite spatial interval $x \in [-L,L]$ using a uniform grid. For $N$ grid points, the grid spacing is 
\[ dx=\frac{2L}{N-1}.\]

The hamiltonian is then diagonalized on this finite grid, giving a discrete set of numerical eigenstates.

For the volcano potential with \(a_1=-0.05, a_2=120, \nu=-2\); there are total 15 bound states ($n=0$ to $n=14$). The positive-energy part of the spectrum represents the continuum. In the numerical calculation, this continuum is box-discretized in exactly the same spatial grid and Hamiltonian diagonalization as the bound states. The continuous spectrum is therefore represented by a discrete set of positive-energy box states (continuum pseudostates), rather than by an explicit energy integral. The OTOC calculation uses the complete numerical energy basis, so these continuum pseudostates are included automatically.

An important consequence is that the continuum contribution depends on the box size $L$. Even when $dx$ is kept fixed, increasing $L$ changes the number of grid points $N$ and therefore changes the number of the discretized continuum states. Thus, the continuum contribution to $C_n(t)$, and associated dephasing, can show a residual dependence on $L$. This dependence is not a change in the physical potential or in the number of true bound states, but a consequence of how the continuum is represented numerically.

\begin{figure}[htbp]
    \centering
    \includegraphics[width=1\linewidth]{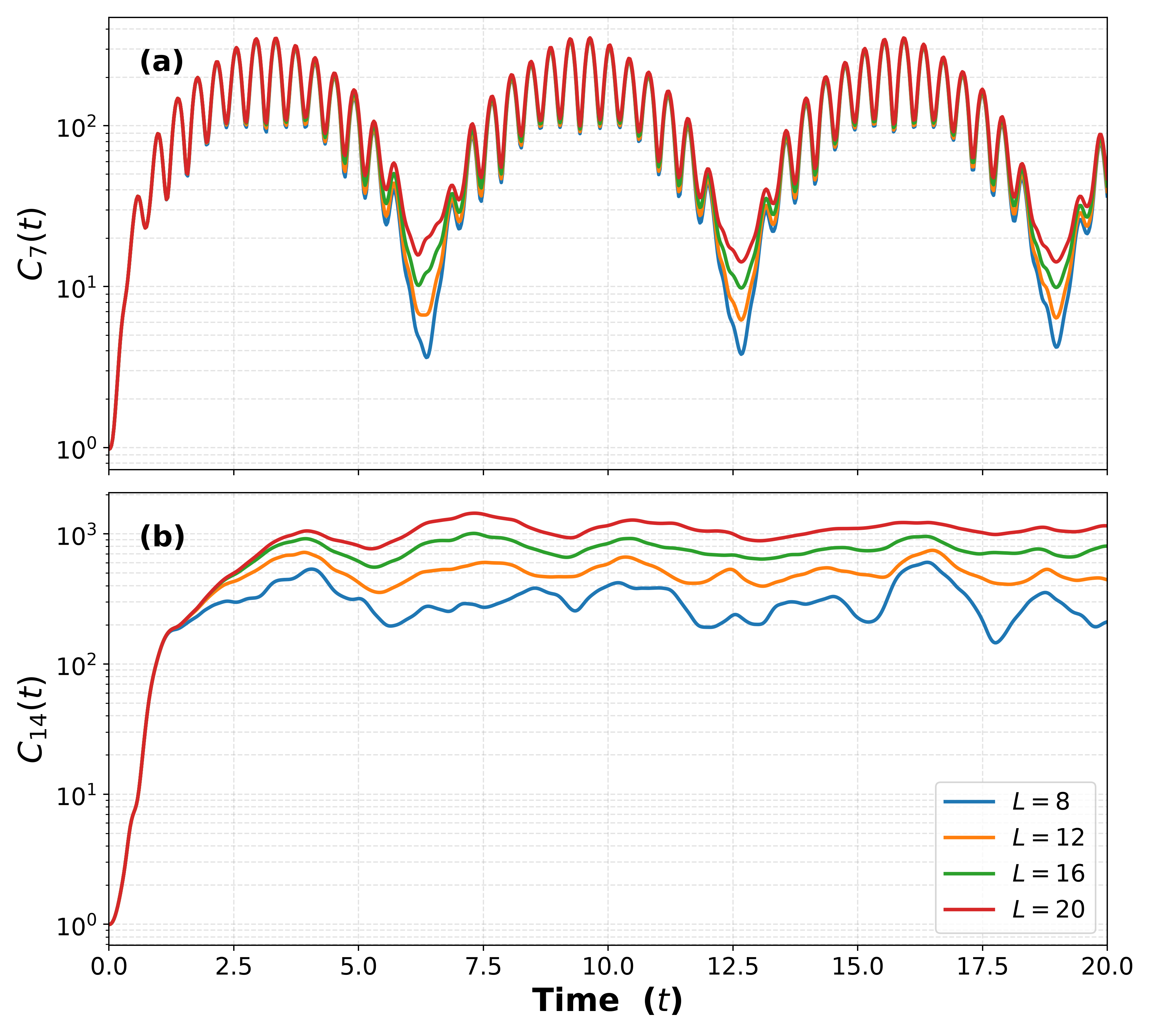}
    \caption{Box-size dependence of the full-spectrum OTOC for the volcano potential with $a_1=-0.05, a_2=120,\nu=-2$; is shown for two bound states (a) $n=7$ and (b) $n=14$.}
    \label{fig:box_size_dependence_of_continuum}
\end{figure}

The effect is more pronounced for state close to dissociation, whose wavefunctions have larger overlap with low-lying continuum states. For this specific case $n=14$ is the highest bound state and lies closest to the continuum threshold. As a result, the continuum contribution and the continuum induced dephasing is more prominent in its OTOC. Therefore, $C_{14}(t)$ shows a stronger dependence on $L$. In contrast, a lower, more deeply bound state such as $n=7$ is less affected by the continuum. Consequently, $C_7(t)$ is much less sensitive to the choice of $L$.

The box-size convergence plots therefore provide a useful numerical robustness check. The OTOC curves for the deeply bound state remain comparatively stable as $L$ is varied, whereas the larger variation observed for the near-dissociation state is consistent with its stronger continuum sensitivity.

Although the early-time OTOC behaviour is well converged across different box sizes $L$, the OTOC evolution of states close to dissociation at longer times cannot be regarded as completely box-independent. Thus, quantitative OTOC results for these states at longer times should be interpreted with caution. In contrast, the early-time regime shows good convergence for all considered $L$, making the extracted short-time behaviour more robust. See FIG.\eqref{fig:box_size_dependence_of_continuum} for some visualization.

This box-size dependence occurs only for the $\nu \le 0 $ cases, where the potential supports a continuum. For $\nu>0$, the potential is confining and the spectrum is entirely discrete. Hence, there is no continuum contribution and no associated box-size dependence of this origin in the $\nu>0$ cases.

\section{Effect of changing potential strength parameters}\label{Appendix_B}

In the numerical analysis presented in main text, we fixed our values of $a_{1,2}$ and varied values of $\nu$ to find regimes with ample number of bound states for our OTOC study.
Now for completeness, in this section, we briefly analyze the OTOC growth fixing the $\nu=1$ case across different $a_{1,2}$ parameter regimes, provided bound states also appear. In the parameter sets examined here, the observed transient growth correlates strongly with whether the Airy turning-region maximum lies in a negative curvature region. Some details of how these potentials behave can be found in Appendix.\eqref{AppA}. As we discussed before, not all cases will be suitable for studying OTOC.

\begin{figure}[htbp]
    \centering
    \includegraphics[width=1\linewidth]{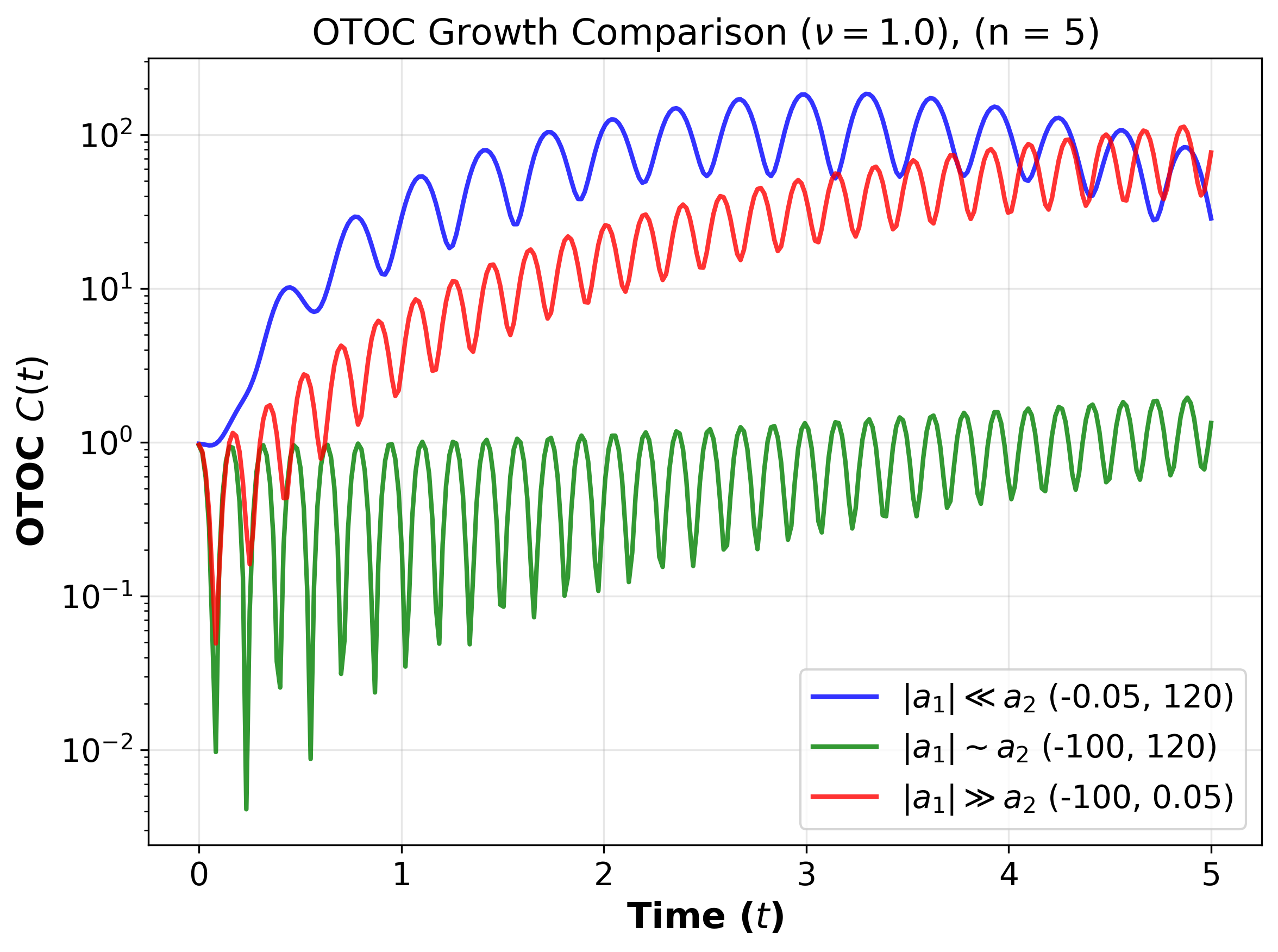}
        
    \caption{Direct comparison of OTOC growth for $\nu=1$ across different parameter regimes in short time frame.}
    \label{fig:otoc_comparison_short_scale_various_a1_a2}
\end{figure}

\begin{figure}[htbp]
    \centering
    \includegraphics[width=1\linewidth]{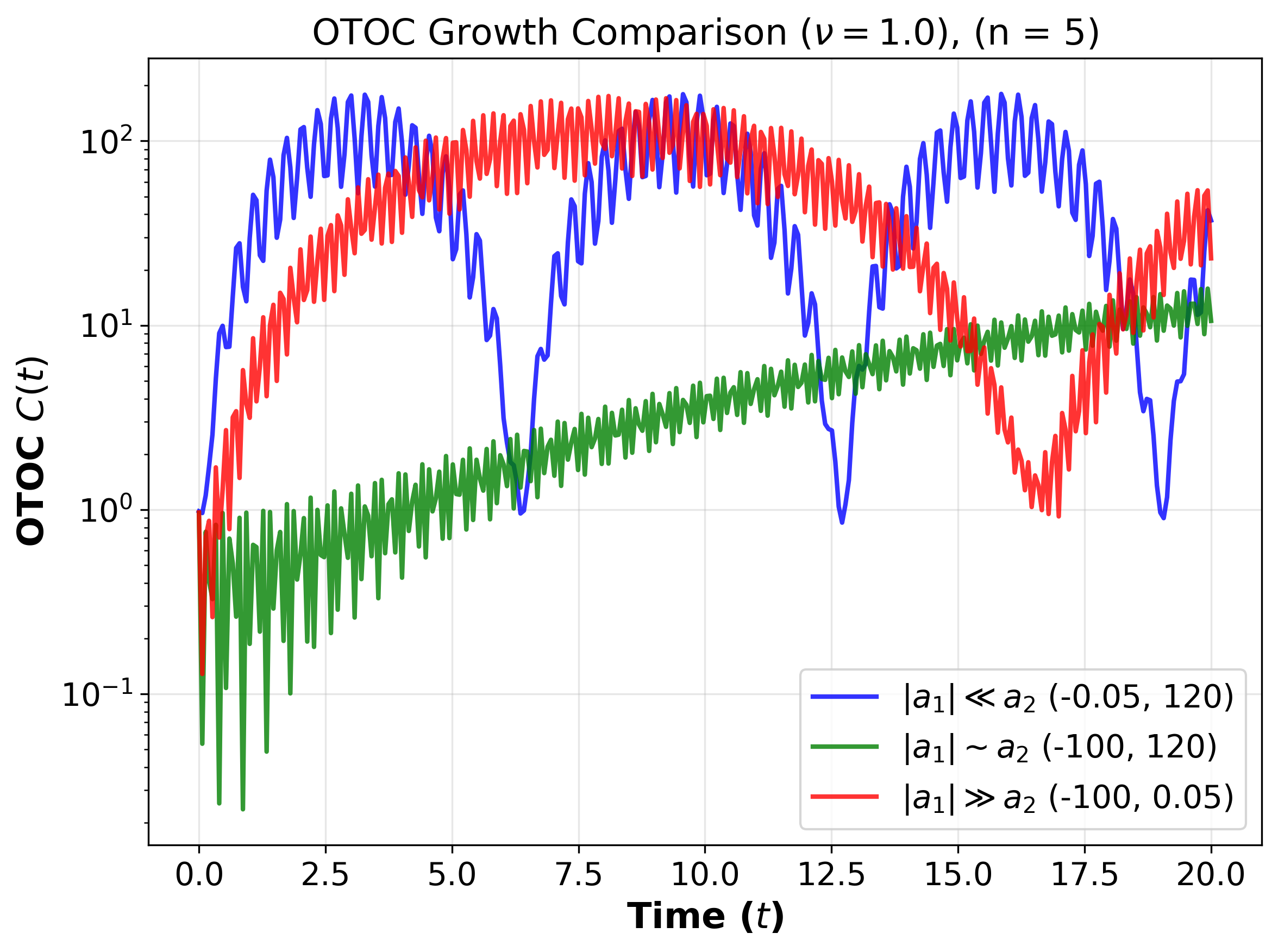}
    \caption{Direct comparison of OTOC growth for $\nu=1$ across different parameter regimes in long time frame.}
    \label{fig:otoc_comparison_long_scale_various_a1_a2}
\end{figure}

For example, we consider three different regimes with $\nu=1$, focusing on the same state $n=5$. In the regime $|a_1|<<a_2$, which is a confining potential, the state $n=5$ has its corresponding $x_m$ located within the negative curvature region. In contrast, for other parameter regimes, such as $|a_1| \sim a_2$ and  $|a_1| >> a_2$ (almost harmonic potentials), the potential does not exhibit a negative curvature region for the same parameter space (see FIG.~\eqref{fig:potential_and_curvature_a1_comp_a2}, ~\eqref{fig:potential_and_curvature_-a1_bigger_a2}). Therefore, the OTOC for the state $n=5$ shows short-time exponential sensitivity at early times only in the $|a_1|<<a_2$ regime, not in other parameter regimes (see FIG.~\eqref{fig:otoc_comparison_short_scale_various_a1_a2}, ~\eqref{fig:otoc_comparison_long_scale_various_a1_a2} for early time and late time growth snapshots for these cases). For all bound states of the first case ($a_1=-0.05, a_2=120)$, comparison between semiclassical, airy-corrected semiclassical sensitivities and early-time quantum OTOC growth rate can be seen in FIG.\eqref{fig:sensitivities_comparison}.

\section{More on unbounded asymptotics}\label{AppE}
For $|x|\to\infty$ with $a_1>0,\nu>0$, we can see that the $\cosh^{2\nu}x$ term dominates:
\begin{equation}
  V(x)
  \sim
  -
  \frac{a_1}{2^{2\nu}}
  e^{2\nu|x|},
  \qquad
   V(x)\to-\infty
   \label{eq:V-infinity}
\end{equation}
So this is the unbounded regime. Our conventions remain same as other Appendices, $\hbar=\mu=1$. For $x\to+\infty$ the Schr\"odinger equation becomes
\begin{equation}
  \psi''
  +
  \frac{2a_1}{2^{2\nu}}
  e^{2\nu x}\psi
  \simeq
  0.
  \label{eq:asymptotic-right}
\end{equation}
Introduce the variable
\begin{equation}
  q_+
  =
  \frac{\sqrt{a_1}}{\nu 2^\nu}
  e^{\nu x}.
  \label{eq:qplus}
\end{equation}
Then $\frac{\dd q_+}{\dd x} = \nu q_+$, and hence \eqref{eq:asymptotic-right} can be written in terms of derivatives with respect to $q_+$:
\begin{equation}
  q_+^2\psi_{q_+q_+}
  +
  q_+\psi_{q_+}
  +
  2q_+^2\psi
  =
  0.
  \label{eq:bessel-zero}
\end{equation}
This is Bessel's equation of order zero in the variable $\sqrt{2}\,q_+$. Thus we can write the asymptotic form of the wavefunctions in terms of Hankel functions:
\begin{equation}
  \psi(x)
  \sim
  C_+H_0^{(1)}(\sqrt{2}\,q_+)
  +
  C_-H_0^{(2)}(\sqrt{2}\,q_+).
  \label{eq:hankel-right}
\end{equation}
For large argument,
\begin{equation}
  H_0^{(1,2)}(z)
  \sim
  \sqrt{\frac{2}{\pi z}}
  \exp\left[
    \pm i\left(z-\frac{\pi}{4}\right)
  \right].
\end{equation}
Therefore, we have the final form
\begin{equation}
  \psi(x)
  \sim
  e^{-\nu x/2}
  \exp\left[
    \pm i
    \frac{\sqrt{2a_1}}{\nu2^\nu}
    e^{\nu x}
  \right].
  \label{eq:asymptotic-wave-right}
\end{equation}
Similarly, at $x\to-\infty$ define the new coordinate:
\begin{equation}
  q_-
  =
  \frac{\sqrt{a_1}}{\nu2^\nu}
  e^{-\nu x}.
\end{equation}
Then the asymptotic solutions behave as
\begin{equation}
  \psi(x)
  \sim
  e^{+\nu x/2}
  \exp(\pm i \sqrt{2}\,q_-).
  \label{eq:asymptotic-left}
\end{equation}
Thus both independent asymptotic oscillatory solutions are
square-integrable for $\nu>0$.
What is the classical dynamics corresponding to such an unbounded situation? The classical equation of motion is
\begin{equation}
\ddot{x} = -V'(x) = \frac{a_1 \nu}{2^{2\nu-1}} e^{2\nu x}.
\end{equation}
Multiplying by $\dot{x}$ and integrating once gives
\begin{equation}
\frac{1}{2}\dot{x}^2 = \frac{a_1}{2^{2\nu}} e^{2\nu x} + \text{const}.
\end{equation}
For large $x$, the constant is negligible, so
\begin{equation}
\dot{x} \simeq \frac{\sqrt{2a_1}}{2^\nu} e^{\nu x} \equiv B e^{\nu x},
\qquad B = \frac{\sqrt{2a_1}}{2^\nu}.
\label{eq:velocity-asymptotic}
\end{equation}
Integrating \eqref{eq:velocity-asymptotic}, we get
\begin{equation}
e^{\nu x(t)}
=
\frac{1}{e^{-\nu x_0}-\nu B t}.
\end{equation}
So, the particle reaches $x = +\infty$ at the finite time
\begin{equation}
t_* = \frac{e^{-\nu x_0}}{\nu B} = \frac{2^\nu e^{-\nu x_0}}{\nu\sqrt{2a_1}}.
\label{eq:tstar}
\end{equation}
That is, the potential is so steep that the particle escapes to infinity in finite time. The deviation $\delta x(t)$ from a central classical trajectory satisfies
\begin{equation}
\delta\ddot{x} = -V''(x_{\text{cl}}(t))\,\delta x = |V''(x_{\text{cl}}(t))|\,\delta x,
\end{equation}
since $V'' < 0$ in this region. The curvature is
\begin{equation}
V''(x) \simeq -\frac{4a_1\nu^2}{2^{2\nu}} e^{2\nu x}.
\end{equation}
Using the classical solution from before, and defining $\tau = t_* - t$, we find
\begin{equation}
|V''(t)| = \frac{4a_1\nu^2/2^{2\nu}}{\nu^2 B^2 \tau^2}
= \frac{4a_1/2^{2\nu}}{B^2 \tau^2}.
\end{equation}
Since $B^2 = 2a_1/2^{2\nu}$, this simplifies to
\begin{equation}
|V''(t)| = \frac{2}{\tau^2}.
\end{equation}
The linearized equation becomes
\begin{equation}
\frac{d^2\delta x}{d\tau^2} = \frac{2}{\tau^2}\,\delta x.
\label{eq:euler}
\end{equation}
This is an Euler equation with solutions $\delta x \propto \tau^\alpha$, where
\begin{equation}
\alpha(\alpha-1) = 2 \quad \Longrightarrow \quad \alpha = 2 \text{ or } \alpha = -1.
\end{equation}
The growing mode is $\delta x \propto \tau^{-1}$. Therefore, the classical sensitivity is
\begin{equation}
\frac{\partial x(t)}{\partial x_0} = \frac{t_*}{t_* - t}.
\label{eq:classical-sensitivity}
\end{equation}
This is a power-law divergence, and the OTOC in the semiclassical limit has the power-law structure
\begin{equation}
C(t) \sim \hbar^2 \left(\frac{\partial x(t)}{\partial x_0}\right)^2
= \hbar^2 \left(\frac{t_*}{t_* - t}\right)^2.
\label{eq:otoc-classical}
\end{equation}
This is a distinctive sign of a finite time singularity.

\section{Critical state scaling for P\"oschl--Teller}\label{appcritical}
We want to expand the Airy-corrected turning point estimate $q_m$ around the classical turning point estimate $q_n$ in the following way
\begin{equation}
    q_m
=
q_n+\delta q
+
\mathcal{O}\!\left((\alpha\Delta(n))^2\right),
\end{equation}
since $\alpha\Delta(n)$ is the dimensionless quantity. Assuming the real line distance between the turning estimates to be small, i.e. $\delta x
=
x_m-x_c
=
-\eta_A\Delta(n),$
the corresponding first-order shift is
\begin{align}
\delta q
=
\left.
\frac{dq}{dx}
\right|_{x_c}
\delta x
=
2\eta_A\alpha\Delta(n)
q_n\sqrt{1-q_n}.
\end{align}
Substituting \eqref{eq:alpha-Delta} gives the first order formula:
\begin{equation}
    \delta q
=
\eta_A\,2^{2/3}
[\Lambda(\Lambda+1)]^{-1/3}
q_n^{2/3}(1-q_n)^{1/3}.
\end{equation}
One may write
\begin{equation}
q_m
\simeq
q_n+
\beta_A
q_n^{2/3}(1-q_n)^{1/3},~~~\beta_A
\equiv
\frac{
\eta_A 2^{2/3}
}{
[\Lambda(\Lambda+1)]^{1/3}
},
\label{eq:qm-beta}
\end{equation}
Now the interesting regime is when $q_n\ll1$ ($n$ approaches $\Lambda$) in the threshold region, which alongwith $q_n=
\frac{\kappa_n^2}
{\Lambda(\Lambda+1)}$ can be used to perturbatively expand the Airy-corrected instability rate $\lambda_{sc}^{(A)}{(n)}$:
\begin{equation}
\lambda_{sc}^{(A)}(n)
\simeq
\sqrt{2}\,
\frac{\hbar\alpha^2}{\mu}
\kappa_n.
\label{eq:classical-nearthreshold}
\end{equation}
We also note that the ratio of the first-order Airy shift to the turning-point value is
\begin{equation}
\frac{\delta q}{q_n}
\simeq
\eta_A
\left(
\frac{2}{\kappa_n}
\right)^{2/3},
\end{equation}
which sets the scale for $\Delta(n)$. Given this, if the for some $n$, the Airy correction dominates over $q_n$, one obtains:
\begin{equation}
\lambda_{sc}^{(A)}(n)
\sim
2^{5/6}\eta_A^{1/2}
\frac{\hbar\alpha^2}{\mu}
\kappa_n^{2/3}.
\label{eq:formal-two-thirds}
\end{equation}
The Airy-dominated condition
$\delta q\gtrsim q_n$ occurs for $\kappa_n$ of order unity, precisely
where the turning-point linearization is becoming marginal. The
$\kappa_n^{2/3}$ behaviour is therefore best regarded as a \emph{formal crossover scaling} of the first-order Airy estimate. This is actually progressively less controllable as we approach the dissociation threshold.

Finally, the critical state crossover, when we see the value of $V''$ at both $x_c$ and $x_m$ are similar:
\begin{align}
    n_c^{\rm cl}
&=
\min
\left\{
n\in\mathbb{N}_0:
n<\Lambda,\;
q_n<\frac23
\right\}\nonumber 
\\ n_{c}^{\rm sc}
&=
\min
\left\{
n\in\mathbb{N}_0:
n<\Lambda,\;
q_m(n)<\frac23
\right\}.
\end{align}
At first order in the Airy displacement thus becomes
\begin{equation}
q_n
+
\beta_A
q_n^{2/3}(1-q_n)^{1/3}
<
\frac23.
\label{eq:critical-linearized}
\end{equation}
One could now put values commensurate with the PT (or shifted PT) wells discussed in the main text to verify semiclassical and Airy-corrected semiclassical crossover states.

\section{Exact matrix elements for P\"oschl--Teller}
\label{appmatrixpt}
Using \eqref{eq:PT-z-wavefunction}, the matrix element involving Gegenbauer polynomial convolution gives
\begin{align}
x_{mn}
&=
\frac{\mathcal{N}_m\mathcal{N}_n}{\alpha^2}
\int_{-1}^{1}
\operatorname{arctanh}z\,
(1-z^2)^{\frac{\kappa_m+\kappa_n}{2}-1}
\nonumber\\
&\hspace{2cm}\times
C_m^{\kappa_m+\frac12}(z)
C_n^{\kappa_n+\frac12}(z)\,dz.
\label{eq:PT-xmn-z}
\end{align}
Here again $\kappa_n = \Lambda-n$.
Using \eqref{eq:PT-N-scaling},
\begin{equation}
\frac{\mathcal{N}_m\mathcal{N}_n}{\alpha^2}
=
\frac{
\mathcal{M}_m(\Lambda)\mathcal{M}_n(\Lambda)
}{\alpha},
\end{equation}
which again explicitly demonstrates the exact scaling
$x_{mn}\propto1/\alpha$. For neighbouring states, $m=n+1$, we have $\kappa_n=\Lambda-n$, hence
$\kappa_{n+1}=\kappa_n-1$,
so that
\begin{align}
x_{n,n+1}
&=
\frac{
\mathcal{M}_n\mathcal{M}_{n+1}
}{\alpha}
\int_{-1}^{1}
\operatorname{arctanh}z\,
(1-z^2)^{\kappa_n-\frac32}
\nonumber\\
&\hspace{1.5cm}\times
C_n^{\kappa_n+\frac12}(z)
C_{n+1}^{\kappa_n-\frac12}(z)\,dz.
\label{eq:PT-xnn1-integral}
\end{align}
Since $C_n^\lambda(-z)
=
(-1)^nC_n^\lambda(z),$ and $\operatorname{arctanh}(-z)
=
-\operatorname{arctanh}z,$, the integrand has parity $(-1)^{m+n+1}$. Then for $m+n$ odd, including $m=n\pm1$, the integral is even and does not vanish. Now consider products like
\begin{equation}
C_m^{\kappa_m+\frac12}(z)
C_n^{\kappa_n+\frac12}(z)
=
\sum_{r=0}^{m+n}
d_r^{(mn)}z^r,
\label{eq:PT-polynomial-product}
\end{equation}
where the coefficients $d_r^{(mn)}$ are finite combinations of Gamma
functions, and have the quite messy structure that one can read off by multiplying the two general finite expansions of Gegenbauer polynomials \cite{Szego1975,DLMF}, using:
\begin{equation}\label{gegenman}
    C_N^\lambda(z)
=
\sum_{j=0}^{\lfloor N/2\rfloor}
(-1)^j
\frac{
2^{N-2j}
\Gamma(N-j+\lambda)
}{
\Gamma(\lambda)\,
j!\,
(N-2j)!
}
z^{N-2j}.
\end{equation}
and collecting powers in \eqref{eq:PT-polynomial-product}:

    \begin{equation}
        \begin{aligned}
d_r^{(mn)}
&=
\frac{
(-1)^{(m+n-r)/2}2^r
}{
\Gamma\left(\Lambda-m+\frac12\right)
\Gamma\left(\Lambda-n+\frac12\right)
}
\\
&\quad\times
\sum_{\substack{
j+k=(m+n-r)/2\\
0\leq j\leq\lfloor m/2\rfloor\\
0\leq k\leq\lfloor n/2\rfloor
}}
\frac{
\Gamma\left(\Lambda-j+\frac12\right)
\Gamma\left(\Lambda-k+\frac12\right)
}{
j!\,k!\,
(m-2j)!\,
(n-2k)!
}.
\end{aligned}
    \end{equation}
Note that 
\begin{equation}
    d_r^{(mn)}=0
\qquad
\text{if $r$ and $m+n$ have opposite parity}.
\end{equation}
Equation \eqref{eq:PT-xmn-z} therefore reduces to
\begin{equation}
x_{mn}
=
\frac{
\mathcal{M}_m\mathcal{M}_n
}{\alpha}
\sum_{r=0}^{m+n}
d_r^{(mn)}
I_r(a_{mn}),
\label{eq:PT-xmn-finite-sum}
\end{equation}
where
\begin{equation}
a_{mn}
\equiv
\frac{\kappa_m+\kappa_n}{2}-1
=
\Lambda-\frac{m+n}{2}-1,
\label{eq:PT-a-mn}
\end{equation}
and
\begin{equation}
I_r(a)
\equiv
\int_{-1}^{1}
z^r(1-z^2)^a
\operatorname{arctanh}z\,dz.
\label{eq:PT-Ir-def}
\end{equation}
By parity, $I_{2j}(a)=0$ and only the odd moments
 $I_{2j+1}(a)$
need to be evaluated. For example, the lowest moment is particularly simple:
\begin{equation}
I_1(a)
=
\int_{-1}^{1}
z(1-z^2)^a
\operatorname{arctanh}z\,dz.
\end{equation}
Using
\begin{equation}
z(1-z^2)^a
=
-\frac{1}{2(a+1)}
\frac{d}{dz}(1-z^2)^{a+1},
\end{equation}
integration by parts gives
\begin{align}
I_1(a)
=
\frac{1}{2(a+1)}
\int_{-1}^{1}
(1-z^2)^a\,dz
=
\frac{1}{2(a+1)}
B\left(\frac12,a+1\right).
\label{eq:PT-I1}
\end{align}
Thus
\begin{equation}
I_1(a)
=
\frac{
B\left(\frac12,a+1\right)
}{
2(a+1)
}.
\end{equation}
More generally, the odd moments can be reduced recursively, using $I_{2j+1}(a)
=
\sum_{s=0}^{j}
(-1)^s
\binom{j}{s}
I_1(a+s).$ Hence, 
\begin{equation}
I_{2j+1}(a)
=
\frac12
\sum_{s=0}^{j}
(-1)^s
\binom{j}{s}
\frac{
B\left(\frac12,a+s+1\right)
}{
a+s+1
}.
\label{eq:PT-Iodd-final}
\end{equation}
Combining
\eqref{eq:PT-xmn-finite-sum} and
\eqref{eq:PT-Iodd-final}, we obtain
\begin{equation}
x_{mn}
=
\frac{
\mathcal{M}_m(\Lambda)
\mathcal{M}_n(\Lambda)
}{\alpha}
\sum_{\substack{r=1\\r\ {\rm odd}}}^{m+n}
d_r^{(mn)}
I_r(a_{mn}),
\label{eq:PT-xmn-closed}
\end{equation}
where $a_{mn}
=
\Lambda-\frac{m+n}{2}-1$,
and the $I_r$ are given explicitly by
\eqref{eq:PT-Iodd-final}. Hence the exact bound-bound position matrix
elements are analytically calculable as finite combinations of Beta
and Gamma functions. 

Further, using the expansion in \eqref{gegenman}, we can also work out the normalization condition for wavefunctions, which becomes
\begin{equation}
1
=
\frac{\mathcal{N}_n^2}{\alpha}
\int_{-1}^{1}(1-z^2)^{\kappa_n-1}
\bigl[C_n^{\kappa_n+\frac12}(z)\bigr]^2\,dz.
\label{eq:norm-xspace}
\end{equation}
The dimensionless constant $\mathcal{M}_n=\mathcal{N}_n/\sqrt{\alpha}$ that appears in the analytic matrix element \footnote{As a consistency check, for $n=0$ and in the deep well approximation this gives
\begin{equation*}
M_0^{\,2}
=
\frac{\Gamma\!\left(\Lambda+\tfrac12\right)}
{\sqrt{\pi}\,\Gamma(\Lambda)}
\xrightarrow{\Lambda\gg1}
\frac{\sqrt{\Lambda}}{\sqrt{\pi}},
\end{equation*}
which matches the harmonic-oscillator ground-state normalization $\sqrt{\omega/\pi}/\alpha$ with $\omega=\alpha^2\sqrt{\Lambda(\Lambda+1)}\approx\alpha^2\Lambda$.}:
\begin{equation}
\mathcal{M}_n^{\,2}
=
\frac{
n!\;\Gamma\!\left(\Lambda-n+\tfrac12\right)\;
\Gamma\!\left(2\Lambda-2n+1\right)
}{
\sqrt{\pi}\;\Gamma(\Lambda-n)\;
\Gamma\!\left(2\Lambda-n+1\right)
}.
\label{eq:Mn-xspace}
\end{equation}
\vspace{0.5cm}

Once we have the exact formula, we use \eqref{eq:OTOC-def} to calculate OTOCs, for example in the diagonal channel case, it simply reads:
\begin{align}
A_{nn}(t)
&=
e^{i(E_n-E_{n+1})t/\hbar}
x_{n,n+1}p_{n+1,n}
\nonumber\\
&\quad+
e^{i(E_n-E_{n-1})t/\hbar}
x_{n,n-1}p_{n-1,n}
\nonumber\\
&\quad-
p_{n,n+1}x_{n+1,n}
e^{i(E_{n+1}-E_n)t/\hbar}
\nonumber\\
&\quad-
p_{n,n-1}x_{n-1,n}
e^{i(E_{n-1}-E_n)t/\hbar}.
\label{eq:PT-Ann-raw}
\end{align}
We could now put $p_{n,n+1}
=
-i\mu\omega_+x_+$ etc. to get the final expression.

\bibliography{custom}

\end{document}